\documentclass[pre,amsmath,aps,showpacs,twocolumn,superscriptaddress]{revtex4}

\usepackage{graphicx,xspace}% Include figure files
\usepackage{dcolumn}% Align table columns on decimal point
\usepackage{bm}% bold math
\usepackage[utf8]{inputenc}
\usepackage[T1]{fontenc}
\usepackage{mathptmx}
\usepackage{float}
\usepackage{amsmath}
\usepackage{amssymb}
\usepackage[normalem]{ulem}
\usepackage[dvipsnames]{xcolor}
\usepackage{footnote}
\usepackage{mathtools}
\usepackage{comment} % to hide text
\input{epsf}

\begin{document}

%\title{Modeling ionic microgel suspensions: Deswelling, structure, and dynamics}
%\title{Modeling ionic microgel suspensions: How deswelling affects thermodynamics, structure, and dynamics}
%\title{Modeling ionic microgel suspensions: How does deswelling affect static and dynamic properties?}
%\title{Modeling ionic microgel suspensions: Influence of deswelling on thermodynamics, structure, and dynamics}
\title{Modeling deswelling, thermodynamics, structure, and dynamics in ionic microgel suspensions}

\author{Mariano E. Brito}
\affiliation{Institute of Complex Systems, ICS-3,
Forschungszentrum J\"ulich GmbH, 52425 J\"ulich, Germany}

\author{Alan R. Denton}
\affiliation{Department of Physics, North Dakota State University, Fargo, ND 58108-6050 USA}

\author{Gerhard N\"agele}
\affiliation{Institute of Complex Systems, ICS-3,
Forschungszentrum J\"ulich GmbH, 52425 J\"ulich, Germany}

\date{\today}

\begin{abstract}
Ionic microgel particles in a good solvent swell to an equilibrium size determined by a balance of electrostatic and elastic forces. When crowded, ionic microgels deswell owing to a redistribution of microions inside and outside the particles. The concentration-dependent deswelling affects the interactions between the microgels, and consequently the suspension properties.
We present a comprehensive theoretical study of crowding effects on thermodynamic, structural, and dynamic properties of weakly cross-linked ionic microgels in a good solvent. The microgels are modeled as microion- and solvent-permeable colloidal spheres with fixed charge uniformly distributed over the polymer gel backbone, whose elastic and solvent-interaction free energies are described using Flory-Rehner theory. Two mean-field methods for calculating the crowding-dependent microgel radius are investigated, and combined with calculations of the net microgel charge characterizing the electrostatic part of an effective microgel pair potential, with charge renormalization accounted for. Using this effective pair potential, thermodynamic and static suspension properties are calculated including the osmotic pressure and microgel pair distribution function. The latter is used in our calculations of dynamic suspension properties, where we account for hydrodynamic interactions.  Results for diffusion and rheological properties are presented over ranges of microgel concentration and charge. We show that deswelling mildly enhances self- and collective diffusion and the osmotic pressure, lowers the suspension viscosity, and significantly shifts the suspension crystallization point to higher concentrations. The paper presents a bottom-up approach to efficiently computing suspension properties of crowded ionic microgels using single-particle characteristics.
\end{abstract}

%\pacs{75.60.Ej, 75.60.Ch, 74.25.Qt}

\maketitle 

\section{Introduction}%%%%%%%%%%%%%%%%%%%%%%%%%%%%%%%%%%%%%%%%%%%%%%%%%%%%%%%%%%%%%%%%%%%%%%%%%%%%%%%%%%%%%%%%%%%%%%%%%%%%%%%%%%
Mesoscopically sized microgel particles are of both fundamental and technological interest owing to their strong sensitivity to external parameters, including temperature (solvent quality), pH, ionic strength, and concentration, which control their equilibrium size and soft interactions \cite{Book_Fernandez_Nieves:2011,Lyon_Fernandez_Nieves:2012}. Because of their profound environmental adaptability, and capacity to partially interpenetrate and host small molecular species, microgels have a multitude of possible applications, e.g., as drug-delivery vehicles, functionalized colloids, switchable membrane filters, and tunable microreactors \cite{Plamper_Richtering:2017}.
%reference?

An important subgroup is formed by ionic microgels, which are typically globular particles consisting of cross-linked polyelectrolyte chains. When dispersed in a polar solvent under good solvent conditions, an ionic microgel particle becomes charged due to the dissociation of counterions from ionizable groups on the polymer backbone. For weak cross-linking, the soft particle swells to an equilibrium size that can be substantially larger than that in the dry state. It has been shown experimentally \cite{Holmqvist_PRL_2012,Nojd_SoftMatter_2018,Gasser_PRE_2019} and theoretically \cite{weitz-jcp2012,Colla_JCP_2014,DentonTang_JCP_2016,Weyer_SoftMatter_2018,Gasser_PRE_2019,Hofzumhaus_SoftMatter_2018,Nojd_SoftMatter_2018} that the particles deswell with increasing concentration, which is most pronounced for low background ionic strength and smaller concentrations well below the threshold where the microgels start to overlap. This  behavior distinguishes ionic microgels from nonionic ones, since for the latter deswelling (or interpenetration and facetting) is observed only at very high concentrations \cite{Urich_SoftMatter_2016,Mohanty:2017}.

In theory-simulation studies by Denton {\em et al.} \cite{DentonTang_JCP_2016,Weyer_SoftMatter_2018}, where the interplay of elastic and electrostatic influences is accounted for using a coarse-grained model, microgel deswelling with increasing concentration is explained by a redistribution of counterions. These ions increasingly permeate the microgel, to maintain balance of the electro-elastic pressure inside and outside the particles. The shrinkage of microgels with increasing concentration is thus accompanied by a decreasing net microgel charge and narrowing of the particle size distribution around the equilibrium mean radius. At low background ionic strength (i.e., low salt content), the mean counterion concentration outside a microgel is much smaller than inside, with an accordingly strong sensitivity of the mean microgel size to concentration changes. With increasing salt content, the strong inside-outside counterion concentration gradient, and the concomitant ion pressure gradient, are flattened out, reducing the sensitivity of the microgel size to concentration variations.

Owing to the counterion-induced deswelling, sufficiently soft ionic microgels can penetrate apertures considerably narrower than their dilute-concentration size, at concentrations below particle overlap \cite{Nir_SoftMatter_2016}. This ability has potentially important applications for drug delivery, microfluidics and filtration. For example in pressure-driven membrane filtration used to concentrate and purify microgels, deswelling can lead to an unwarranted enhanced clogging of membrane pores. On the other hand, deswelling reduces the formation of a fluid concentration-polarization layer and a solid filter cake layer of particles accumulated on the membrane surface, which cause additional (effective) hydraulic resistance. The formation of these inhomogeneous layers is determined by concentration-dependent transport properties of crowded microgel suspensions in conjunction with osmotic pressure effects, namely by the collective diffusion coefficient and the suspension viscosity \cite{Roa_SoftMatter_2015,Roa_SoftMatter_2016,Park:2019}.

The filtration example illustrates the demand for studying diffusion and rheological transport properties of soft ionic microgel suspensions in general, and the effects of counterion-regulated deswelling in particular. In this paper, we present a comprehensive theoretical exploration of dynamic and equilibrium microstructural properties 
of fluid-phase suspensions of ionic microgels in the swollen state. Being of interest in their own right, microstructural properties such as the radial distribution function (rdf) and static structure factor are 
also required as input in the calculation of dynamic suspension properties, including generalized sedimentation and collective diffusion coefficients and the high-frequency and zero-frequency viscosities. Following earlier work by Denton {\em et al.} \cite{Denton_PRE_2003,Hedrick_JCP_2015,DentonTang_JCP_2016,Weyer_SoftMatter_2018}, we model the ionic microgels in a coarse-grained way as microion- and solvent-permeable monodisperse elastic colloidal spheres, with the charged sites of the cross-linked polymer gel backbone described by a uniform charge distribution. This description is reasonable, under the proviso that the cross-linker density does not vary strongly along the particle radius. We describe the elastic and solvent-interaction free energy contributions of a microgel using Flory-Rehner theory \cite{flory-rehner1943-I,flory-rehner1943-II,FloryBook} for uniform cross-linker distribution. For calculating the electrostatic semi-grand free energy contribution of microgels in a concentrated suspension, 
in Donnan equilibrium with a 1:1 strong electrolyte reservoir, we use two different mean-field methods, namely, the spherical Poisson-Boltzmann cell model (PBCM) approach of Denton and Tang \cite{DentonTang_JCP_2016} and a first-order thermodynamic perturbation theory (TPT) method of Weyer and Denton \cite{Weyer_SoftMatter_2018}, based on a multi-center linear-response approach. The equilibrium microgel radius is obtained from minimizing the total suspension free energy, equivalent to enforcing the balance of total pressure inside and outside a particle. In combination with an effective electrostatic pair potential expression for ionic microgels derived from the multi-center approach \cite{Denton_PRE_2003,Gottwald_JCP_2005,Riest_ZPhysChem_2012}, we determine the pressure and osmotic compressibility of microgel suspensions as well as the microgel pair distribution function and static  
structure factor using the hypernetted chain (HNC) and thermodynamically self-consistent Rogers-Young (RY) integral-equation methods \cite{Hansen-McDonald}. For non-overlapping particles, the effective pair potential is of a screened-Coulomb form, akin to the potential for ion-impermeable charge-stabilized colloidal particles, 
but with a coupling strength that decreases with increasing concentration and ionic strength of the suspension.  
The net microgel charge likewise decreases with increasing concentration. For overlapping microgels, 
the effective electrostatic potential remains finite, and is augmented in our model by a soft Hertz potential accounting for elastic repulsion at modest overlap \cite{LandauLifschitzElasticity,Rovigatti_HertzModel2019}. Van der Waals attraction between the weakly cross-linked microgels can be neglected due to their high solvent content. 

The microgel pair distribution function is used as input to our calculations of dynamic suspension properties. Semi-analytic methods are used to calculate dynamic properties, whose good performance has been established, by comparison with elaborate dynamic computer simulations, 
for a variety of colloidal model systems describing globular proteins, impermeable charge-stabilized colloids, and non-ionic spherical microgels. These methods account for the salient hydrodynamic particle interactions mediated by intervening solvent flow. In our assessment of deswelling effects, the results obtained for various static and dynamic suspension properties are compared with the ones for a (fictitious) reference suspension of constant-sized microgels. 

The paper is structured as follows: In Sec.~\ref{Sec:OCM}, we review the derivation of the effective one-component model of ionic microgels by integrating out the microion and particle-internal polymer degrees of freedom, resulting in a state-dependent effective pair potential and volume pressure contribution.
Section~\ref{Sec:Deswelling} gives the essentials of the linear TPT and nonlinear PBCM methods used for calculating the equilibrium microgel radius  as a function of concentration, backbone charge, and reservoir ionic strength, and further describes how the net microgel charge and the electrostatic screening constant characterizing the pair potential are obtained. The methods used for calculating structural and thermodynamic properties 
are discussed in Sec.~\ref{sec:StructureThermodynamics}. The diffusion and rheological properties explored in this work are summarized in Sec.~\ref{Sec:Dynamics}, together with the analytic methods for their calculation within the one-component model framework. In Sec.~\ref{Sec:Results}, 
results are presented for static properties, including the microgel swelling ratio, pair distribution function, and osmotic pressure, as well as for dynamic properties, including the hydrodynamic function, collective diffusion coefficient, and low- and high-frequency suspension viscosities. Finally, Sec.~\ref{Sec:Conclusions} summarizes our conclusions.

\section{Effective One-Component Model}\label{Sec:OCM}

We describe here the employed microgel model and the essential steps of tracing out the microion and polymer-backbone monomeric degrees of freedom, leading to an effective one-component suspension description of pseudo-microgels interacting via a state-dependent effective pair potential \cite{DentonTang_JCP_2016,Urich_SoftMatter_2016,Weyer_SoftMatter_2018}. This potential determines the equilibrium microstructure of the microgels and, in conjunction with a structure-independent volume grand free energy contribution, also the (osmotic) thermodynamic properties of the whole suspension, including its phase behavior. For conciseness, we use the compact notation of \cite{Weyer_SoftMatter_2018}, to which we refer for further details, focusing here on the physical aspects.    

The distributions of backbone polyelectrolyte monomers, cross-linkers, and backbone charges of ionic microgels depend on the synthesis method. For simplicity, we assume here uniform distributions \cite{Weyer_SoftMatter_2018}. The charged microgel backbone polymers and cross-linkers coexist with polymer-released counterions and salt ions dissolved in the solvent. For temperatures $T>T_\text{cr}$ higher than the lower critical solution temperature (LCST) $T_\text{cr}$ of the corresponding polymer solution, the microgels are collapsed into a dry state, characterized by a dry radius $a_0<a$, where $a$ is the microgel equilibrium radius of the swollen microgel at a temperature lower than the LCST. The swollen microgel radius depends, in addition to temperature and solvent quality, on the elastic properties of the backbone network and the backbone charge, and furthermore on the microgel concentration and background (reservoir) ionic strength. Two methods used in this paper to calculate the swelling ratio $\alpha=a/a_0$ are described in the next section (Sec.~\ref{Sec:Deswelling}). Assuming that a single microgel consists of a uniform polymer network with $N_\text{mon}$ monomers, the dry microgel radius $a_0$ is well approximated by $a_0\approx (N_\text{mon}/\phi_\text{rcp})^{1/3}a_\text{mon}$, where $\phi_\text{rcp}=0.64$ is the volume fraction for random close-packing of spherical monomers and $a_\text{mon}$ is the monomer radius. It is assumed here that random close-packing is the unstressed polymer backbone structure in the collapsed state \cite{flory-rehner1943-I,flory-rehner1943-II,FloryBook,Colla_JCP_2014}.
	
We consider in this paper a monodisperse microgel suspension formed by $N$ spherical microgels, each of negative backbone charge $-Z e$, with $e$ the proton charge, dispersed in a volume $V$ of water at room temperature $T$. The suspension is assumed to be in (Donnan) osmotic equilibrium with a 1:1 strong electrolyte reservoir of ion concentration $2n_\text{res}$, via an ideal membrane permeable to the microions and solvent only. The counterions dissociated from the backbones are likewise taken as monodisperse. The microgel concentration (number density) $n=N/V$ determines the volume fraction $\phi_0 =  4\pi a_0^3n/3$ of dry microgels and the volume fraction $\phi =  4\pi a^3n/3$ of swollen microgels. The dry volume fraction should be thought of as a non-dimensionalized microgel concentration. For simplicity, the backbone valence $Z>0$ is assumed to be constant, independent of concentration, ionic strength, and equilibrium radius, thus disregarding possible chemical charge regulation effects. Here, $Z$ should be viewed as 
net backbone valence, already accounting for the possibility of Manning counterion condensation on polymer sites. Global electroneutrality implies $ZN = \langle N_{+}\rangle-\langle N_{-}\rangle$, where $N_\text{s}=\langle N_{-}\rangle$ is the equilibrium number of monodisperse coions in the system, equal to the number $N_\text{s}$ of salt ion pairs, and $\langle N_{+}\rangle$ is the equilibrium number of monovalent counterions. The concentration (number density) $n_\text{s}=N_\text{s}/V$ of salt ion pairs in the suspension is determined by the equality, $\mu_\pm=\mu_\text{res}$, of the microion chemical potentials of cations and anions, $\mu_\pm$, in the suspension and the microion chemical potential, $\mu_\text{res}=k_\text{B}T \ln \left(\Lambda_0^3 n_\text{res}\right)$, in the reservoir, assuming equal thermal de Broglie wavelength $\Lambda_0$ for all microions. In Donnan equilibrium, the salt pair concentration $n_\text{s}$ in the suspension is determined by the given reservoir salt pair concentration (number density) $n_\text{res}$. A closed suspension of given salt content can be straightforwardly mapped to an equivalent Donnan equilibrium system using an accordingly selected salt concentration $n_\text{res} \geq n_\text{s}$.    
	
Our starting point in deriving the one-component model of pseudo-microgels is a semi-grand canonical description of uniform-backbone spherical microgels with the solvent degrees of freedom already integrated out. This amounts to describing the solvent statically as a dielectric continuum of dielectric constant $\epsilon$ and Bjerrum length $\lambda_B=e^2/(\epsilon k_\text{B}T)$, and dynamically as a Newtonian solvent of shear viscosity $\eta_0$. In this McMillan-Mayer implicit solvent picture, the semi-grand canonical partition function of the suspension reads     
\begin{equation}
\Xi=\langle\langle\langle \text{e}^{-\beta(K+U_\text{m}+U_\text{mm}+U_{\text{m}\mu}+U_{\mu\mu})}\rangle_\text{p}\rangle_\mu\rangle_\text{m}\,.
\end{equation}
Here $\beta=1/(k_\text{B}T)$, $K$ is the total kinetic energy of all polymeric and ionic suspension constituents, and the angular brackets denote canonical traces over polymer (p) and center-of-mass microgel (m) coordinates, and grand-canonical traces over the microion ($\mu$) coordinates. The polymer coordinates are particle-internal degrees of freedom associated with the motion of segments and associated fixed charges constituting the cross-linked polymer chains. In the Boltzmann factor, $U_\text{m}$ is the single-microgel energy, comprising both polymeric and electrostatic self energies, $U_\text{mm}$ incorporates polymeric and electrostatic energies of interaction between the microgels, and $U_{\text{m}\mu}$ and $U_{\mu\mu}$ account, respectively, for microgel–microion and microion–microion interactions.

Performing the trace over polymer coordinates, one obtains
\begin{equation} 
\Xi=\text{e}^{-\beta(U_\text{e}+F_\text{p})}\langle\langle e^{-\beta(K_{\text{m},\mu}+U_\text{mm}+U_{\text{m}\mu}+U_{\mu\mu})}\rangle_\mu\rangle_\text{m}
\label{coarse-grainedPM}
\end{equation}
where $U_\text{e}$ is the sum of the electrostatic self energies of the $N$ microgels, which for uniformly distributed backbone charges is
\begin{equation}
U_\text{e}(a)=\sum_{i=1}^N u_\text{e}(a)= N\left(\frac{3}{5}\frac{Z^2e^2}{\epsilon a}\right)\,,
\end{equation}
with the equilibrium radius $a$ of swollen microgels. Furthermore, $K_{\text{m},\mu}$ is the translational kinetic energy associated with the center-of-mass microgel (m) and microion ($\mu$) coordinates. 

The free energy associated with the non-electrostatic polymeric degrees of freedom of the $N$ microgels is
\begin{equation}
F_\text{p}=\sum_{i=1}^N f_\text{p}(a)\,.
\end{equation}
We use Flory-Rehner theory \cite{flory-rehner1943-I,flory-rehner1943-II,FloryBook} to approximate the polymer free energy per microgel, $f_\text{p}(a)$, for a particle network with uniformly distributed cross-linkers that is divided into $N_\text{ch}$ chains, i.e.,
\begin{eqnarray}
\beta f_\text{p}(a) &=& N_\text{mon}[(\alpha^3-1)\ln(1-\alpha^{-3})+\chi(1-\alpha^{-3})]+\nonumber\\
&+&\frac{3}{2}N_\text{ch}(\alpha^2-\ln\,\alpha-1)\,,
\label{fee-energy_FloryRehner}
\end{eqnarray}
where $\chi$ is the Flory solvency parameter, $\alpha$ the swelling ratio, and $N_\text{mon}$ the total number of polymer monomers in a microgel. The first term on the right-hand side is the ideal mixing entropy of microgel monomers and solvent molecules. The second term accounts for polymer-solvent interactions in a mean-field approximation, by neglecting interparticle correlations. The last term accounts for the elastic free energy for isotropic stretching of the microgel network, with the polymers treated as Gaussian coils. As argued in \cite{Urich_SoftMatter_2016,Weyer_SoftMatter_2018}, the approximations employed here for the microgel backbone self-energies are reasonable for loosely cross-linked, uniformly structured microgels.   

Tracing out in a second step the microion degrees of freedom for fixed configuration of microgels leads to the expression \cite{Weyer_SoftMatter_2018}  
\begin{equation}
\Xi= \langle \text{e}^{-\beta H_\text{eff}}\rangle_\text{m}\,,
\end{equation}
with the effective Hamiltonian of pseudo-microgels,
\begin{equation}
H_\text{eff} = K_\text{m}+U_\text{e}+F_\text{p}+E_\text{V}(n)+U_\text{eff}(n)\,,
\label{EffHamilt}
\end{equation}
where $K_\text{m}$ accounts for the translational kinetic energy of the microgels, $E_\text{V}(n)$ is the microgel configuration-independent volume energy, and $U_\text{eff}(n)$ is the configuration-dependent effective $N$-particle interaction energy of pseudo-microgels. The latter, which incorporates electrostatic screening by the traced-out microions, consists of the bare interaction energy, $U_\text{mm}$, comprising the concentration-independent Coulomb and elastic inter-microgel interactions, and a concentration- and temperature-dependent contribution, related to the free energy of microions in the presence of the microgels.

With $F= -k_\text{B}T \ln \Xi$ denoting the semi-grand suspension free energy, the pressure, $p$, of the multi-component suspension, consisting of polymer networks with charged sites and microions, is then determined by the generalized one-component virial equation \cite{Hedrick_JCP_2015},
\begin{eqnarray}\label{eq:OsmoticPressure}
 p\!\!&=&\!\! -\left(\frac{\partial F}{\partial V}\right)_{\text{res}}\nonumber\\
 \!\!&=&\!\!p_\text{V} + p_\text{se} + n k_\text{B} T 
 -\frac{1}{3V}\big< \sum_{i=1}^N {\bf r}_i\cdot\frac{\partial U_\text{eff}}{\partial {\bf r}_i}\big>_\text{eff} - \big<\frac{\partial U_\text{eff}}{\partial V}\big>_\text{eff}\,, \nonumber\\ 
\end{eqnarray}
invoking an extra volume derivative term due to the concentration dependence of $U_\text{eff}\left(X;n\right)$. Here, $X=\{{\bf r}_1,\cdots,{\bf r}_N\}$ are the center-of-mass positions of microgels and $p_\text{V} = -\partial E_\text{V}/\partial V$ is the pressure contribution of the volume energy. There is an another pressure contribution, $p_\text{se} = n^2 \partial [u_e(a) + f_\text{p}(a)]/\partial n$, originating from the electrostatic and polymeric self-energies (se) per particle, owing to their implicit concentration dependence via the equilibrium radius $a(n)$. This contribution is absent for incompressible particles. The angular brackets $\langle\cdots\rangle_\text{eff}$ denote the canonical average with respect to the equilibrium distribution function, $P_\text{eq}(X) \propto\exp[-\beta U_\text{eff}(X)]$, of pseudo-microgels, not to be confused with the canonical microgel trace $\langle\cdots\rangle_\text{m}$ over microgel center positions and momenta. The volume derivative of $F$ in Eq.~(\ref{eq:OsmoticPressure}) is for fixed reservoir ion chemical potentials and, hence, fixed $n_\text{res}$. The generalized virial equation does not suffer from ambiguities introduced when state-dependent pair potentials are combined in an {\em ad hoc} manner with the compressibility and virial equation of state expressions for one-component simple liquids \cite{Louis2002,Hoffmann_JCP_2004}.

In Donnan equilibrium, the reduced microgel osmotic compressibility can be expressed via the Kirkwood-Buff (KB) relation \cite{Kirkwood_JCP_1951,Dobnikar2006},
\begin{equation}\label{eq:OsmoticCompressibility}
 k_\text{B} T \left( \frac{\partial n}{\partial p}\right)_{\text{res}}
=1 + n\int d^3r \left[g(r;n)-1\right] =S(q\to 0;n)\,,
\end{equation}
solely in terms of the solvent-averaged microgel radial distribution function $g(r;n)=g_\text{mm}(r;n)$, which in turn is solely determined by the effective interaction energy $U_\text{eff}$. In contrast, $p$ is not determined by $U_\text{eff}(n)$ alone, since it has also a pressure contribution arising from the volume energy. The KB relation follows from the isothermal differential Gibbs-Duhem relation in Donnan equilibrium, $dp = n\;\!d\mu$, where $\mu$ is the microgel chemical potential, in conjunction with the relation $\beta S(0;n)= \left(\partial \ln n /\partial \mu\right)_{T,\mu_\pm}$ for the zero-wavenumber structure factor of the microgels confined to the suspension. The radial distribution function is basically the inverse Fourier transform of the microgel structure factor \cite{Naegele_PhysRep_1996},
\begin{equation}
S(q;n)=1+4\pi n\int_0^\infty\,dr\,r^2\,[g(r;n)-1]\frac{\sin(qr)}{qr}\,,
\end{equation}
determined in static scattering experiments as a function of the scattering wavenumber $q$. The zero-wavenumber limit of $S(q;n)$ is proportional to the osmotic compressibility.

The volume and concentration derivatives, taken in Eqs.~(\ref{eq:OsmoticPressure}) and (\ref{eq:OsmoticCompressibility}) respectively, are for fixed reservoir properties, i.e., fixed $n_\text{res}$ and $T$, so that $\partial p/\partial n=\partial\pi_\text{os}/\partial n$, where
\begin{equation}
\pi_\text{os}=p-2 k_\text{B}T n_\text{res}
\end{equation}
is the osmotic pressure of the suspension, measured relative to the reservoir pressure $p_\text{res}$. 
Non-ideality contributions to the reservoir pressure are negligible for the considered reservoir ionic strengths of {\it monovalent} electrolyte ions. Notice that both the generalized virial equation and the KB relations are valid also for a non-pairwise additive $U_\text{eff}$.

For weak particle overlap, it is reasonable to assume pairwise additive elastic forces which for swollen volume fractions $\phi \lesssim 1$ can be reasonably modeled \cite{Rovigatti_HertzModel2019} by the Hertz pair potential \cite{LandauLifschitzElasticity},
\begin{equation}
\beta u_\text{H}(r) = \begin{cases} \epsilon_\text{H}\left(1-\frac{r}{2a}\right)^{5/2}\,, & r\le 2a \\ 0\,, & r>2a \end{cases}\,,
\end{equation}
where $r$ is the center-to-center separation of two particles. The Hertz soft particle radius is identified with the equilibrium (swollen) radius $a$. The reduced interaction (softness) parameter $\epsilon_\text{H}$ is determined by the single-particle elastic moduli, independent of temperature and particle volume \cite{Riest_ZPhysChem_2012}, and it scales linearly with $N_\text{ch}$. The pairwise-additive bare microgel interaction energy is thus
\begin{equation}
U_\text{mm}(X) = \sum_{i<j}^N\left[u_\text{H}(r_{ij})+u_\text{C}(r_{ij}) \right]\,,
\end{equation}
where $u_\text{C}(r_{ij})$ is the Coulomb interaction energy between microgels $i$ and $j$ at center-to-center distance $r_{ij}$, modeled as the electrostatic energy of two uniform spherical charge clouds, 
each of charge $-Ze$ and radius $a$. 

To obtain the effective potential energy, $U_\text{eff}(X;n)$, of $N$ pseudo-microgels, the concentration-dependent free energy contribution due to the traced out microions needs to be calculated. Assuming weak perturbation of the microion distribution by the uniform microgel backbone charges, this can be done using the linear-response approximation method of Denton \cite{Denton_PRE_2003,Gottwald_JCP_2005,Riest_ZPhysChem_2012}. This method invokes a random phase approximation for the static response functions of a reference plasma of pointlike-assumed microions, resulting in a linear superposition of isotropic coion and counterion concentration profiles (orbitals), $n_\pm(|{\bf r}-{\bf r}_i|)$, centered at the respective microgel positions ${\bf r}_i$. The effective $N$-microgel interaction energy has then only two-body contributions, so that
\begin{equation}
U_\text{eff}(X;n)= \sum_{i<j}^N\left[u_\text{H}(r_{ij})+u_\text{eff}(r_{ij};n)\right]\,.
\end{equation}
The effective electrostatic pair potential in the linear-response approximation is of different functional form for overlapping and non-overlapping microgels, i.e.,
\begin{equation}
	u_\text{eff}(r;n) = \begin{cases} u_\text{Y}(r;n)\,, & r>2a \\ u_\text{ov}(r;n)+u_\text{H}(r)\,, & r\le2a \end{cases}\,.
\label{eff_pot}
\end{equation}
For non-overlapping pairs, the effective electrostatic pair potential has the functional form of a screened-Coulomb potential \cite{Denton_PRE_2003,Hedrick_JCP_2015,Gottwald_JCP_2005},
\begin{equation}\label{Yukawa_part}
	\beta u_\text{Y}(r;n)=\lambda_\text{B}[Z_\text{net}(n)]^2\left(\frac{e^{\kappa a}}{1+\kappa a}\right)^2 \frac{e^{-\kappa r}}{r}\,,
\end{equation}
with net microgel valence 
\begin{equation}\label{eq:NetCharge}
Z_\text{net}=Z\;\!\frac{3\;\!\left(1+\kappa a\right)}{\left(\kappa a\right)^2\;\!e^{\kappa a}}\;\!\left[\cosh(\kappa a) 
- \frac{\sinh(\kappa a)}{\kappa a}\right]\,,
\end{equation}
depending on the product, $\kappa a$, of the Debye screening constant $\kappa$ and the swollen radius $a$. The net microgel valence is obtained from 
\begin{equation}
	Z_\text{net}(n)=Z-4\pi\int_0^{a}[n_+(r)-n_-(r)]r^2dr\,,
\label{Def_Znet}
\end{equation}
using the linear-response prediction for the equilibrium counterion and coion concentration orbitals $n_\pm(r)$. 
The Debye screening constant has the form
\begin{equation}\label{eq:Debye}
\kappa^2 = 4\pi\lambda_\text{B}(Zn+2n_\text{s})\,,
\end{equation}
with dependence on the microgel concentration $n$ and salt pair concentration $n_\text{s}$ in the suspension. 
The somewhat lengthy expression for the repulsive effective electrostatic potential of overlapping microgels, $u_\text{ov}(r;n)$, which depends on $Z$ and $\kappa a$, is given elsewhere \cite{Denton_PRE_2003,Hedrick_JCP_2015,Gottwald_JCP_2005}. As an ultrasoft potential, $u_\text{ov}(r;n)$ is bounded with zero slope (no repulsion) for full overlap of two spherical microgels and connects smoothly with $u_\text{Y}(r;n)$ at $r=2a$, where the first derivatives are equal. Since $u_\text{H}(r)$ and its first derivative are zero at contact distance, also $u_\text{eff}(r)$ crosses over smoothly at $r=2a$.

Owing to the pairwise additivity of $U_\text{eff}(n)$ 
in the invoked linear-response electrostatic plus Flory-Hertz elasticity approximations, 
the generalized virial equation for the suspension pressure reduces to \cite{Hansen-McDonald}
\begin{eqnarray}\label{eq:PressureTwoBody}
	\beta p\!\!&=&\!\!n-\frac{2\pi}{3}n^2\int_{0}^{\infty}dr r^3g(r)\frac{\partial\beta u_\text{eff}(r)}{\partial r}\nonumber\\
	\!\!&+&\!\!\!2\pi n^3\!
	\!\int_{0}^{\infty}\!\!dr r^2 g(r)\frac{\partial\beta u_\text{eff}(r)}{\partial n}
	+\beta p_\text{V}+\beta p_\text{se}\!\,.
\end{eqnarray}
From the explicit expression for the volume energy $E_\text{V}(n)$ in the linear-response approximation, the corresponding pressure contribution is \cite{Hedrick_JCP_2015}
\begin{eqnarray}\label{eq:VolumePressure}
 \beta p_\text{V} &=& n^2\left(\frac{\partial\beta \varepsilon_\text{V}}{\partial n}\right)_\text{res}
 = Zn+2n_\text{s}\nonumber\\
&+& \frac{3Z^2}{2}\frac{\lambda_\text{B}}{a} n \left[ -\frac{1}{\tilde{\kappa}^2} +
 \frac{9}{4\tilde{\kappa}^3} - \frac{15}{4\tilde{\kappa}^5}\right. 
 \nonumber\\
 &+&\left.\left( \frac{3}{2\tilde{\kappa}^2}+\frac{21}{4\tilde{\kappa}^3}+\frac{15}{2\tilde{\kappa}^4}
 +\frac{15}{4\tilde{\kappa}^5} \right)e^{-2\tilde{\kappa}} \right]\,,
\end{eqnarray}
where $\tilde{\kappa}=\kappa a$, and  $\varepsilon_\text{V}=E_\text{V}/N$ is the volume energy per particle. Note that the (reduced) kinetic pressure of the microions, $Zn + 2\;\!n_\text{s}$, is included in $p_\text{V}$.

In calculations of the osmotic compressibility, one can take advantage of a theorem by Henderson \cite{Henderson_PhysLettA_1974}, asserting that for a one-component system with only pairwise interactions, for each considered thermodynamic state (concentration $n$) there is a one-to-one correspondence between $g(r;n)$ and the underlying pair potential, up to an irrelevant additive constant for the latter. 
As thoroughly discussed in \cite{Hoffmann_JCP_2004,Dobnikar2006}, at given concentration $n$ and temperature, the osmotic compressibility can thus be obtained also from the concentration derivative of the suspension pressure, $p^\text{OCM}$, for a fictitious system with state-independent pair potential $u(r) = u_\text{eff}(r;n,T)$. Explicitly,
\begin{eqnarray}\label{eq:Consistency}
 \left( \frac{\partial p}{\partial n}\right)_{\text{res}} = \left( \frac{\partial p^\text{OCM}_\text{}}{\partial n}\right)_{u_\text{eff}}\,,
\end{eqnarray}
where $p^\text{OCM}$ is the one-component model (OCM) pressure of the fictitious system, given by the right-hand side of Eq.~(\ref{eq:PressureTwoBody}) without volume pressure $p_\text{V}$
and without the integral invoking the concentration derivative of $u_\text{eff}(r;n)$. The concentration derivative on the right-hand side of Eq.~(\ref{eq:Consistency}) is taken for fixed $u_\text{eff}$, by discarding any concentration dependence of the effective pair potential, which amounts to keeping $\kappa a$ and $Z_\text{net}$ fixed to their values at the considered concentration. A consistency test of the approximations used in calculating $u_\text{eff}(r;n)$, the equilibrium radius $a$, and $g(r;n)$ follows from comparing numerical values for the reduced osmotic compressibility identified by $S(q\to 0;n)$ with values obtained from Eq.~(\ref{eq:Consistency}).

%{\bf (Alan: it would be nice to illustrate Eq. (\ref{eq:Consistency}) by a numerical example checking its %degree of consistency).} 

\section{Microion-Induced Deswelling}\label{Sec:Deswelling}
	
So far, we have treated the microgel radius as a given quantity. However, as pointed out in the introduction, it is actually a state-dependent thermodynamic variable whose equilibrium mean value, $a$, for temperatures $T < T_\text{cr}$ in the swollen state, is determined from minimization of the semi-grand free energy $F(a_\text{t})$ of the suspension with respect to trial radius values $a_\text{t}$. The necessary condition for determining $a$ is thus  
\begin{equation}
\left.\frac{\partial F}{\partial a_\text{t}}\right|_{N,Z,\text{res}}=0
\label{eq:min-cond}
\end{equation}
at $a_\text{t}=a$. In addition to the dry radius $a_0$ and the electrostatic parameters $Z$ and $n_\text{res}$, the elasticity-related Flory-Rehner and Hertz potential parameters, $\chi$, $N_\text{m}$, $N_\text{ch}$, and $\epsilon_\text{H}$, are kept constant in taking the size derivative. In this way, $a$ is determined as a function of the control parameters $n$, $Z$, and $n_\text{res}$ for fixed temperature and microgel elastic properties.   

The minimization of $F$ with respect to $a_\text{t}$ is equivalent to the mechanical requirement that the intrinsic pressure difference \citep{Colla_JCP_2014,DentonTang_JCP_2016},
\begin{equation}\label{eq:PInrinsic}
\Pi(a_\text{t}) = \Pi_\text{g}(a_\text{t})+\Pi_\text{e}(a_\text{t})\,,
\end{equation}
between the interior and exterior of a single microgel is zero at thermodynamic equilibrium, where $a_\text{t}=a$. The radius-dependent gel-elasticity pressure contribution, $\Pi_\text{g}$, due to solvency, elasticity, and mixing entropy of individual microgel networks and the Hertz elastic pair interactions, is given in the present microgel model by
\begin{equation}\label{eq:Pig}
\Pi_\text{g}(a_\text{t}) = -\frac{\partial}{\partial v_\text{t}}\left[f_p(a_\text{t}) + \frac{n}{2}\big< u_\text{H}(r;a_\text{t})\big>_\text{eff} \right]\,, 
\end{equation}
where $v_\text{t}=4\pi a_\text{t}^3/3$ is the microgel trial volume. The electrostatic pressure contribution to $\Pi(a_\text{t})$ is
\begin{equation}\label{eq:Pie}
\Pi_\text{e}(a_\text{t}) = -\frac{\partial}{\partial v_\text{t}}\left[u_e(a_\text{t}) + \varepsilon_\text{V}(a_\text{t})
+\frac{n}{2}\big< u_\text{eff}^\text{el}(r;a_\text{t}) \big>_\text{eff} \right]\,, 
\end{equation}
where $u_e(a_\text{t})$ is the electrostatic self energy of the uniform backbone microgel charge, and $u_\text{eff}^\text{el}(r;a_\text{t})$ is the effective electrostatic pair potential [Eq.~(\ref{eff_pot})]. For conditions where overlap distances are very unlikely, the Hertz potential energy does not contribute to $\Pi(a_\text{t})$ and the canonical average $\langle\cdots\rangle_\text{eff}$ over the center positions of pseudo-microgels of radius $a_\text{t}$ is determined alone by the Flory-Rehner and electrostatic parameters. The equilibrium radius is determined by the competition between $\Pi_\text{g}$, which is negative for $a_\text{t}$ sufficiently larger than $a_0$ favoring deswelling, and the positive-valued $\Pi_{e}(a_\text{t})$ favoring swelling. Physically, the microion distribution in the microgel interior and the self-repulsion of the charged sites of the polymer backbone network generate an outward electrostatic pressure that swells the macroion. This swelling is limited by the inward elastic restoring forces due to the cross-linked polymer gel. In equilibrium, the balance between these opposing pressures determines the microgel size.

The microgel surface plays here the role of a {\em mobile} semi-permeable membrane, permeable to microions and solvent only where the outer and inner pressures balance to $\Pi(a)=0$ at mechanical equilibrium. 
This single-particle osmotic pressure should be distinguished from the non-zero suspension osmotic pressure, $\pi_\text{os} = p-p_\text{res}$, acting across a (mentally pictured) {\em fixed} semi-permeable membrane separating the suspension from the microion reservoir.    

In the following, we describe and contrast two methods used for calculating the state-dependent equilibrium radius $a$ as a function of microgel concentration, backbone valence, and reservoir salt concentration. The first method makes direct use of Eq.~(\ref{eq:min-cond}) and of the one-component multi-center picture of pseudo-macroions interacting electrostatically by the linear-response effective pair potential $u_\text{eff}(r)$, using a thermodynamic perturbation theory (TPT) approximation for the semi-grand free energy. The second method invokes a spherical cell model (CM) approximation for the semi-grand free energy of a single macroion with nonlinear Poisson-Boltzmann (PB) distributions of microions, referred to accordingly as the PBCM method. The two methods differ in the manner in which they treat inter-microgel electrostatic interactions and correlations.

\subsection{Thermodynamic Perturbation Theory}

In the thermodynamic perturbation theory (TPT) method, the equilibrium radius $a$ is obtained by minimizing the semi-grand free energy per microgel \cite{Weyer_SoftMatter_2018},
\begin{equation}\label{eq:FperParticle}
\frac{F(a_\text{t},n)}{N}=u_\text{e}(a_\text{t})+\varepsilon_\text{V}(a_\text{t})+f_\text{p}(a_\text{t})+f_\text{ex}(a_\text{t},n)\,,
\end{equation}
with respect to trial radius values $a_\text{t}$. We have disregarded here the kinetic (ideal gas) free energy contribution, $\ln \left(\Lambda_\text{m}^3 n\right)-1$, to $F/N$ where $\Lambda_\text{m}$ 
denotes the thermal de Broglie wavelength of microgels, since it is independent of $a_\text{t}$. The excess semi-grand free energy per microgel, $f_\text{ex}(a_\text{t},n)$, is due to the effective interactions between the pseudo-microgels. Provided $U_\text{eff}(X;n)$ is pairwise additive, $f_\text{ex}(a_\text{t},n)$ is exactly given by the charging-process ($\lambda$-integration) expression \cite{Hansen-McDonald,Hoffmann_JCP_2004}
\begin{eqnarray}
f_\text{ex}(a_\text{t},n)&=&\frac{n}{2}\int\!\!d^3r\left[u_\text{H}(r;a_\text{t})+u_\text{eff}(r;a_\text{t},n)\right]\nonumber\\
&&\;\;\times\int_0^1\!\!d\lambda\;\!g_\lambda(r;a_\text{t},n)\,,
\end{eqnarray}
irrespective of whether the pair potential is state-dependent or not. Here, $g_\lambda(r;a_\text{t},n)$ is the rdf corresponding to the pair potential $\lambda\left[u_\text{H}(r)+u_\text{eff}(r)\right]$ at charging fraction $\lambda$, which ranges from $g_\lambda(r)=1$ for $\lambda=0$ to the rdf of the actual suspension for $\lambda=1$. In principle, the above two-step integral expression can be used in Eq.~(\ref{eq:FperParticle}) to determine $a$ by minimization of $F(a_\text{t},n)$. Moreover, it provides another route to determine the suspension pressure $p$ and the osmotic compressibility from the first and second volume derivatives of $F(a,n)$.  

To avoid the cumbersome double integration involving the calculation of a large number of rdfs for different values of $\lambda$, we approximate $f_\text{ex}(a_\text{t},n)$ instead using the first-order perturbation expression \cite{Hansen-McDonald} given by the right-hand side of
\begin{eqnarray}\label{eq:Perturbation}
f_\text{ex}(a_\text{t},n)&\leq&\min_{(d)}\left.\Big\{ f_\text{EHS}(d,n)\right.\nonumber\\
+ 2\pi n&&\!\!\!\!\!\!\!\!\!\!\!\! \int_d^\infty dr r^2 g_\text{EHS}(r;d,n)\;\!u_\text{eff}(r;a_\text{t},n)\Big\},
\end{eqnarray}
which invokes a reference system of effective hard spheres (EHS) of diameter $d$, rdf $g_\text{EHS}(r;d,n)$, and free energy per particle $f_\text{EHS}(d,n)$. The EHS free energy, $f_\text{EHS}$, is accurately described by the analytic Carnahan-Starling free energy expression, and the EHS rdf by the semi-analytic Percus-Yevick result \cite{Henderson-PY:2009} with Verlet-Weis correction \cite{Hansen-McDonald}. The above perturbation expression provides an upper bound to the actual excess free energy $f_\text{ex}(a_\text{t},n)$ for all values of the effective diameter $d$, as follows from the Gibbs-Bogoliubov inequality \cite{Hansen-McDonald}. The equilibrium radius $a$ results from the (double) minimization of $F(a_\text{t},n)/N$ in Eq.~(\ref{eq:FperParticle}) with respect to $a_\text{t}$, after substitution of the right-hand side of Eq.~(\ref{eq:Perturbation}) for the excess free energy minimized with respect to $d>0$. For $u_\text{eff}$ and $\varepsilon_\text{V}$, we use the analytic linear-response expressions of Denton {\em et al.} \cite{Denton_PRE_2003,Hedrick_JCP_2015}, and for $f_\text{p}(a_\text{t})$ the Flory-Rehner expression given in Eq.~(\ref{fee-energy_FloryRehner}).

In the TPT, the suspension pressure can be computed from $f(a,n)=F(a,n)/N$ using the thermodynamic relation,
\begin{eqnarray}\label{eq:ThermoRelation}
 p = n^2 \left(\frac{\partial f(a,n)}{\partial n}\right)_\text{res}\,,
\end{eqnarray}
where the concentration dependence of $a(n)$ must be accounted for, giving rise, in particular, to the extra pressure contribution $p_\text{se}$. In taking the concentration derivative, the electroneutrality condition $n_\text{s}=\langle N_{+}\rangle/V-n Z$ must be maintained for given $Z$. The suspension salt pair concentration $n_\text{s}$, which affects 
$\kappa(n,n_\text{s})$, and hence the range of the effective pair potential in the TPT expression for $f_\text{ex}(a,n)$ in Eq.~(\ref{eq:Perturbation}), is determined, in turn, from equating the microion chemical potentials in suspension and reservoir, using $\langle N_{-}\rangle=N_\text{s}$, according to
\begin{eqnarray}
 \frac{\partial }{\partial n_\text{s}} \Big[n\left(\varepsilon_\text{V}(a) + f_\text{ex}(a,n)\right)\Big]_n= k_\text{B} T \ln\left(\Lambda_0^3\;\!n_\text{res} \right)\,. 
\end{eqnarray}

The TPT method was successfully tested in earlier works for deswelling ionic microgels \cite{Weyer_SoftMatter_2018}, incompressible ionic microgels \cite{Hedrick_JCP_2015}, and impermeable charged colloids \cite{Denton_PRE_2006}. The method self-consistently incorporates effective microgel pair interactions for low to moderately high $Z$, 
where linear-response theory can be used.

\subsection{Poisson-Boltzmann Cell Model}

The PBCM applies to suspensions of ionic microgels, where on average around each microgel there is a region void of others \cite{CollaLevinTrizacJCP2009}. This condition requires sufficiently strong and long-ranged electrostatic repulsion between the microgels and concentrations small enough that particle overlap is unlikely. In this case, a Wigner-Seitz (WS) cell tessellation can be used, with each WS cell subsequently approximated by an overall electroneutral spherical cell of radius $R=\left(3/4\pi\right)^{1/3} n^{-1/3}$, containing a single spherical microgel of radius $a_\text{t}$ at its center. In Donnan equilibrium, the cell is in osmotic contact with a 1:1 strong electrolyte reservoir of salt pair concentration $2\;\!n_\text{res}$. In the PBCM, the radially symmetric concentration profiles $n_\pm(r)$ of the pointlike monovalent microions dissolved in a structureless dielectric solvent of Bjerrum length $\lambda_\text{B}$ are described in a mean-field way by the Boltzmann distributions, $n_\pm(r)=n_\text{res}e^{\mp \Phi(r)}$, where $\Phi(r)=\psi(r)e/k_\text{B}T$ is the reduced form of the total electrostatic potential $\psi(r)$ due to all charges in the cell. As in the TPT method, polarization and image charge effects are disregarded, which can be justified by the high solvent content of weakly cross-linked, swollen microgels. While the cell model focuses on only a single microgel, with the semi-grand suspension free energy being $N$ times that of the cell, 
the presence of other microgels is implicitly accounted for through the cell radius $R$ and the associated (trial) volume fraction $\phi_\text{t}=\left(a_\text{t}/R\right)^3$. 
		
Assuming, as in the TPT, a uniform backbone charge distribution inside each microgel, the electrostatic potential in the cell region $0<r<R$ is obtained from solving the nonlinear Poisson-Boltzmann (PB) equation,
\begin{equation}
\Phi''(r)+\frac{2}{r}\Phi'(r)=\begin{cases} \kappa_\text{res}^2\sinh\Phi(r)
+\displaystyle{\frac{3Z\lambda_\text{B}}{a_\text{t}^3}}\,,& 0<r\leq a_\text{t} \\ \kappa_\text{res}^2\sinh\Phi(r)\,,& a_\text{t}<r\leq R \end{cases}\,,
\label{non-lin_PBeq_CM}
\end{equation}
where $\kappa_{\text{res}}^2=8\pi\lambda_{\text{B}}n_\text{res}$ is the square of the reservoir Debye screening constant. The solution for $\Phi(r)$ is uniquely determined by the boundary conditions, $\Phi'(0)=0=\Phi'(R)$, on the electric field at the cell center and edge, and by the continuity conditions, $\Phi(a_\text{t}^-)=\Phi(a_\text{t}^+)$ and $\Phi'(a_\text{t}^-)=\Phi'(a_\text{t}^+)$, at the microgel surface. 
%		\begin{equation}
%		n_\text{s}=\int_0^Rn_-(r)d{\bf r}.
%		\end{equation}
Once $\Phi(r)$, and hence the microion concentration distributions, are determined by numerically solving Eq.~(\ref{non-lin_PBeq_CM}) for given boundary conditions and microgel trial radius $a_\text{t}$, the intrinsic osmotic pressure $\Pi(a_\text{t})$ in the PBCM follows from Eqs.~(\ref{eq:Pig}) and (\ref{eq:Pie}) taken for $\langle u_\text{H}\rangle_\text{eff}=0=\langle u_\text{eff}\rangle_\text{eff}$, and for $\varepsilon_\text{V}$ replaced by $\langle u_{\text{m}\mu}(r)\rangle_\mu$, i.e., by the electrostatic interaction energy between the uniform central microgel backbone charge and pointlike microions, weighted by the microion number density profiles $V_\text{R} n_\pm(r)$ and averaged over the cell volume $V_\text{R}=(4\pi/3)R^3$. Considering the variation of the electrostatic component of the free energy with respect to the microgel radius leads to an exact statistical mechanical relation for the electrostatic pressure \cite{DentonTang_JCP_2016}:
\begin{equation}
\beta\Pi_\text{e}(a_\text{t}) v_\text{t} = \frac{Z\lambda_\text{B}}{2a_\text{t}}\left( \frac{2}{5}Z-\langle N_+\rangle+\langle N_-\rangle+\frac{\langle r^2\rangle_+-\langle r^2\rangle_-}{a_\text{t}^2}\right)\,,
\label{elec_OsmPress}
\end{equation}
where 
\begin{equation}
\langle N_\pm\rangle=4\pi\int_0^{a_\text{t}}n_\pm(r)r^2 dr
\end{equation}
and
\begin{equation}
\langle r^2\rangle_\pm=4\pi\int_0^{a_\text{t}}n_\pm(r)r^4 dr
\end{equation}
are the mean numbers of internal microions and the second moments of the interior microion number density profiles, respectively.
		
Using Eq.~(\ref{fee-energy_FloryRehner}), the polymer gel contribution to the intrinsic osmotic pressure for trial radius $a_\text{t}$ is \cite{DentonTang_JCP_2016}
\begin{eqnarray}\label{poly_OsmPress}
\beta \Pi_g(a_\text{t}) v_\text{t}= &-&N_\text{m}[\alpha^3\ln(1-\alpha^{-3})+\chi\alpha^{-3}+1]\nonumber\\
&-&N_\text{ch}(\alpha^2-1/2)\,.
\end{eqnarray}
According to Eq.~(\ref{eq:PInrinsic}), the equilibrium microgel radius is obtained from setting the sum of the intrinsic pressure contributions in Eqs.~(\ref{elec_OsmPress}) and (\ref{poly_OsmPress}) equal to zero.

Once the microgel equilibrium radius is determined, the pressure in the cell model due to the mobile microions, $p_{\mu}$, follows from the contact theorem \cite{Wennerstrom1982}
\begin{equation}\label{eq:Contact}
\beta p_{\mu} = n_+(R;a)+n_-(R;a)\,,
\end{equation}
i.e., from the microion concentrations at the cell edge, where the electric field vanishes due to overall electroneutrality. In addition to the kinetic (ideal gas) microgel pressure $n k_\text{B} T$, there are microgel-correlation-induced pressure contributions to the suspension pressure, $p$, not accounted for in the cell model. Therefore, $p_{\mu}$ can differ significantly from $p$, except for relatively low reservoir salt concentrations, where the dominant number of backbone-released counterions ($ZN \gg N_\text{s}$) contribute most to $p$ \cite{Dobnikar2006}. As shown elsewhere \cite{Brito_tosubmit_2019}, in this counterion-dominated regime, where $Zn \gg 2\;\!n_\text{res}$, the dominant contribution to $p$ in Eq.~(\ref{eq:PressureTwoBody}) stems from the volume energy-related pressure $p_\text{V}$. Akin to the cell model pressure $p_\mu$, the pressure contribution $p_\text{V}$ arises from the microions in the presence of fixed microgels. The positive-valued OCM pressure on the right-hand side of Eq.~(\ref{eq:PressureTwoBody}) is nearly compensated at low salinity by the negative-valued pressure contribution from the concentration derivative of $u_\text{eff}(r;n)$. While this compensation is observed for non-permeable charge-stabilized colloids \cite{Brito_tosubmit_2019}, it likely holds also for ionic microgels. One should not infer from this compensation, however, the practical identity of $p_\mu$ and $p_\text{V}$ in the counterion-dominated concentration region, since the underlying models, i.e., spherical cell versus multi-center model, and the respectively invoked approximations (linear-response theory versus PB aproximation) in the pressure calculations are distinctly different.        
		
In the cell model, the net microgel valence $Z_\text{net}$ is calculated by means of Eq.~(\ref{Def_Znet}) using the microion number density profiles $n_\pm(r;a)$, and the suspension salt pair concentration $n_\text{s}$, 
by integrating the coion (anion) profile over the cell volume according to 
\begin{equation}\label{eq:ns}
n_\text{s}=\frac{4\pi}{V_\text{R}}\int_0^R\!\! n_{-}(r;a)\;\!r^2\;\!dr\,.
\end{equation}

While TPT is self-consistently linked to the effective pair potential $u_\text{eff}(r)$ in Eq.~(\ref{eff_pot}), characterized by $Z_\text{net}$ and $\kappa a$ for given backbone valence $Z$, such  a direct link does not exist in the single-microgel PBCM, which does, however, incorporate a nonlinear electrostatic response of the microions that is neglected in the TPT. 
However, an {\em ad hoc} link between PBCM and the linear-response $u_\text{eff}(r)$ is readily established, for given $Z$, by identifying $Z_\text{net}$ and $\kappa a$ in the no-overlap Yukawa potential in Eq.~(\ref{Yukawa_part}) with the PBCM-calculated values $Z_\text{net}^\ast$ and $\kappa^\ast a^\ast$, respectively, where 
\begin{equation}
 (\kappa^\ast)^2 = 4\pi\lambda_\text{B}(n Z^\ast+2n_\text{s}^\ast)
\label{eq:kappa_ast} 
\end{equation}
and the asterisk labels PBCM-calculated properties. An apparent backbone valence $Z^\ast$ is defined here as a function of $Z_\text{net}^\ast$ and $\kappa^\ast a^\ast$ by
\begin{equation}
Z_\text{net}^\ast=Z^\ast\;\!\frac{3\;\!\left(1+\kappa^\ast a^\ast\right)}{\left(\kappa^\ast a^\ast\right)^2\;\!e^{\kappa^\ast a^\ast}}\;\!\left[\cosh(\kappa^\ast a^\ast)
-\frac{\sinh(\kappa^\ast a^\ast)}{\kappa^\ast a^\ast}\right]\,,
\label{eq:Zast}
\end{equation}
which when used in the expression for the overlap electrostatic potential $u_\text{ov}(r)$, according to the substitution $\{Z,a,\kappa\} \to \{Z^\ast,a^\ast,\kappa^\ast\}$, maintains the continuity of the effective potential and its first derivative at $r=2 a$. Substitution of $Z^\ast$ into Eq.~(\ref{eq:kappa_ast}) gives an implicit equation for $\kappa^\ast$, which can be solved iteratively. For lower backbone valences $Z\leq 200$, $Z^\ast$ is close to $Z$, so that the latter can be used instead as input in Eq.~(\ref{eq:kappa_ast}).     

Most results presented here are for ionic microgel systems with electrostatic coupling strengths $\Gamma_\text{el}\equiv Z_\text{net}\lambda_\text{B}/a \lesssim 5$, where nonlinear electrostatic effects by the microions are negligible or small, so that both TPT and PBCM can be directly used in conjunction with the linear theory effective pair potential in Eq.~(\ref{Yukawa_part}). For stronger electrostatic couplings, experience gained with rigid charged colloids suggests that the Yukawa form of the effective potential in Eq.~(\ref{Yukawa_part}) is still applicable, but now for renormalized values of $Z$ and $\kappa$, which can be obtained, e.g., from linearization of the potential $\Phi(r)$ in the cell model with respect to its value at the cell boundary \cite{Colla_JCP_2014}, or with respect to the cell volume averaged potential value. While microgel charge-renormalization is not in the focus of this paper, in the framework of PBCM we use it to assess the concentration shift at a fluid-solid freezing transition caused by the deswelling of strongly charged microgels.   

%Notice that $\kappa^\text{CM}$ is computed with $Z$ instead of $Z^\ast$, since we are in the low charge limit for permeable colloids, $Z_\text{net}\lambda_\text{B}/a\lesssim5$.

\section{Thermodynamics and Structure}
\label{sec:StructureThermodynamics}

Once the mean radius $a$ is determined for given system parameters $n$, $Z$, $n_\text{res}$, $\chi$, $N_\text{mon}$, $N_\text{ch}$ and $\epsilon_\text{H}$, we are in the position to calculate thermodynamic, structural, and dynamic properties of the one-component suspension of pseudo-microgels interacting via the effective pair potential in Eq.~(\ref{eff_pot}). As we show below in the Results section (Sec.~\ref{Sec:Results}), the TPT and PBCM predictions for $a$ are quantitatively different, as reflected in the calculated static and dynamic properties. 

Our methods for calculating dynamic properties of the microgel suspension require the static structure factor, $S(q)$, of microgels and the associated radial distribution function, $g(r)$, as the only input. Since $u_\text{eff}(r;n)$ is purely repulsive, we can use the thermodynamically self-consistent Rogers-Young (RY) integral-equation scheme \cite{Hansen-McDonald} for calculating these structural properties. This hybrid scheme, which uses a closure mixing function interpolating between the hypernetted chain (HNC) and Percus-Yevick (PY) integral-equation schemes \cite{Hansen-McDonald}, is known from comparisons with computer simulation data to make accurate structural predictions for a variety of repulsive interaction potentials, including the screened-Coulomb potential \cite{Banchio_JCP_2008,Banchio_JCP_2018} used to model non-overlapping ionic microgels. 
The mixing parameter $\alpha$ in the RY mixing function is determined self-consistently from enforcing equality of the microgel osmotic compressibility obtained from the one-component compressibility and virial equation of states, respectively, i.e., from demanding
\begin{equation}
\frac{1}{S(q=0;\alpha)} = \beta \left(\frac{\partial p^\text{OCM}(\alpha)}{\partial n}\right)_{u_\text{eff}}
\end{equation}
in accord with Eqs.~(\ref{eq:OsmoticCompressibility}) and (\ref{eq:Consistency}). 

With $\alpha$ determined self-consistently, the pressure $p$ can be calculated in the TPT using the thermodynamic relation in Eq.~(\ref{eq:ThermoRelation}). Alternatively, the pressure can be calculated from Eq.~(\ref{eq:PressureTwoBody}) using the RY-$g(r)$ as input, in conjunction with the volume energy-related pressure contribution $p_\text{V}$ in Eq.~(\ref{eq:VolumePressure}). Differences in the predictions for $p$ by the two routes reflect the accuracies of the approximations going into the TPT and RY methods.

\section{Dynamic Properties}\label{Sec:Dynamics}	

\subsection{General Theory}
	
The employed methods for calculating dynamic properties of microgel suspensions are based on the one-component model of pseudo-microgels interacting by the state-dependent effective pair potential in Eq.~(\ref{eff_pot}). With regard to dynamic properties, different colloidal time regimes need to be distinguished \cite{Naegele_PhysRep_1996,Naegele_Varenna2013}.

We focus mainly on the colloidal short-time regime, characterized by correlation times $t$ for which $\tau_\text{B}\ll t\ll\tau_\text{I}$ holds, i.e., for times $t$ well separated from the long-time regime where $t \gg \tau_\text{I}$. Here, $\tau_\text{B}=M/(6\pi\eta_0 a_\text{h})$, with $M$ the particle (microgel) mass and $a_\text{h}$ the hydrodynamic particle radius, is the particle momentum relaxation time characterizing the time range where momentum changes (i.e. inertia) matters. Moreover, $\tau_\text{I}=a^2/d_0$ is the structural relaxation time, where $d_0=k_\text{B}T/(6\pi\eta_0a_\text{h})$ is the Stokes-Einstein-Sutherland translational free diffusion coefficient of a spherical colloidal particle.  
Moreover, $\eta_0$ is the shear viscosity of the suspending low-molecular-weight Newtonian solvent (i.e., water). Owing to the low hydrodynamic permeability of (ionic) microgels \cite{Riest_SoftMatter_2015}, we identify for simplicity the hydrodynamic radius $a_\text{h}$ of the microgels with the equilibrium radius $a$ determined in the TPT and PBCM, respectively. 

During times $t \ll \tau_{I}$, over which particle displacements by Brownian motion are minuscule compared to the particle radius, short-time dynamic properties are influenced solely by the inter-microgel hydrodynamic interactions (HIs), which are quasi-instantaneously transmitted 
by intervening solvent-flow perturbations. Short-time transport properties can thus be calculated 
as genuine equilibrium averages of configuration-dependent hydrodynamic mobilities. The non-dynamic interactions embodied in $u_\text{eff}(r;n)$ are only indirectly influential through their effect on the equilibrium microstructure encoded in $g(r)$ and $S(q)$. Long-time transport properties, such as the zero-frequency, steady-shear suspension viscosity $\eta$ and the long-time self-diffusion coefficient $d_\text{l}$, with the latter coefficient being proportional to the long-time slope of the particle mean-squared displacement, are influenced additionally by $u_\text{eff}(r;n)$ via non-instantaneous caging (i.e., memory) effects, whose description requires, in general, more elaborate calculations.

The short-time diffusion of microgels is commonly probed experimentally by measuring the $q$-dependent dynamic structure factor $S(q,t)$ using dynamic light scattering. At short times, $S(q,t)$ decays exponentially according to  \cite{Naegele_PhysRep_1996,Banchio_JCP_2018,Pamvouxoglou_JCP_2019}
\begin{equation}\label{eq:EXP}
S(q,t\ll\tau_\text{I})=S(q)\exp\{-q^2 D(q)\;\!t\}\,,
\end{equation}
where $D(q)$ is the wavenumber-dependent short-time diffusion function characterizing the decay of concentration fluctuations of wavelength $2\pi/q$. A statistical-mechanical expression for $D(q)$ follows from the generalized Smoluchowski diffusion equation of interacting Brownian particles in the form of the ratio \cite{Naegele_PhysRep_1996,Banchio_JCP_2018,Pamvouxoglou_JCP_2019},
\begin{equation}
	 D(q)=d_0\frac{H(q)}{S(q)}\,,
\label{diff_function}
\end{equation}
where $H(q)$ is the so-called hydrodynamic function given by the equilibrium average \cite{Naegele_PhysRep_1996},
\begin{equation}
H(q)
\!=\!\left<\!\frac{1}{N\mu_0 q^2}\!\sum_{l,j=1}^N\!{\bf q}\cdot \bm{\mu}_{lj}(X)\cdot{\bf q}\;\!\displaystyle{e^{i{\bf q}\cdot\left({\bf r}_l\!-\!{\bf r}_j\right)}}\!\right>_\text{eff}\,,
\label{Def_H}
\end{equation}
over the positional configurations $X$ of the microgels.

Here, $k_\text{B}T\mu_0=d_0$ and $\bm{\mu}_{lj}(X)$ are the translational $N$-sphere mobility tensors linearly relating the hydrodynamic force on a sphere $j$ to the instant velocity change of sphere $l$ caused by the solvent-transmitted HIs. These tensors depend on the instantaneous configuration, $X$, of the $N$ microgel centers through the specified hydrodynamic surface boundary conditions.
	The positive-valued function $H(q)$ is a measure of the influence of HIs on short-time diffusion over the length scale $\sim 1/q$. In the (hypothetical) case of hydrodynamically non-interacting particles, $H(q)\equiv 1$, independent of $q$ and the particle concentration. Deviations of $H(q)$ from the infinite dilution value of one thus hallmark the influence of HIs.
	  
According to
\begin{equation}\label{eq:HDist}
H(q)= \frac{d_\text{s}}{d_0}+H_\text{d}(q)\,,
\end{equation}
the hydrodynamic function is the sum of a self-part equal to the short-time self-diffusion coefficient $d_\text{s}$ (in units of $d_0$), quantifying the initial slope of the particle mean-square displacement, and a wavenumber-dependent distinct part, $H_\text{d}(q)$, accounting for hydrodynamic cross correlations between the microgels. The latter part decays to zero at large $q$. 
For large $qa\gg 1$, the hydrodynamic function becomes thus equal to $d_\text{s}/d_0$, while for small wavenumbers $qa\ll 1$ it reduces to the (short-time) sedimentation coefficient $K(n) = H(q\to 0;n)$  of a homogeneous suspension subjected to a weak (gravitational) force field. The associated short-time collective diffusion coefficient,
\begin{equation}\label{eq:dcShort}
d_\text{c}(n)=d_0(n)\frac{K(n)}{S(q\to 0;n)}=\frac{d_0(n)\;\!K(n)}{k_\text{B} T\left(\partial n/\partial p \right)_\text{res}}\,,
\end{equation}
is even for a concentrated suspension only slightly larger (by a few percent) than the long-time collective diffusion coefficient appearing in the macroscopic Fickean constitutive law, which linearly relates the particles current 
to the concentration gradient \cite{Banchio_JCP_2018}. This behavior should be distinguished from self-diffusion, where $d_\text{l} \approx 0.1\times d_\text{s}$ right at the fluid-crystal freezing transition point of a three-dimensional colloidal system \cite{Loewen_PRL_1993,Naegele_MolecPhys2002}.

An important feature distinguishing (ionic) microgels from impermeable solid particles is that $d_0(n)= k_\text{B} T/(6\pi\eta_0 a(n)) = d_0^\text{dry}/\alpha(n)$ depends on concentration. Here, $d_0^\text{dry}$ is the Stokes-Einstein diffusion coefficient of collapsed (dry) microgels, and $\alpha(n)=a(n)/a_0$ the swelling ratio at concentration $n$. In our calculations of diffusion and rheological properties, we identify  the hydrodynamic microgel radius for simplicity with the thermodynamic mean particle radius $a(n)$ as obtained by the TPT/PBCM methods. While on first sight this appears to be a severe approximation owing to the solvent permeability of weakly cross-linked (ionic) microgels, calculations show that the hydrodynamic penetration depth related to the Darcy permeability of microgels is actually quite small so that solvent-permeability effects can be disregarded, as they play a noticeable role only at high concentrations \cite{Riest_SoftMatter_2015}.
	
A non-diffusional, rheological short-time property characterizing the microgel suspension as a whole is the high-frequency viscosity, $\eta_\infty$, for low shear rates. This property linearly relates the average deviatoric suspension shear stress to the applied rate of strain in a low-amplitude, oscillatory shear experiment at frequencies $\omega\gg 1/\tau_\text{I}$, where shear-induced perturbations of the microstructure away from the equilibrium spherical symmetry are negligible. Experimentally, $\eta_\infty$ can be determined using a torsional rheometer operated at high frequencies and low amplitudes. The high-frequency viscosity is a purely hydrodynamic property, whose statistical physics expression is given, owing to isotropy, by (see, e.g., \cite{AbadeVisc_JCP_2010})
\begin{equation}\label{eq:EtaInfty}
	\eta_\infty=\eta_0+
\lim_{q\to 0}\sum_{\alpha,\beta=1}^3 \Big<\frac{1}{10V}\sum_{l,j=1}^N\mu_{lj,}^{dd}\textsubscript{$\alpha\beta\beta\alpha$}(X)\;\!\displaystyle{e^{i {\bf q}\cdot\left({\bf r}_l-{\bf r}_j\right)}}\Big>_\text{eff}\,,
\end{equation}
where $\mu_{lj,}^{dd}\textsubscript{$\alpha\beta\beta\alpha$}$ are the Cartesian components of the fourth-rank dipole-dipole hydrodynamic tensor $\boldsymbol{\mu}_{lj}^{dd}$ relating the symmetric hydrodynamic force dipole moment tensor of microgel sphere $l$ to the rate of strain tensor evaluated at the center of a sphere $j$. The zero-wavenumber limit is taken after the ensemble averaging over a macroscopic system, guaranteeing in this way convergence of the integrals following from the averaging over the spatially slowly decaying hydrodynamic tensors \cite{Szymczak_JStatMech_2008}.

As an important colloidal long-time property, we compute also the low shear rate, zero-frequency viscosity $\eta >\eta_\infty$, measured in a suspension subjected to steady-state weak shear flow. The viscosity $\eta$ is the sum \cite{Naegele_Visco:1998},
\begin{equation}\label{eq:ViscoSum}
	\eta = \eta_\infty+\Delta\eta\,,
\end{equation}
of $\eta_\infty$ and a shear stress relaxation contribution denoted $\Delta\eta$. The latter contribution is related to the additional dissipation in the suspension originating from stress relaxations of the shear-perturbed next-neighbor particle cages formed around each microgel, and it is influenced both by direct and hydrodynamic interactions. The viscosity part $\Delta\eta$ can be calculated based on an exact Green-Kubo relation for the time integral of the equilibrium stress time auto-correlation function where HIs are included \cite{Naegele_Visco:1998}.

In the employed one-component model of ionic pseudo-microgels, electro-kinetic effects due to a non-instantaneous dynamic response of the microion clouds formed inside and outside the microgels are disregarded. These effects tend to lower $d_\text{c}$ and $d_\text{l}$, and to increase $\eta$, but in general by only small amounts. Electrokinetic effects on diffusion and rheology are of secondary importance, in particular, when non-dilute suspensions are considered and when the microions are small compared to the microgels, which is commonly the case. 

\subsection{Methods of Calculation}\label{Subsec:Methods}
	
For the calculation of $H(q)$, we use the well-established analytic BM-PA scheme \cite{Heinen_Rheo_JCP_2011}. This scheme is a hybrid of the second-order Beenakker-Mazur method (BM), used here for the wavenumber-dependent distinct part $H_\text{d}(q)$, and the hydrodynamic pairwise-additivity approximation (PA) used for the $q$-independent self part $d_\text{s}/d_0$. The BM-PA scheme combines the advantages of the BM and PA methods. It requires the microgel $S(q)$ and $g(r)$ as its only input, for which the RY results based on $u_\text{eff}(r)$, and the TPT/PBCM results for $a(n)$ and hence for $d_0(n)$, are used. The overall good accuracy of the BM-PA scheme was assessed by the comparison with elaborate dynamic simulation results, where many-particles HIs are accounted for, and with experimental $H(q)$ data, for a variety of colloidal model systems, including solvent-permeable hard spheres (non-ionic microgels), charge-stabilized rigid particles, and globular proteins exhibiting short-range attraction and long-range repulsion \cite{Banchio_JCP_2008,Heinen_Rheo_JCP_2011,Das_SoftMatter_2018,Banchio_JCP_2018}. For details about the employed BM-PA method, we refer to \cite{Heinen_Rheo_JCP_2011}.
	
For the here considered low-salinity microgel suspensions, which show counterion-induced deswelling, we calculate the high-frequency viscosity $\eta_\infty$ using a modified Beenakker-Mazur mean-field method described in \cite{Heinen_Rheo_JCP_2011}. In this semi-analytic method invoking one-dimensional integrals only, many-particles HIs are approximately accounted for, but lubrication is disregarded. Lubrication is irrelevant, however, for solvent-permeable microgels. Just as the BM-PA scheme for $H(q)$, the modified BM method for $\eta_\infty$ has $S(q)$ as its only input. The modified BM expression for $\eta_\infty$ is \cite{Heinen_Rheo_JCP_2011}
\begin{equation}\label{eq:BM-MOD}
	\frac{\eta_\infty}{\eta_0}=1+\frac{5}{2}\phi(1+\phi)-\frac{1}{\lambda_0}+\frac{1}{\lambda_0+\lambda_2}\,,
\end{equation}
with so-called zeroth and second-order BM viscosity contributions, $\lambda_0(\phi)$ and $\lambda_2(\phi)$, respectively, whose explicit forms are given in \cite{Heinen_Rheo_JCP_2011}. As explained in detail in this reference, the invoked modification of the standard BM expression for $\eta_\infty/\eta_0$ is the subtraction of the structure-independent BM part $1/\lambda_0$, and its replacement by the structure-independent pairwise additive viscosity contribution $1+2.5\phi(1+\phi)$, which is known to give the dominant contribution at low salinity and small volume fractions. The modified BM expression is in very good agreement with Stokesian Dynamics simulation data for the high-frequency viscosity of low-salinity charge-stabilized suspensions, even up to the freezing transition concentration.

As noted above, the calculation of the shear relaxation contribution $\Delta\eta$ to the zero-frequency viscosity, 
$\eta=\eta_\infty +\Delta\eta$, is more demanding since it 
is explicitly influenced by direct and hydrodynamic interactions. Starting from an exact but formal Green-Kubo relation for $\Delta\eta$, mode-coupling theory (MCT) integro-differential equations with HIs included have been derived for its approximate calculation, whose numerical solution is quite involved. We use therefore a simplified MCT theory expression for $\Delta\eta$, constituting the first-iteration step in the self-consistent numerical solution of the MCT equations. This simplified MCT expression is \cite{Naegele_Visco:1998}
\begin{equation}\label{eq:MCT}
\frac{\Delta\eta}{\eta_0}=\frac{1}{40\pi}\int_0^\infty dy\,y^2\,\frac{(S'(y))^2}{S(y)}\frac{1}{H(y)}\,,
\end{equation}
where $y=2qa$ and $S'(y)=dS(y)/dy$. HIs enter here only through the dynamic structure factor $S(q,t)$, which in turn is approximated by its short-time form given by the right-hand side of Eq.~(\ref{eq:EXP}) involving $H(q)$. Since for correlated particles $S(q,t)$ decays more slowly than exponentially at longer times, $\Delta\eta$ is somewhat underestimated by Eq.~(\ref{eq:MCT}), as compared to the fully self-consistent MCT viscosity solution. 
This underestimation becomes more pronounced at higher $\phi$.

\section{Results}\label{Sec:Results}

To analyze the influence of counterion-induced deswelling on thermodynamic, structural, and dynamic properties of ionic microgel suspensions, and to make contact with a recent study by Weyer {\it et. al.} \cite{Weyer_SoftMatter_2018}, in which TPT results for the mean microgel radius were compared against computer simulations for salt-free systems, we use the following system parameters, corresponding to aqueous suspensions at lower salinity: solvent Bjerrum length $\lambda_\text{B} = 0.714\;\!\text{nm}$ (i.e., water at temperature $T = 293\;\!K$), backbone valences $Z =100,\,200$, and $500$, dry microgel radius $a_0 = 10$ nm, monomer number per microgel $N_\text{mon} = 2\times10^5$, polymer chain number per microgel $N_\text{ch} = 100$, solvency parameter $\chi = 0.5$, and Hertz softness parameter $\epsilon_\text{H} = 1.5\times10^4$. For the 1:1 electrolyte reservoir concentration, we use $c_\text{res} = 100\,\,\mu$M, if not stated otherwise, so that $n_\text{res}=c_\text{res}\,N_\text{A}$, where $N_\text{A}$ is the Avogradro number. Values of the dry volume fraction $\phi_0=(4\pi/3)n_0 a_0^3$ in the range from $2\times 10^{-4}-5\times 10^{-2}$ are considered.

Note here that $\phi_0\propto n$ has the meaning of a dimensionless microgel concentration. The Debye screening length, $1/\kappa$, in Eq.~(\ref{eq:Debye}) attains values from $40 - 4.4$ nm, for (reduced) concentration values $\phi_0$ varied from $0.001 - 0.05$. It is noteworthy that, for most of the considered suspensions, $\kappa$ is determined by the mean concentration, $Zn$, of monovalent counterions released from the microgel polymer backbone, which is significantly higher than the salt pair concentration $n_\text{s}$.  

The electrostatic repulsion between the microgels is quantified by the reduced electrostatic coupling strength 
$\Gamma_\text{el}\equiv Z_\text{net}\lambda_\text{B}/a$, which in the present study is in the range of $1$-$9$, comparable to values for typical ionic microgel systems \cite{Holmqvist_PRL_2012,Braibanti_PRE_2016,Nojd_SoftMatter_2018}.
The electrostatic repulsion between the microgels is here strong enough that configurations of microgels that are in contact or overlapping are very unlikely, such that $g(r \leq 2a) \approx 0$. On the other hand, nonlinear screening effects are in most cases weak enough that the linear TPT method can be used for determining $a$, in addition to the PBCM method.

%The screening constant has the form $\kappa a=\sqrt{3\phi Z\lambda_\text{B}/a+8\pi\lambda_\text{B}\tilde{n}_\text{s}/a}$, where $\tilde{n}_\text{s}=a^3n_\text{s}$ is the reduced system salt concentration. This is computed from the coion density distribution within CM approximation.

\subsection{Equilibrium Radius Predictions}

In the following, we analyze TPT and PBCM predictions for the concentration-dependent microgel swelling ratio $\alpha(\phi_0)=a(\phi_0)/a_0$, the suspension salt concentration $n_\text{s}(\phi_0)$, and 
the electrostatic coupling strength $\Gamma_\text{el}(\phi_0)$, 
where $\phi_0\propto n$ is the non-dimensional microgel concentration.  

\begin{figure}[h!]
\includegraphics[width=8.5cm]{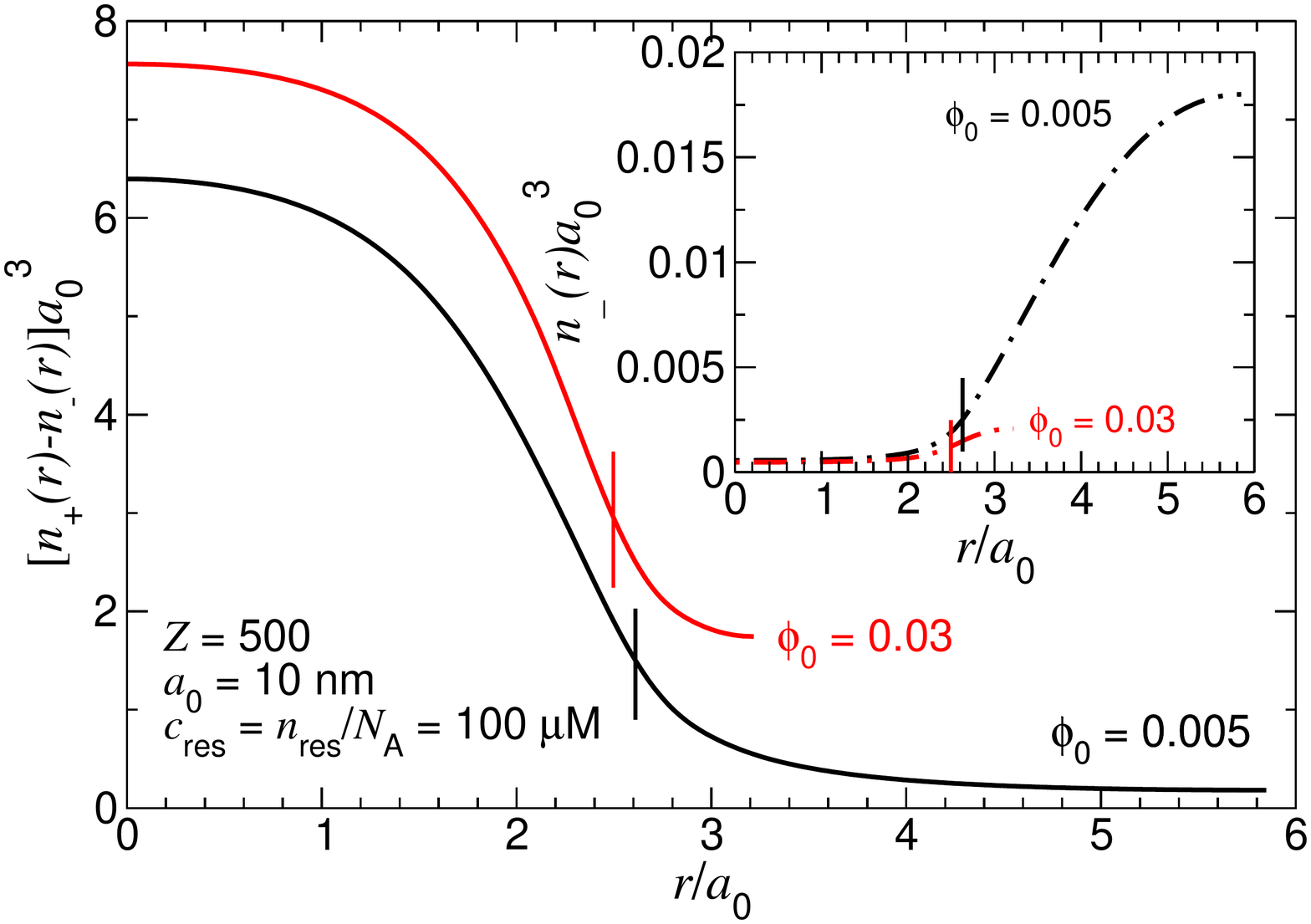}
\caption{PBCM predictions for radial profile of microion charge density $\rho^\text{el}(r)$ in units of $1/(a_0^3\,e)$, versus radial distance $r$ from center of cell (units of dry radius $a_0$) for microgel valence $Z=500$, dry radius $a_0=10$ nm, reservoir salt concentration $c_\text{res}=100\,\mu$M, and dry volume fractions $\phi_0=0.03$ and $0.005$ (red and black solid curves). Inset: Reduced coion number density $n_{-}(r)a_0^3$ (dashed-dotted curves). All curves terminate at the cell radius $r=R=a_0/\phi_0^{1/3}$. Vertical line segments indicate equilibrium swelling ratio, $\alpha=a/a_0$, computed from zero balance of intra-particle pressure contributions in Eqs.~(\ref{elec_OsmPress}) and (\ref{poly_OsmPress}).}
\label{figure3b}
\end{figure}

The physical mechanism leading to counterion-induced deswelling with increasing concentration can be reasoned on the basis of Fig.~\ref{figure3b}, showing PBCM results, at two different concentrations for the radial dependence of the (reduced) total microion charge density, $\rho^\text{el}(r)=\left[n_{+}(r)-n_{-}(r)\right]e$, and of the coion concentration $n_{-}(r)$ (displayed in the inset) inside and outside of a negatively charged microgel centered at $r=0$. For both considered concentrations $\phi_0$, the counterions constitute the dominant microion species where $n_{+}(r) \gg n_{-}(r)$, and hence $\rho^\text{el}(r) \approx n_{+}(r)e$ holds for the total microgel charge concentration inside the cell up to its boundary at radius $R=a_0/\phi_0^{1/3}$, where the curves in Fig.~\ref{figure3b} terminate. 

One clearly notices that both the equilibrium microgel radius $a$, marked by the vertical line segments in the figure, and the cell radius $R$ decrease with increasing concentration. The counterion concentration profile rises with increasing system concentration, while the coion concentration profile falls. With increasing concentration, the volume exterior to the microgels is reduced, making it less favorable (entropically) for counterions to reside outside the oppositely charged microgel backbone region. Consequently, a fraction of the outside counterions permeates into the backbone region, thereby lowering the expansive intrinsic PBCM pressure contribution $\Pi_\text{e}$ [Eq.~(\ref{elec_OsmPress})]. In response, the microgel deswells until a new equilibrium with the contractile polymer gel pressure contribution $\Pi_\text{g}$ is established at a smaller equilibrium radius. The enhanced counterion permeation of microgels with increasing concentration is reflected in the lowering of the net microgel valence $Z_\text{net}$, defined in Eq.~(\ref{Def_Znet}), which for the backbone valence $Z=500$ is given by $Z_\text{net}=223$ at $\phi_0=0.005$ and by $Z_\text{net}=154$ at $\phi_0=0.03$. The counterion-induced deswelling becomes weaker with increasing salt concentration, which causes a flattening of the microion concentration profiles across the microgel surface.  

\begin{figure}[h!]
\includegraphics[width=8.5cm]{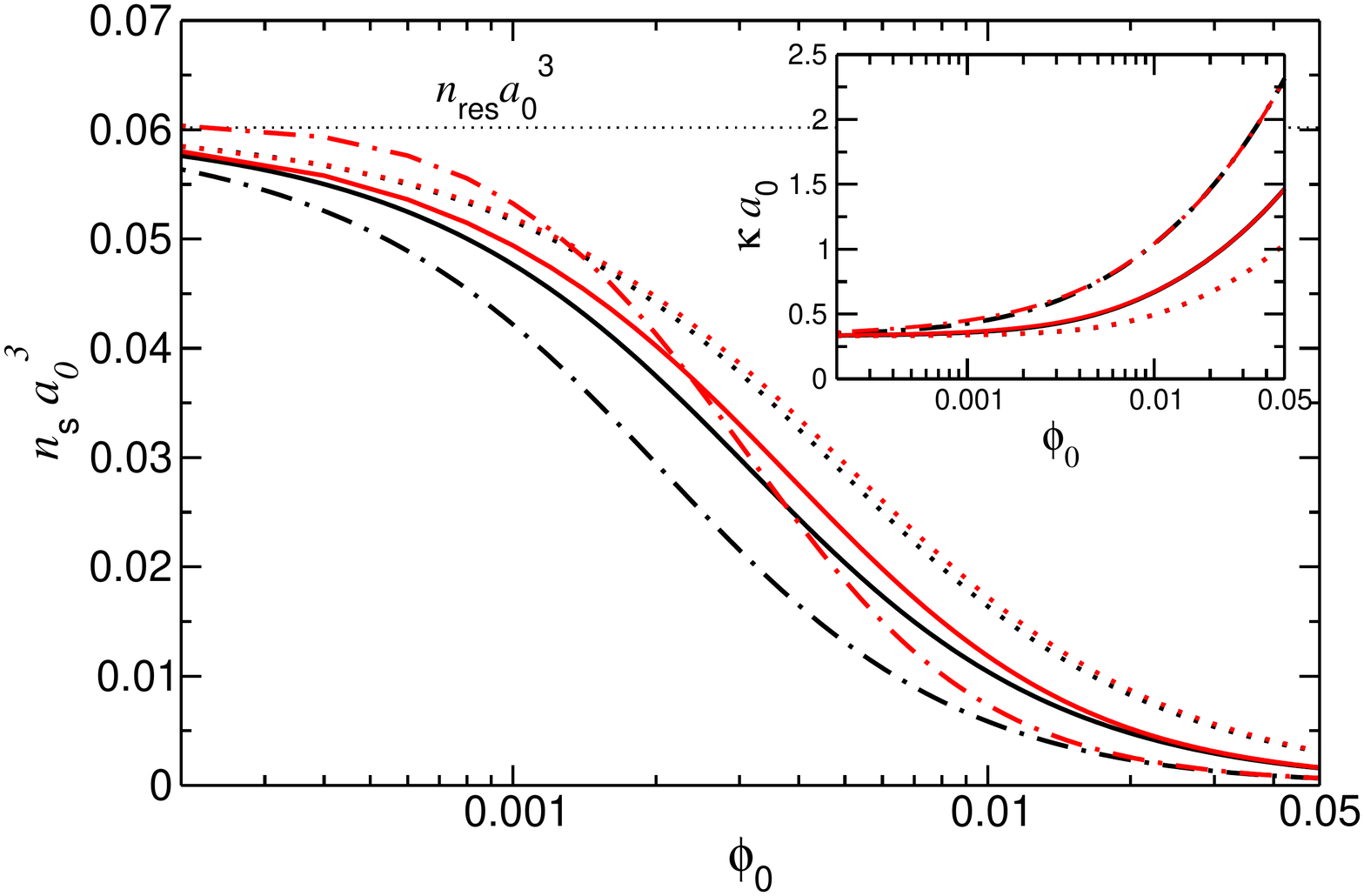}
\caption{Reduced suspension salt concentration $n_\text{s}a_0^3$ versus microgel concentration $\phi_0$. Inset: Reduced Debye screening constant $\kappa a_0$. TPT predictions are in red, and PBCM predictions in black for backbone valence $Z=100$ (dotted), $200$ (solid), and $500$ (dash-dotted). Reservoir salt concentration is $c_\text{res}= n_\text{res}/N_\text{A}=100\,\mu$M.}
\label{figure3}
\end{figure}

In the PBCM method, the mean salt concentration $n_\text{s}$ in the suspension is obtained by integrating the coion concentration profile over the cell volume according to Eq.~(\ref{eq:ns}). In the TPT method, $n_\text{s}$ is computed using the equality of the chemical potentials of the microions in the suspension and reservoir. In Donnan equilibrium, $n_\text{s}$ is a state-dependent quantity. The TPT and PBCM predictions for the concentration dependence of $n_\text{s}$ are depicted in Fig.~\ref{figure3} (red and black curves, respectively), for reservoir microion concentration $c_\text{res}= 100\;\!\mu$M and backbone valences $Z=100$, $200$, and $500$. The monotonic decrease of $n_\text{s}$ with increasing $\phi_0$, and hence with increasing number of backbone-released counterions, is due to an increasing expulsion of salt ion pairs into the reservoir, necessitated to maintain global electroneutrality in the suspension. 

At high dilution, $\phi_0\rightarrow 0$, where the concentration of salt counterions greatly exceeds the concentration of backbone-released counterions, the exact limit $n_\text{s}\rightarrow n_\text{res}$ is recovered by both methods. For the moderately high valences $Z=100$ and $Z=200$ considered here, the TPT and PBCM curves for $n_\text{s}(\phi_0)$ in Fig.~\ref{figure3} lie close to each other, but with a slightly stronger salt expulsion predicted in the PBCM. Pronounced differences are observed for the high valence $Z=500$ and intermediate $\phi_0$, where the concentration, $Zn$, of backbone-released counterions is comparable to the salt-counterion concentration. While the PBCM predicts a decreasing $n_\text{s}$ with increasing $Z$, in accord with physical expectation, this trend is reversed for $\phi_0 \lesssim 0.05$ by the TPT curve for $Z=500$. We attribute this reversal to the disregard in the linear TPT of nonlinear electrostatic effects, which come into play at high valences and low $\phi_0$. The PBCM accounts for nonlinear electrostatic effects, but not for inter-microgel correlations, which the TPT accounts for on a linear level.  

The inset of Fig.~\ref{figure3} displays TPT and PBCM predictions for the Debye screening constant $\kappa$ in Eq.~(\ref{eq:Debye}), which on the scale of the inset are practically equal. In dimensionless form, the screening constant is 
\begin{equation}\label{eq:KappaTwo}
(\kappa a_0)^2=(\kappa_\text{c} a_0)^2+(\kappa_\text{s} a_0)^2=3\phi_0 \frac{Z\lambda_\text{B}}{a_0}+8\pi\lambda_\text{B}n_\text{s}a_0^2\,.
\end{equation}
The first term on the right-hand side is the contribution by the backbone-released counterions (subscript $\text{c}$). The second term, proportional to $n_\text{s}$, is the salt-ion contribution (subscript $\text{s}$). This splitting of $\kappa^2$ into released-counterion and salt-ion contributions allows to identify the counterion-dominated regime by the condition $\kappa_\text{c}\gg \kappa_\text{s}$ and the salt-dominated regime by $\kappa_\text{c}\ll\kappa_\text{s}$. At very low microgel concentrations, i.e., in the salt-dominated regime where $\kappa\approx \kappa_\text{s}$, the TPT and PBCM predictions for $\kappa$ differ due to differing values for $n_\text{s}$. However, these differences are not visible in the inset. At higher concentrations in the counterion-dominated regime where $\kappa \approx \kappa_\text{c}$, both methods predict practically the same $\kappa\approx \kappa_\text{c}$, determined by $Z$ and $\phi_0$.  With increasing backbone valence, $\kappa_\text{c}$ increases while $\kappa_\text{s}$ decreases, owing to increased salt expulsion. The total screening constant $\kappa$ increases monotonically with increasing concentration, more steeply so for higher $Z$.
\begin{figure}
\includegraphics[width=8.5cm]{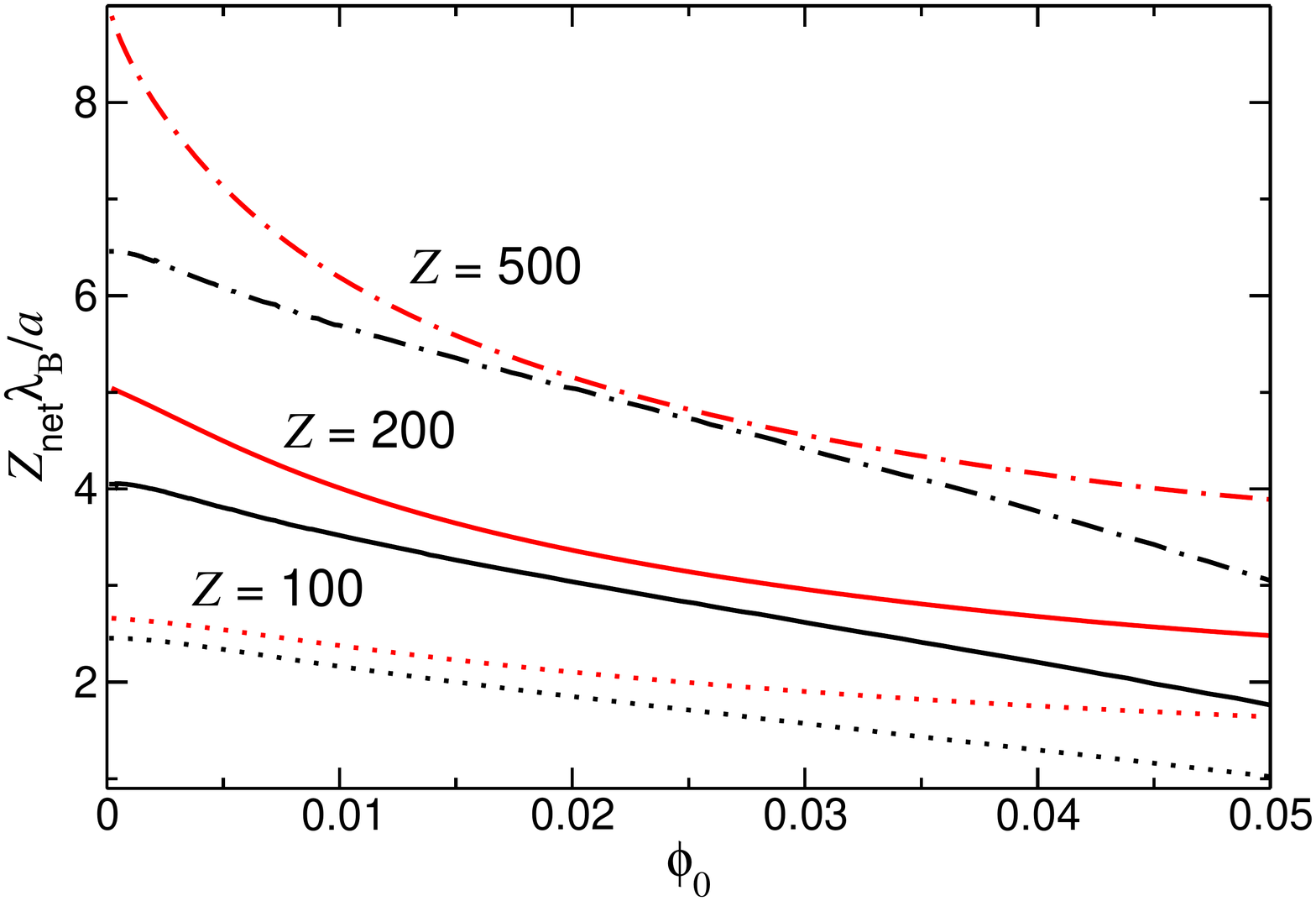}
\caption{Electrostatic coupling parameter $\Gamma_\text{el}=Z_\text{net}\lambda_\text{B}/a$ versus microgel concentration $\phi_0$, where $Z_\text{net}$ is the net microgel valence. System parameters, colors, and linetypes are same as in Fig.~\ref{figure3}.}
\label{figure1b}
\end{figure}

Figure~\ref{figure1b} shows the electrostatic coupling strength $\Gamma_\text{el}$ as a function of $\phi_0$. Notice that $\Gamma_\text{el}$ depends on the equilibrium radius $a$ and net microgel valence $Z_\text{net}$, both of which are monotonically decreasing with increasing $\phi_0$. The decrease of $Z_\text{net}$ due to inside-permeated counterions is more pronounced than the decrease of $a$ with increasing concentration, which explains the monotonic decrease of $\Gamma_\text{el}$. The overall behavior of the coupling strength as function of concentration and backbone valence is similar in the TPT and PBCM, but the TPT consistently predicts a stronger electrostatic coupling than PBCM. The greatest differences are visible for low concentrations and for the highest considered backbone valence $Z=500$ where $\Gamma_\text{el}>5$, such that nonlinear electrostatic effects, not accounted for in the linear TPT, come into play \cite{Denton_JPCM_2008,Denton_JPCM_2010,Brito_tosubmit_2019}.
We stress here that, in contrast to suspensions of impermeable, solid particles, a reduction in the concentration of permeable, compressible particles results in a strengthening of the electrostatic interparticle repulsion.
\begin{figure}[h!]
\includegraphics[width=8.5cm]{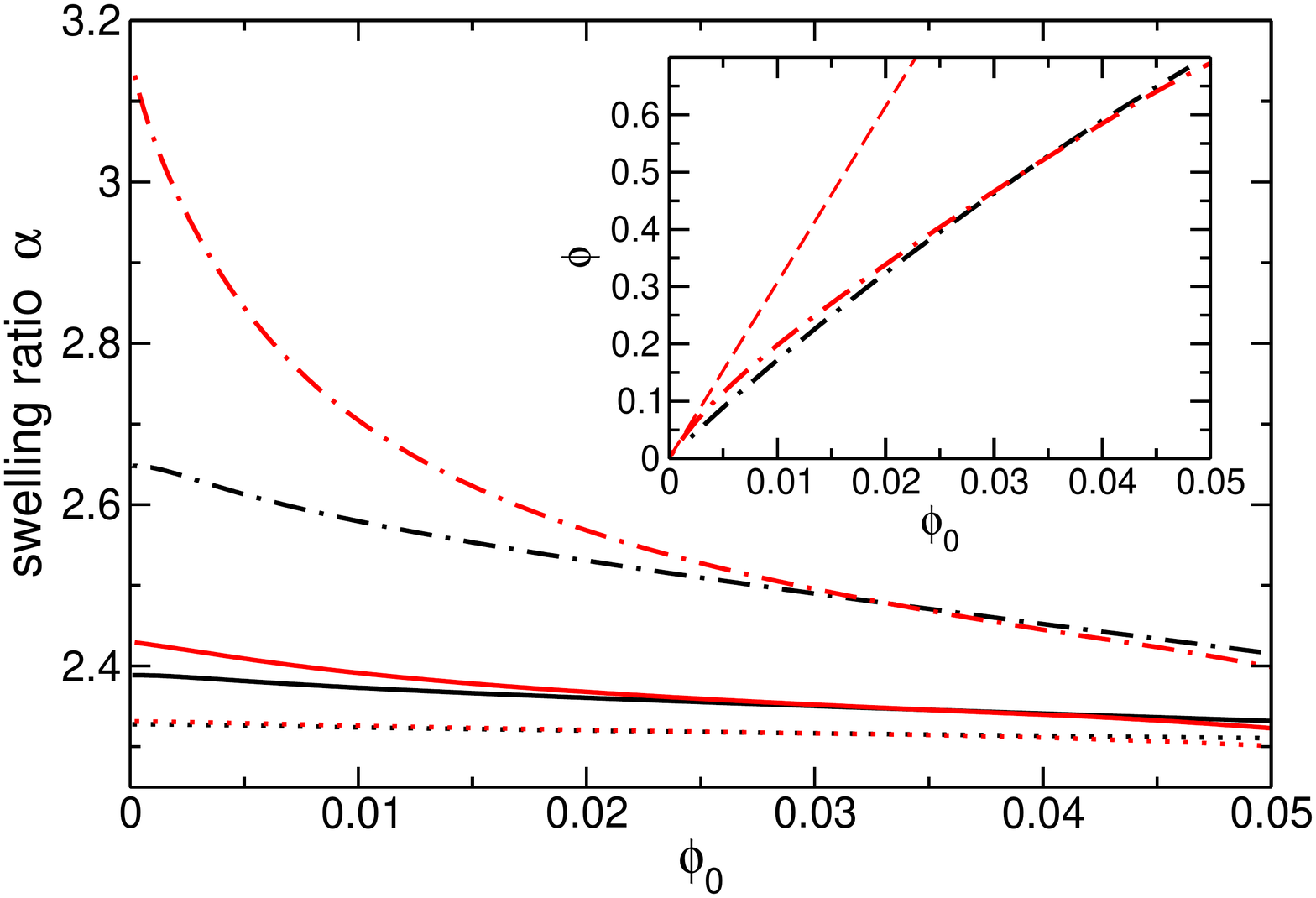}
\caption{Swelling ratio $\alpha=a/a_0$ versus reduced concentration $\phi_0$ for backbone valence $Z=500$ (dashed-dotted), $200$ (solid), and $100$ (dotted) at $c_\text{res}=100\;\!\mu$M. Inset: Swollen microgel volume fraction $\phi = \phi_0\;\!\alpha^3$ versus $\phi_0$. The straight dashed line in the inset depicts  
$\phi = \phi_0\;\!\alpha\!\left(\phi_0^\ast\right)^3$ for a fixed TPT microgel radius taken at $\phi_0^\ast=2.0\times 10^{-4}$ and backbone valence $Z=500$. Other system parameters are same as in Fig.~\ref{figure3}.}
\label{figure1}
\end{figure}

Figure~\ref{figure1} depicts the concentration dependence of the equilibrium microgel swelling ratio, $\alpha=a/a_0$, for three different backbone valences. At a given $\phi_0$, the swelling ratio increases with increasing $Z$, owing to an enhanced electrostatic repulsion between the $Z$ monovalently charged backbone sites for a constant reservoir salt concentration $c_\text{res}=100\;\mu$M. Deswelling in the counterion-dominated regime displayed in the figure is most pronounced at smaller $\phi_0$, and $a$ decreases here more strongly for higher $Z$. For $Z=500$, the TPT predicts distinctly higher swelling ratios than the PBCM, and a distinctly steeper decay of $\alpha$ with increasing $\phi_0$. 

An important quantity characterizing the swollen microgels is the volume fraction $\phi=\phi_0\;\!\alpha^3$, whose concentration dependence is shown in the inset for $Z=500$. Due to deswelling, $\phi$ increases sublinearly with increasing $\phi_0$. Differences between the TPT and PBCM predictions for $\phi$ are small except for small concentrations, where nonlinear electrostatic coupling is significant.

To assess quantitatively the effect of deswelling on structural and dynamic properties, it is useful to compare findings for the actual suspension of deswelling microgels with those for a fictitious reference suspension of non-swelling particles. We select the system parameters of the reference system to be the same as in the actual one, except for the microgel radius $a_\text{ref}$, which is fixed to the equilibrium value of the deswelling system at the lowest considered concentration, $\phi_0^\text{ref}$, where nonlinear screening by the microions can still be disregarded. Explicitly, we set $a_\text{ref}=a(\phi_0^\text{ref})$ using $\phi_0^\text{ref}=0.005$, a reservoir concentration fixed to $c_\text{res}=100\;\mu$M, and backbone valences restricted to values $Z \leq 200$.
 
\begin{figure}[h!]
\includegraphics[width=8.5cm]{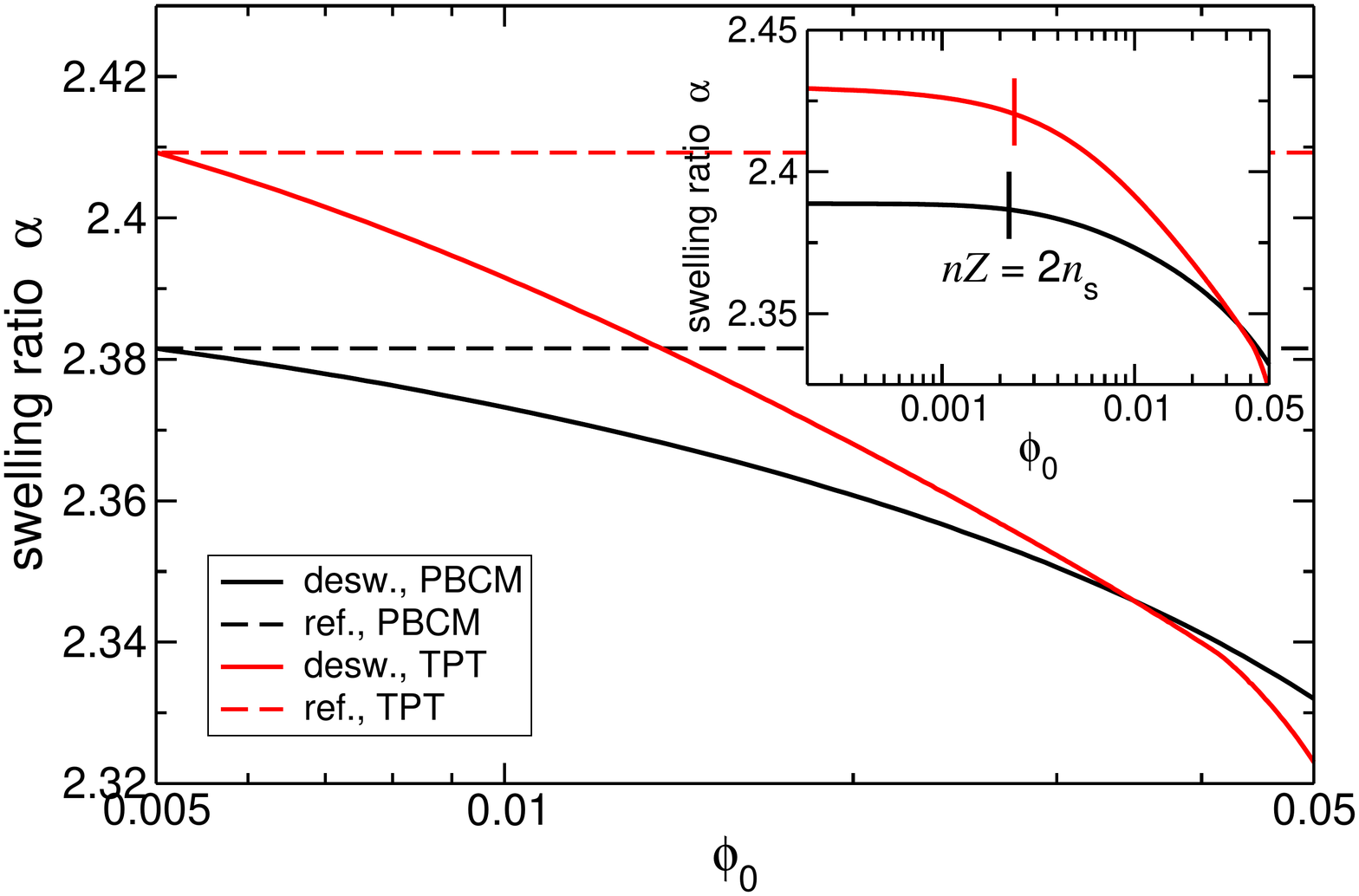}
\caption{Predictions of TPT and PBCM for swelling ratio $\alpha=a/a_0$ versus microgel concentration $\phi_0$ compared with corresponding reference system fixed value (dashed lines). 
Inset: Swelling ratio $\alpha$ for low concentrations where salt-dominated regime is resolved. 
System parameters: $Z=200$ and $c_\text{res}=100\,\mu$M.}
\label{figure4}
\end{figure}

%{\bf (Remark Mariano: This concentration value is selected based on the performance of TPT concerning $n_\text{s}$ %calculations, and the limitations of PBCM approximation in comparison to PM-cell model simulations, where the %predictions for the microion profiles have been tested. For $\phi_0<$ 0.005 and $Z\lesssim$ 200, the methods give %approximative result which properly reflect the expected physical behavior, however without robust accuracy)}.
%
Figure~\ref{figure4} shows the swelling ratio $\alpha$ predicted by the two methods, compared with the respective constant value $\alpha(\phi_0=0.005)$ (dashed horizontal lines) for the reference system. Note that the reference system microgel radius is different for the two methods, namely, $a_\text{ref}\approx 24.1$ nm in the TPT and $a_\text{ref}\approx 23.8$ nm in the PBCM. The transition from salt-ion to counterion domination occurs at very small concentrations, resolved in the inset of Fig.~\ref{figure4}. The vertical line segments mark here the microgel concentration where $Zn=2n_\text{s}$ and hence $\kappa_\text{c}=\kappa_\text{s}$. At very small concentrations where $\kappa_\text{c} < \kappa_{s}$, $\alpha$ changes only little with concentration.      

It was shown in \cite{Trizac2003,Pianegonda2007,Denton_JPCM_2008,CollaLevinTrizacJCP2009,Denton_JPCM_2010,Boon_PNAS_2015} that nonlinear electrostatic coupling, which comes into play for $\Gamma_\text{el} \gtrsim 5$, can be incorporated into linear Yukawa-type effective pair potentials, such as in Eq.~(\ref{Yukawa_part}), by using renormalized values of the particle (backbone) charge and of the Debye screening constant. Different renormalization schemes were developed for this purpose for charge-stabilized suspensions of impermeable particles \cite{Trizac2003,Pianegonda2007,Denton_JPCM_2008,CollaLevinTrizacJCP2009,Denton_JPCM_2010,Boon_PNAS_2015}, but considerably less applications to ionic microgels were reported so far \cite{Baulin_SoftMatter_2012,Colla_JCP_2014,Braibanti_PRE_2016,Nojd_SoftMatter_2018,PhysRevE.100.032602}.

In the remainder of this paper, we study fluid-phase suspensions mostly for conditions $\Gamma_\text{el}<5$ where the TPT and PBCM can be directly compared without the need to invoke microgel charge renormalization. For higher $Z_\text{net}$, this condition limits us to concentrations $\phi_0 >\phi_0^\text{ref}=0.005$ 
in the counterion-dominated regime where deswelling is most pronounced.
%	
%\textcolor{red}{We have also explored variations in the dry particle size. Comment on that?}\\
%
	
\subsection{Potential Parameters and Pressure Contributions}

Having introduced the reference system of constant-size microgels, we analyze next the 
parameters characterizing the effective pair potential of deswelling microgels, 
in comparison with the reference system values. For the considered system parameters, the likelihood of particle overlap is small. The microgel interaction is thus determined by the non-overlapping (Yukawa) part of the effective pair potential, $u_\text{Y}(r;n)$, in Eq.~(\ref{Yukawa_part}). The Yukawa potential, which is characterized by $Z_\text{net}$ and $\kappa$, can be expressed in the form
\begin{equation}\label{eq:AY}
	\beta u_\text{Y}(r;n) = 2a_0 A_\text{Y}\frac{e^{-\kappa r}}{r}\,,
\end{equation}
where $A_\text{Y}=\beta u_\text{Y}(2a_0;n)\exp(2\kappa a_0)$ is an interaction strength parameter. 

\begin{figure}[h!]
\includegraphics[width=8.5cm]{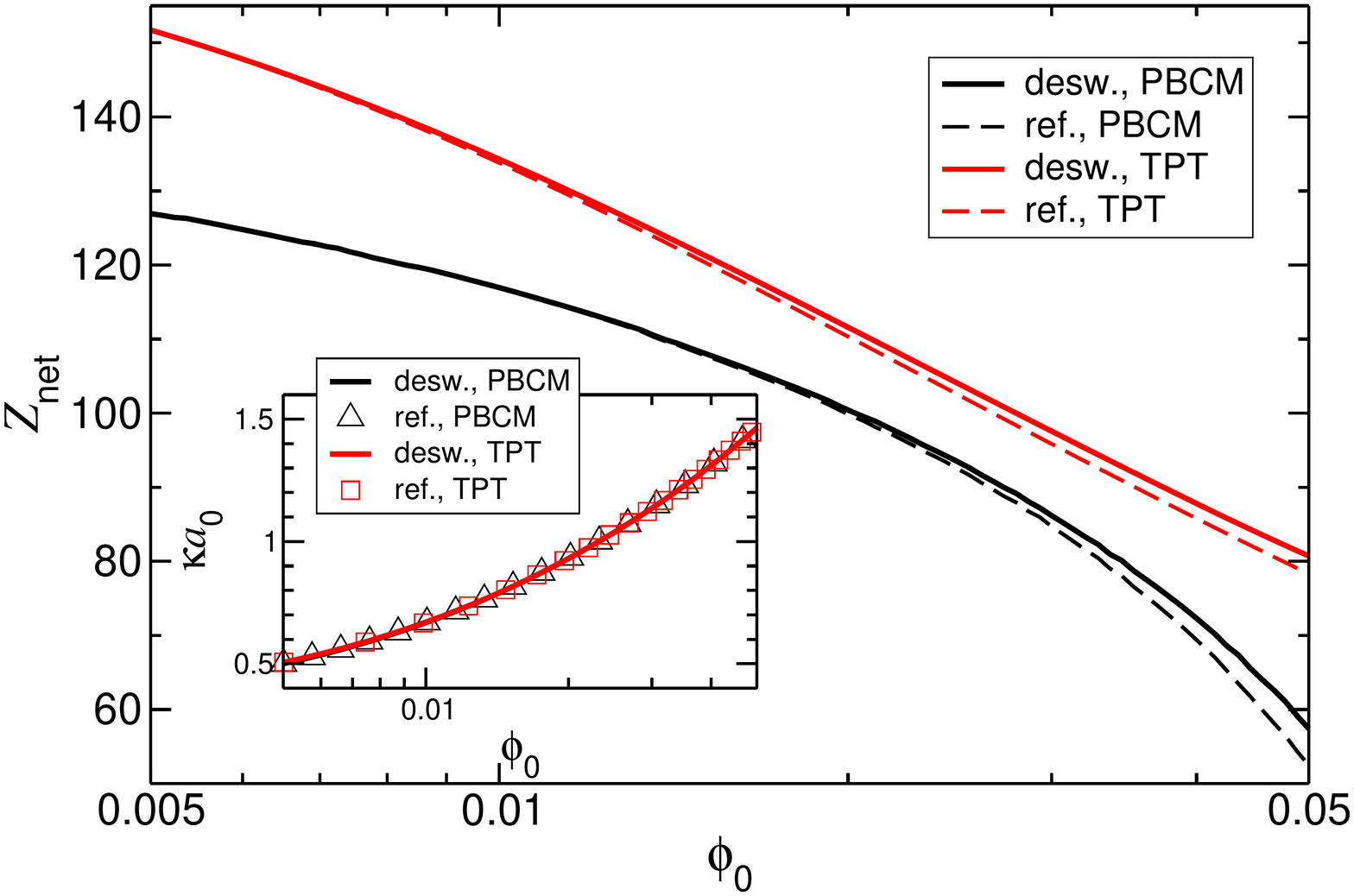}
\caption{Concentration dependence of net microgel valence $Z_\text{net}$. Inset: Reduced Debye screening constant versus $\phi_0$. Predictions of TPT and PBCM for deswelling systems (solid red and black lines, respectively) are compared with constant-size reference system predictions (dashed lines). System parameters: $Z=200$ and $c_\text{res}=100\;\mu$M.}
\label{figure5}
\end{figure}
\begin{figure}[h!]
\includegraphics[width=8.5cm]{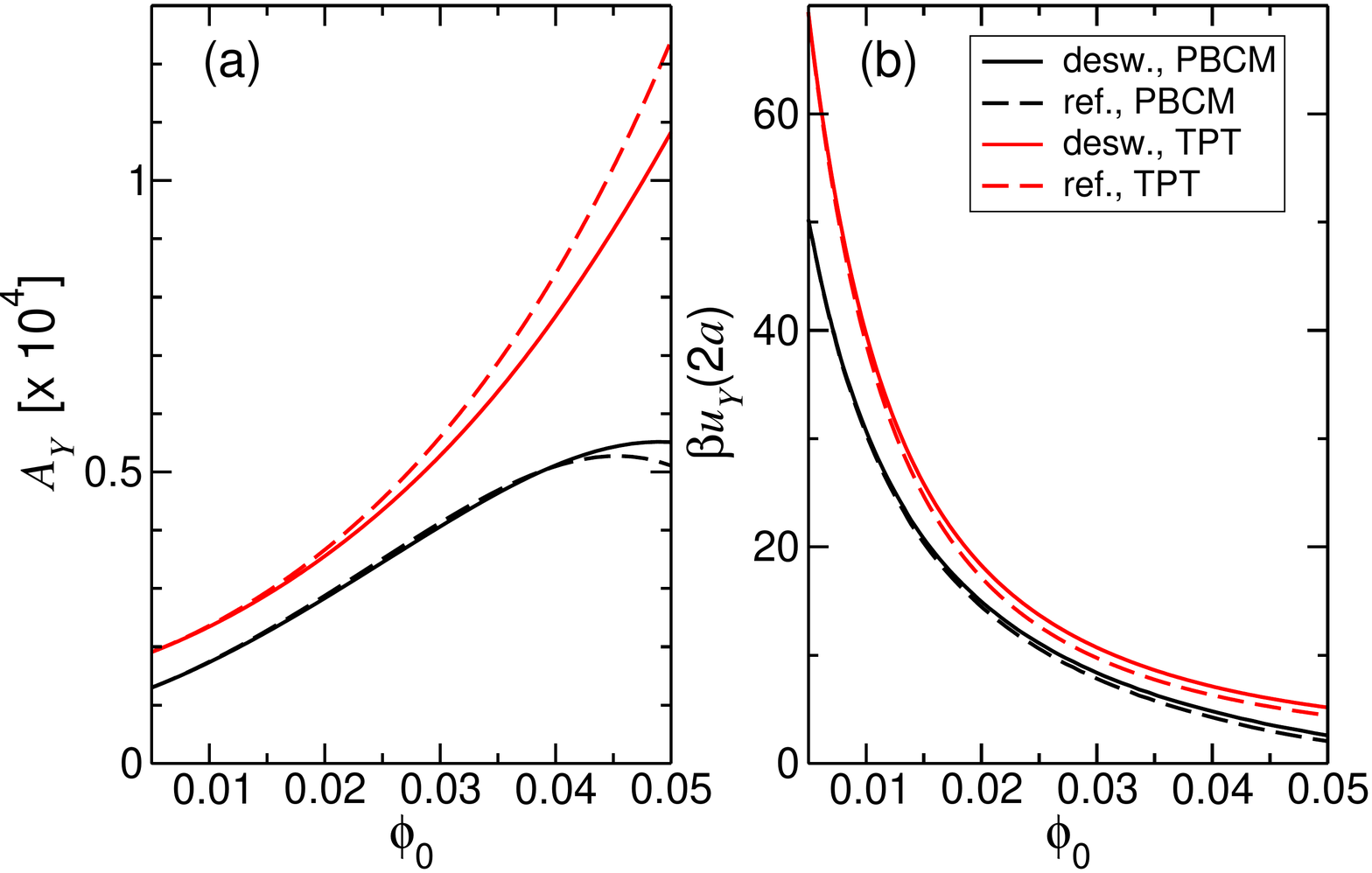}
\caption{Influence of deswelling on interaction strength parameters (a) $A_\text{Y}=\beta u_\text{Y}(2a;n)\exp(2\kappa a_0)$ and (b) $\beta u_\text{Y}(2a;n)$ of effective Yukawa pair potential for non-overlapping microgels, as predicted by TPT. System parameters: $Z=200$ and $c_\text{res}=100\;\mu$M.}
\label{figure7b}
\end{figure}

In Fig.~\ref{figure5}, the concentration dependence of the net microgel valence $Z_\text{net}$ and the Debye screening constant $\kappa$ of deswelling particles are compared with the reference system predictions. While $Z_\text{net}$ decreases with increasing concentration, $\kappa$ monotonically increases. This trend can be attributed to the associated increase in the number of counterions inside the microgels. 
Deswelling slightly increases $Z_\text{net}$, but has almost no effect on $\kappa$, which in the counterion-dominated regime is determined solely by $Z$ and $\phi_0$, independent of $n_\text{s}$ [see Eq.~(\ref{eq:Debye})]. 
At low $\phi_0$, the $Z_\text{net}$ curves merge with those of the reference system, since $a_\text{ref}$ becomes at $\phi_0=0.005$ equal to the radius $a$ of the deswelling microgels. 
Deswelling enlarges the volume available to the microions outside the microgels by a factor $V\left(\phi-\phi_\text{ref} \right)$, where $\phi_\text{ref}=\phi_0\left(a_\text{ref}/a_0\right)^3$ is the volume fraction of the reference system. The resulting gain in entropy for counterions leaving the deswelling microgels is nearly compensated by a greater work required to expel these ions, as the backbone charge density of opposite sign is increased by a factor $\left(a_\text{ref}/a\right)^3$. The net effect is an only slightly increased $Z_\text{net}$ for the deswelling microgel system. Both TPT and PBCM predict such a slight enhancement of $Z_\text{net}$ at higher $\phi_0$, but with consistently higher values for TPT.         

According to Figs.~\ref{figure7b}(a) and (b), $A_\text{Y}$ grows with increasing concentration, while $\beta u_\text{Y}(2a)$ decreases. The order relation $Z_\text{net}(\phi_0) \geq Z_\text{net}^\text{ref}(\phi_0)$ is valid, which implies the order relation $\beta u_\text{Y}(2a) \geq \beta u_\text{Y}^\text{ref}(2a)$ for the effective potential at contact distance $2a$. The opposite order $A_\text{Y} \leq A_\text{Y}^\text{ref}$ holds for the interaction parameter $A_\text{Y}$ in Eq.~(\ref{eq:AY}).  To understand these relations, recall with Eq.~(\ref{eq:NetCharge}) that $A_\text{Y}$ is proportional, in addition to $Z_\text{net}^2$, to a geometric factor depending on $\kappa a$,  
and this factor is higher for the reference system. Fig.~\ref{figure7b}(b) quantifies the aforementioned peculiarity of ionic microgel systems that, with decreasing concentration, 
the electrostatic coupling strength measured at contact distance is increased. 

\begin{figure}[h!]
\includegraphics[width=8.5cm]{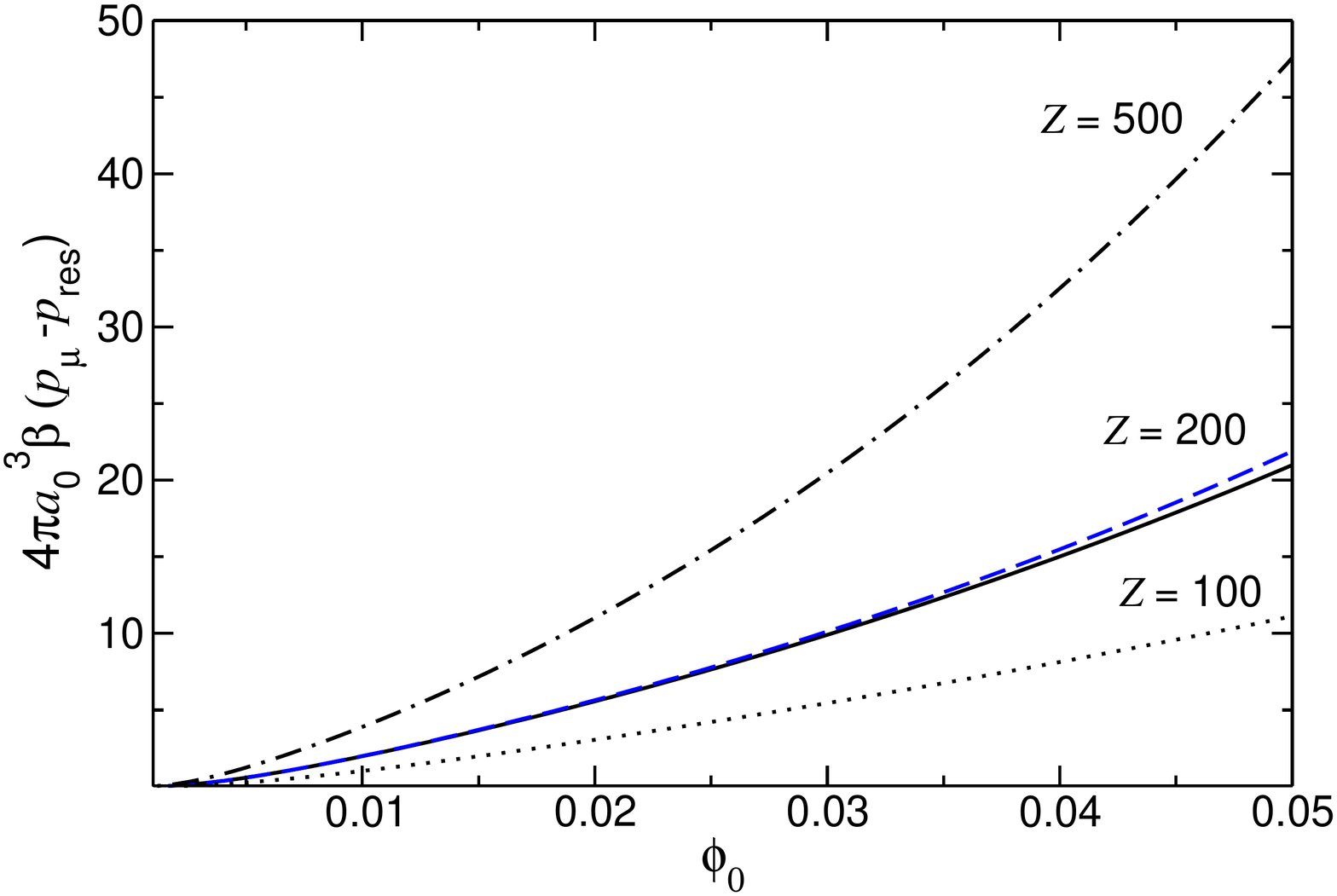}
\caption{PBCM prediction for microion osmotic pressure $p_\mu-p_\text{res}$ (in reduced units) versus $\phi_0$ for backbone valences $Z$ as indicated and reservoir pressure $p_\text{res}=2 n_\text{res} k_\text{B}T$. For $Z=200$, microion osmotic pressure of reference system is also shown (blue dashed line). Other parameters as in Fig.~\ref{figure3}.}
\label{extra6}
\end{figure}
Having assessed how the effective pair potential is affected by deswelling, we address next various pressure contributions. Figure~\ref{extra6} displays PBCM results for the microion pressure $p_\mu$, calculated using the contact theorem in Eq.~(\ref{eq:Contact}). As expected, for given $\phi_0$,
$p_\mu$ grows rapidly with increasing backbone valence. For $Z=200$, the microion pressure of the reference system slightly exceeds the pressure for deswelling particles, essentially due to the higher volume fraction, $\phi_\text{ref}>\phi$, of the reference system.

\begin{figure}[h!]
\includegraphics[width=8cm]{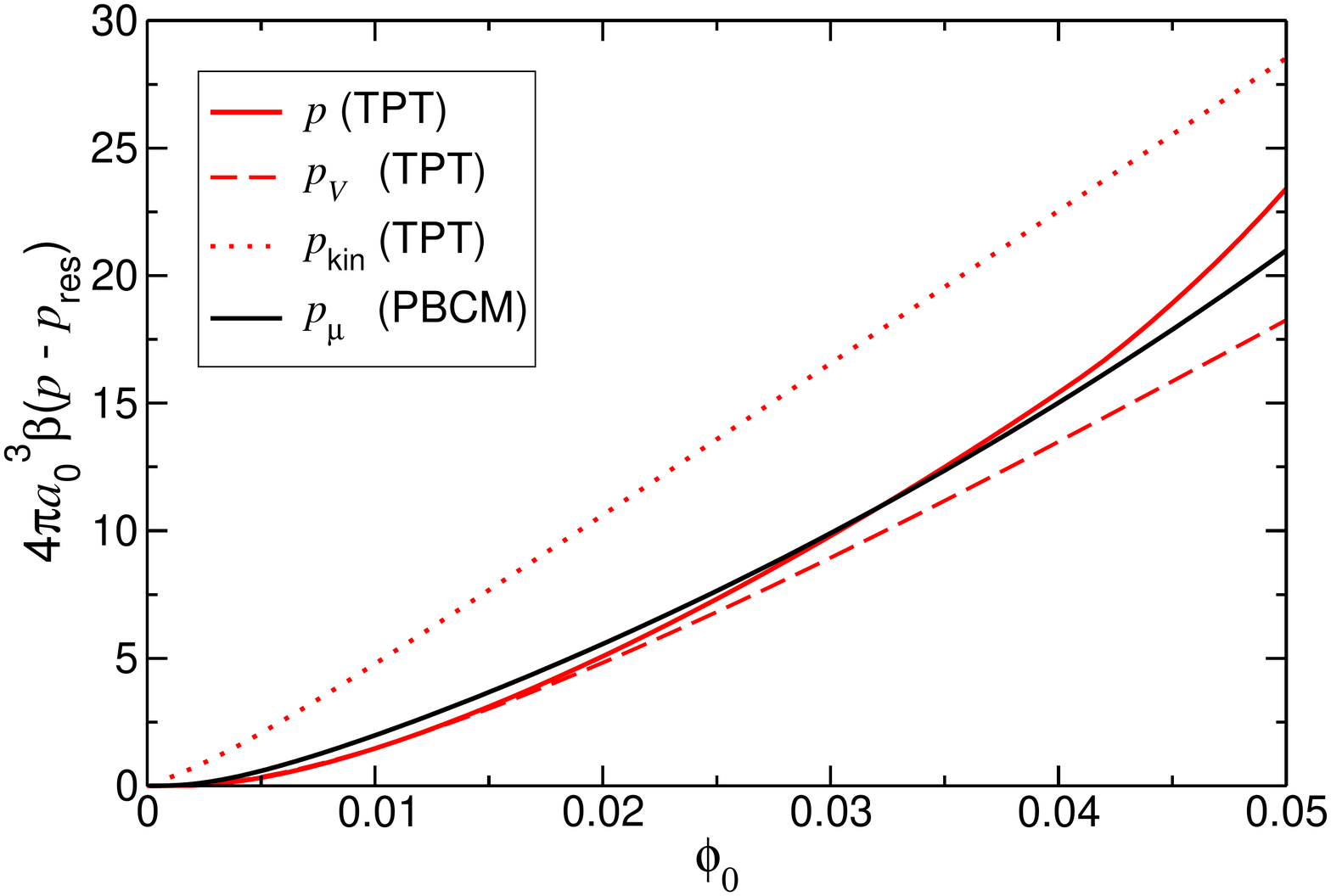}
\caption{Reduced pressure $p$ of microgel suspension from TPT [Eqs.~(\ref{eq:Perturbation}) and (\ref{eq:ThermoRelation})], volume energy contribution $p_\text{V}$ [Eq.~(\ref{eq:VolumePressure})], and PBCM pressure $p_\mu$ [Eq.~(\ref{eq:Contact})] versus microgel concentration $\phi_0$. 
All pressures are relative to reservoir pressure $p_\text{res}$. System parameters are $Z=200$ and $c_\text{res}=100~\mu$M. Also shown is the TPT prediction for the kinetic (ideal gas) pressure, $p_\text{kin}= Zn + 2n_\text{s}$, where $n_\text{s}$ is system salt density (nearly identical to PBCM prediction).
%{\bf Alan: we suggest to replace the black dotted curve for the kinetic pressure with $n_\text{s}$ in the PBCM by the one using TPT (now as red dotted line). Text is already updated.}
}
\label{fig:PressureComparison}
\end{figure}

It is instructive to compare the PB cell model pressure $p_\mu$ with the total suspension pressure $p$ from TPT and the volume energy-derived contribution $p_\text{V}$. Such a comparison is shown in Fig.~\ref{fig:PressureComparison} for a system with $Z=200$ and $c_\text{res}=100\;\mu$M. All pressures are measured relative to the reservoir osmotic pressure $p_\text{res}$. Here, $p_\mu$ is calculated according to Eq.~(\ref{eq:Contact}) using the PBCM microion concentrations $n_\pm(R)$ at the cell edge, $p$ according to Eq.~(\ref{eq:ThermoRelation}) with TPT input for $f(a,n)$ [Eq.~(\ref{eq:Perturbation})], and $p_\text{V}$ according to the linear-response expression in Eq.~(\ref{eq:VolumePressure}). Also shown is the kinetic microion pressure $p_\text{kin}=\left(Zn+2\;\!n_\text{s}\right)k_\text{B}T$, with $n_\text{s}$ calculated from TPT. In the dilute limit ($\phi_0 \to 0$), all pressure terms converge to $p_\text{res}$, and the system salt concentration $n_\text{s}$ tends to $n_\text{res}$. 
The dry volume fraction at which $Zn = 2n_\text{s}$ is $\phi_0 \approx 0.002$. The displayed pressure curves hence represent the counterion-dominated regime.

As seen from comparing $p$ to $p_\text{V}$, inter-microgel correlation contributions to $p$ are significant for $\phi_0 \gtrsim 0.04$ where $p$ becomes distinctly higher than $p_\text{V}$. This comparison shows further 
that the pressure contribution $p_\text{se}$, generated by the concentration dependence of $a(n)$ in the single-particle energies $u_\text{e}(a)$ and $f_\text{p}(a)$ in Eq.~(\ref{eq:FperParticle}), is negligible at lower concentrations. The PBCM pressure $p_\mu$ exceeds $p_\text{V}$ for non-zero concentrations and is overall close to $p$, except at high $\phi_0$. At this relatively low salt concentration, the kinetic microion pressure difference $p_\text{kin}-p_\text{res}$ (dotted curve) is practically equal to the reduced ideal gas pressure of counterions, $Znk_\text{B}T$ (or $3Z\phi_0$ in reduced units), up to a small negative correction proportional to $2\left(n_\text{s}-n_\text{res}\right)$, owing to the salt expulsion (Donnan) effect (cf. Fig.~\ref{figure3}). While in the concentration range of Fig.~\ref{fig:PressureComparison} the counterions contribute most strongly to the suspension osmotic pressure, due to the electrostatic attraction of the fixed backbone charge they behave distinctly non-ideal, which is reflected in the non-constant, radially decaying counterion concentration profile $n_{+}(r)$ (see Fig.~\ref{figure3b}). 
This is why $p_\text{kin}$ is higher than $p$.  

\subsection{Structural Properties and Charge Renormalization}

As explained in Sec.~\ref{sec:StructureThermodynamics}, using $u_\text{eff}(r;n)$ with associated values for the equilibrium radius $a$, net valence $Z_\text{net}$, and screening constant $\kappa$, one can calculate the microgel radial distribution function (rdf) $g(r)$ and static structure factor $S(q)$ characterizing pair correlations in real and Fourier space, respectively. For this purpose, we use the Rogers-Young (RY) integral-equation scheme. This thermodynamically self-consistent scheme is known, from comparisons with computer simulation results, to be very accurate for fluids of charge-stabilized particles interacting via a repulsive Yukawa-type potential below concentrations where the suspension crystallizes \cite{Gapinski_JCP2012,Gapinski_JCP_2014,Banchio_JCP_2018}.

\begin{figure}[h!]
\includegraphics[width=8.5cm]{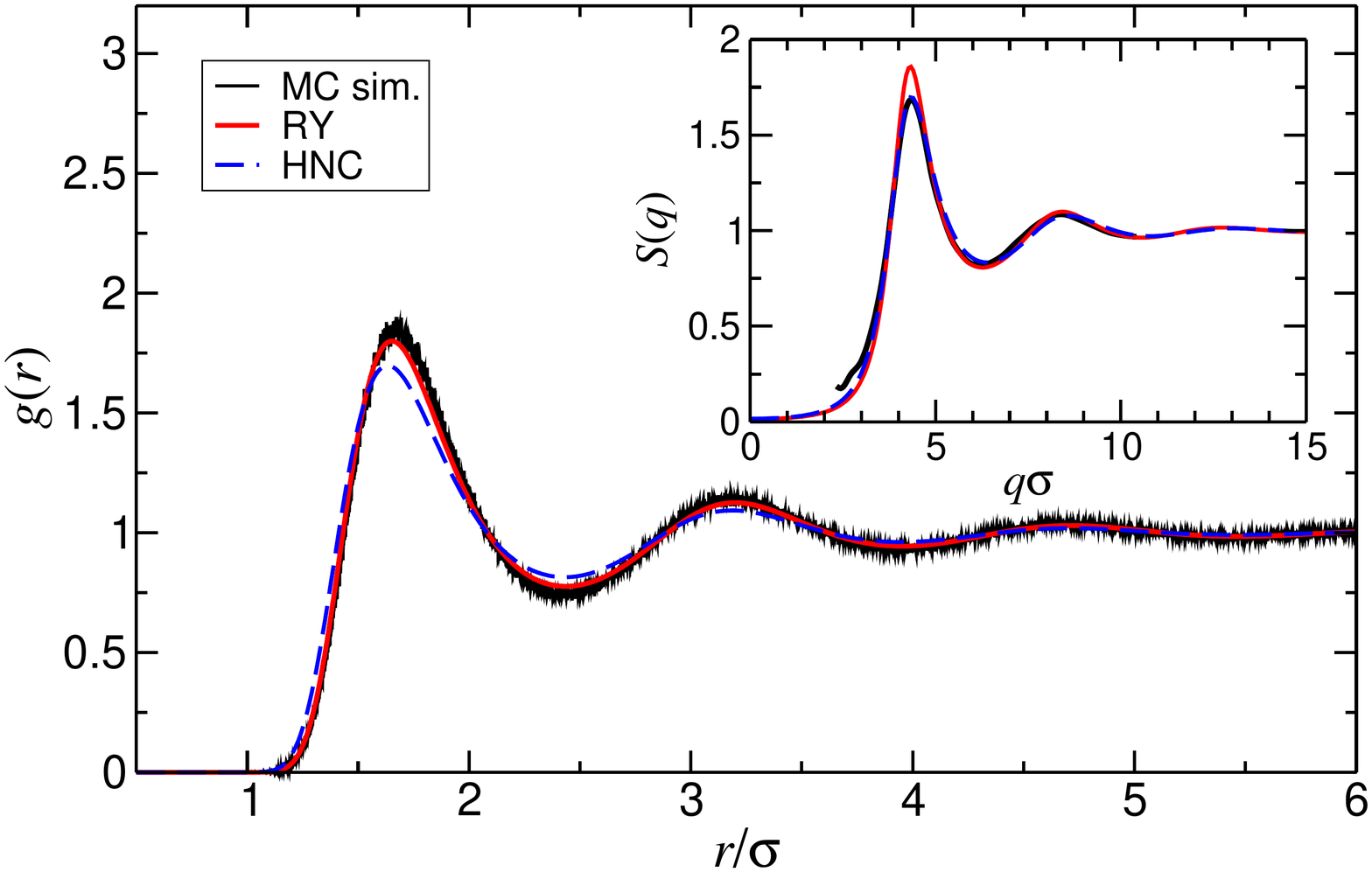}
\caption{Results of RY, HNC, and MC for radial distribution function, $g(r)$, and static structure factor, $S(q)$ (inset), of a salt-free suspension with $\phi_0=0.01$, $Z=100$, $\alpha=2.327$, and $\kappa a_0=0.463$. Swelling ratio $\alpha$ is computed using TPT. Length unit is microgel diameter $\sigma=2a$.}
\label{figure7}
\end{figure}

To illustrate the high accuracy of the RY method for microgel particles interacting via the pair potential in Eq.~(\ref{eff_pot}), in Fig.~\ref{figure7} the RY results for $g(r)$ and $S(q)$ are compared with Monte-Carlo (MC) simulation data obtained using the method in \cite{Weyer_SoftMatter_2018} for a salt-free suspension with $Z=100$. The RY predictions are also compared with results from the numerically faster, but thermodynamically not self-consistent, hypernetted chain (HNC) integral-equation scheme. There is overall good agreement between the RY and MC data for $Z=100$, while the real-space pair correlations are underestimated by the HNC scheme. 

%{\bf The state point of the $Z=200$ system is located inside the fluid phase region of the known phase diagram for %particles with Yukawa-type interactions \cite{Gapinski_JCP_2014}, we conclude that the MC simulation data at $Z=200$ %for $g(r)$ and $S(q)$ describe a system not fully equilibrated. We have skipped the figure part 8 (b)?)}.  
%

\begin{figure}[h!]
\includegraphics[width=8.5cm]{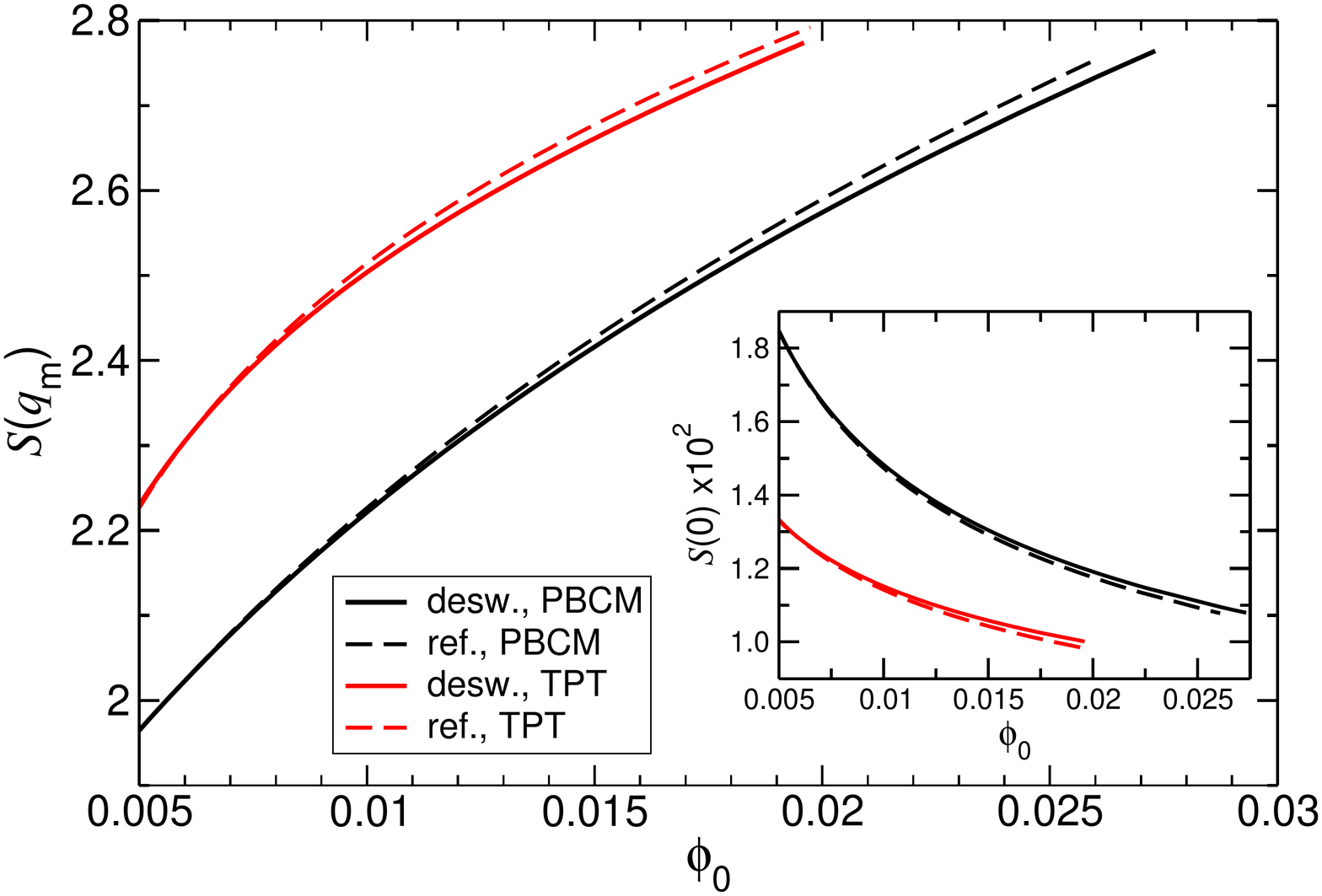}
\caption{RY structure factor peak height, $S(q_\text{m})$, and osmotic compressibility factor, $S(q\to 0)$ (inset), versus $\phi_0$ for $Z=200$ and $c_\text{res}=100\;\mu$M. Results are presented for deswelling microgels with radius $a$ computed in TPT and PBCM and compared with corresponding results for reference system (dashed curves). Other parameters as in Fig.~\ref{figure4}.}
\label{figure8}
\end{figure}

Rogers-Young results for the concentration dependence of the structure factor peak height $S(q_\text{m})$ and the osmotic compressibility factor $S(0)$ (inset) are displayed in Fig.~\ref{figure8}. The equilibrium radius $a$ in the potential $u_\text{eff}(r;n)$, on which the RY calculations are based, is determined here using the TPT and PBCM. The contact value, $g(\sigma)$, of the associated rdf remains small in the considered concentration range, showing that the no-overlap potential, $u_\text{Y}(r)$, essentially determines the microstructure of the microgels. The overlap potential, $u_\text{ov}(r)+u_\text{H}(r)$ in Eq.~(\ref{eff_pot}), comes into play only at high concentrations. From comparison with the reference system peak height predictions, one notices that $S(q_\text{m})$ is reduced when deswelling is accounted for, though only slightly, since for $Z=200$ the decrease in the microgel radius relative to the reference value remains small even at higher concentrations (cf. Fig.~\ref{figure4}).

For a given concentration, the TPT predicts a more structured system than the PBCM, as reflected by the higher values of $S(q_\text{m})$. This difference originates from the higher net charge and larger microgel radius in the TPT, as discussed already in relation to Figs.~\ref{figure1b} and \ref{figure1}.

In discussing the Kirkwood-Buff relation [Eq.~(\ref{eq:OsmoticCompressibility})], we noted that, for a monodisperse suspension in osmotic equilibrium with a salt reservoir, $S(0)=S(q \to 0)$ equals the osmotic compressibility factor. 
Rogers-Young predictions for the concentration dependence of $S(0)$ (inset of Fig.~\ref{figure8}) show that deswelling slightly increases the osmotic compressibility. The increase of $S(0)$ predicted by both methods 
follows from the reduced volume fraction of deswelling particles, which is lower by the factor $\left(a/a_\text{ref}\right)^3$ than that of the reference system. In its effect on $S(0)$, this reduction in volume fraction overcompensates the small deswelling-induced increase of $Z_\text{net}$ (see Fig.~\ref{figure5}). 
The PBCM yields distinctly higher compressibilities than the TPT, 
since it predicts smaller equilibrium radii and net charges. 
  
The peak value of the structure factor, $S(q_\text{m})$, can be used as an indicator for the proximity of a fluid suspension to a freezing transition. The frequently cited empirical Hansen-Verlet criterion, $S(q_\text{m})=2.85$, applies only to the freezing of a hard-sphere fluid. It does not apply to suspensions with longer-range, soft inter-particle repulsion. As shown in detail in \cite{Gapinski_JCP_2014}, 
a somewhat higher freezing indicator value, $S(q_\text{m})=3.1$, should be used for suspensions with 
long-range Yukawa-type repulsion, where overlap configurations are unlikely.  
An alternative indicator of freezing in these systems, applicable for very low salinity only, where $\kappa n^{-1/3} \lesssim 7$ and freezing into a bcc lattice takes place, is the value $g(r_\text{m})\approx 2.6$ for the principal rdf peak height at radial distance $r_\text{m}$ \cite{Gapinski_JCP_2014}.

To illustrate how the freezing transition concentration is determined using the citerion $S(q_\text{m})=3.1$, 
we consider a strongly charged microgel suspension with $Z=500$ and $c_\text{res}=50\;\mu$M, for which 
$g(2a)\approx 0$ holds to excellent accuracy up to the freezing transition concentration. For such a strongly coupled system, 
it is necessary to renormalize the (net) microgel charge and suspension screening constant, so as to incorporate nonlinear response of the microions to the strong electric field of the microgel backbone. To determine these renormalized parameters in the framework of the PBCM, we follow Colla {\it et al.} \cite{Colla_JCP_2014} in linearizing the PB equation [Eq.~(\ref{non-lin_PBeq_CM})] around the nonlinear potential value $\Phi_\text{R}=\Phi(R)$ at the cell boundary. This procedure leads to a linear PB equation,
\begin{equation}
\Delta\Phi_\text{l}(r)=\kappa_\text{eff}^2
\left[\Phi_\text{l}(r)-\Phi_\text{R}+\gamma_\text{R}\right]+\frac{3 \lambda_\text{B} Z^\text{ren}}{a^3}\;\!\Theta(a-r)\,,
\label{lin_PBeq_CM}
\end{equation}
%\begin{eqnarray}
%\Delta\Phi_\text{l}(r)&=& \kappa_\text{eff}^2
%\left[\Phi_\text{l}(r) -\Phi_\text{R}+\gamma_\text{R} \right]\nonumber\\
%&&+ \frac{3 \lambda_\text{B} Z^\text{ren}}{a^3}\;\!\Theta(a-r)\,,
%\label{lin_PBeq_CM}
%\end{eqnarray}
%    
for $r \leq R$, where $\kappa_\text{eff}^2 = \kappa_\text{res}^2 \cosh\left(\Phi_\text{R}\right)$, and $\gamma_\text{R}=\tanh\left(\Phi_\text{R}\right)$, with the latter quantity being negative due to the negative backbone charge. 
 Here, $Z^\text{ren}$ is the yet unknown renormalized backbone valence, and $\Theta(r)$ is the unit step function.

The unique solution for the linearized potential, $\Phi_\text{l}(r)$, inside and outside the microgel sphere can be obtained analytically using the boundary 
conditions $\Phi_\text{l}(R)=\Phi_\text{R}$, $\Phi_\text{l}'(R)=0$, and $\Phi_\text{l}'(0)=0$, 
in conjunction with the continuity of $\Phi_\text{l}(r)$ and its first derivative at $r=a$. These five conditions determine $Z^\text{ren}$, together with the four integration constants arising from the integration of the linearized PB equation [Eq.~(\ref{lin_PBeq_CM})] inside and outside a microgel sphere. Input parameters are here $\Phi_\text{R}$ 
and the equilibrium radius $a$, determined independently. Note that the linearized potential $\Phi_\text{l}(r)$ gives rise to the same potential and electric field values at the cell boundary as the nonlinear PBCM potential. 
We refrain from quoting the somewhat lengthy analytic expressions for $\Phi_\text{l}(r)$ and $Z^\text{ren}$ given in \cite{Colla_JCP_2014}. As discussed in \cite{Colla_JCP_2014} (see also \cite{Levin:2002}), due to the monotonic increase of the microgel radius with increasing bare backbone valence $Z$, the renormalized valence $Z^\text{ren} \leq Z$ does not reach a saturation value beyond the linear regime, as it does for non-permeable rigid colloids \cite{Alexander1984,Trizac_Langmuir:2003}. Instead, $Z^\text{ren}$ grows monotonically with increasing $Z$, showing only a slight indication of a plateau behavior in the regime of intermediately high $Z$ and low suspension salinity.     

In Donnan equilibrium, the renormalized {\it net} microgel charge number, $Z_\text{net}^\text{ren}$, is obtained in the PBCM as 
\begin{eqnarray}
Z_\text{net}^\text{ren}&=&-\frac{a^2 \Phi'_\text{l}(a)}{\lambda_\text{B}}\nonumber\\
&=&\frac{\tanh(\Phi_\text{R})}{\kappa_\text{eff}\lambda_\text{B}}
\Big[\kappa_\text{eff}(a\!-\!R)\cosh(\kappa_\text{eff}(a\!-\!R))
\nonumber\\
&+&(\kappa_\text{eff}^2aR\!-\!1)\sinh(\kappa_\text{eff}(a\!-\!R))\Big],\,
\label{Z_net^ren}
\end{eqnarray}
%\begin{eqnarray}
%  &&Z_\text{net}^\text{ren}=-\frac{a^2 \Phi'_\text{l}(a)}{\lambda_\text{B}}=\frac{\tanh(\Phi_\text{R})}{\kappa_\text{eff}\lambda_\text{B}}\times\nonumber\\
%  &&\left[\kappa_\text{eff}(a\!-\!R)\cosh(\kappa_\text{eff}(a\!-\!R))\!+\!(\kappa_\text{eff}^2aR\!-\!1)\sinh(\kappa_\text{eff}(a\!-\!R))\right],\nonumber\\ 
%\label{Z_net^ren}
%\end{eqnarray}
% 
which follows alternatively from Eq.~(\ref{Def_Znet}), wherein  
$Z$ and $n_\pm(r)$ on the right-hand side are replaced, respectively, by $Z^\text{ren}$ and the linearized microion profiles
\begin{eqnarray}
  n_{\text{l},\pm}(r)=n_\text{res} e^{\mp \Phi_\text{R}}\left[1\mp \left(\Phi_\text{l}(r)-\Phi_\text{R}\right)
  \right]\,, 
\label{eq:n_pm_lin}
\end{eqnarray}
whose values at the cell boundary match the non-linearized ones.  

To implicitly account for nonlinear effects, we use $Z_\text{net}^\text{ren}$ given by Eq.~(\ref{Z_net^ren}) as the input for the net valence in the linear-response, no-overlap Yukawa potential 
$u_\text{Y}(r)$ in Eq.~(\ref{Yukawa_part}). In addition, $\kappa_\text{eff}$ might be 
identified as the renormalized input for the screening constant in $u_\text{Y}(r)$, 
as done based on the original description by Alexander {\it et al.} \cite{Alexander1984} for the case of non-permeable rigid spheres \cite{Trizac_Langmuir:2003,Boon_PNAS_2015}. However, to recover the PBCM screening constant in Eq.~(\ref{eq:kappa_ast}) in the limiting case of low backbone charges, where charge renormalization is not operative, 
we determine the renormalized screening constant, $\kappa^\text{ren}$, to be substituted into $u_\text{eff}(r)$, 
in a manner that maintains the smoothness of the effective potential at $r=a$ for the nonlinear case. 
Explicitly, we determine $\kappa^\text{ren}$ as 
\begin{eqnarray}
 \left(\kappa^\text{ren}\right)^2=4\pi \lambda_\text{B} \left(n Z^\text{ren}_\text{app} + 2\;\!n_\text{s}^\text{ren} \right)\,, 
\label{eq:kappa_ren}
\end{eqnarray}
where 
\begin{eqnarray}
 n_\text{s}^\text{ren} = \frac{4\pi}{V_\text{R}}\int_0^R\;\!n_\text{l,-}(r)\;\!r^2\;\!dr\,.
\end{eqnarray}
The apparent renormalized backbone valence, $Z^\text{ren}_\text{app}$, is defined by Eq.~(\ref{eq:Zast}) using the substitutions $Z^\ast \to Z^\text{ren}_\text{app}$, $\kappa^\ast \to \kappa^\text{ren}$, and $Z^\ast_\text{net}\to Z^\text{ren}_\text{net}$. Notice that $\kappa^\text{ren}$ is given here only implicitly, 
so that an iteration procedure with starting seed $\kappa_\text{eff}$ is used for its calculation. For $\kappa_\text{eff}$, we obtain the expression
\begin{eqnarray}
\left(\kappa_\text{eff}\right)^2=\frac{4\pi \lambda_\text{B}}{1-\gamma_R}
  \left[n Z^\text{ren} + \frac{2\;\!n_\text{s}^\text{ren}}{1+\gamma_R} \right]\,,
\label{eq:kappa_eff}
\end{eqnarray}
identical to the one for non-permeable charged colloids \cite{Trizac_Langmuir:2003}. The screening constants $\kappa_\text{eff}$, $\kappa^\text{ren}$, and $\kappa$ mutually differ, except in the limit $Z \to 0$, where $\gamma_\text{R} \to 0$ and $\{ n_\text{s}, n_\text{s}^\text{ren} \} \to n_\text{res}$, in which case all three quantities then equal the reservoir screening constant, $\kappa_\text{res}$. 
\begin{figure}[h!]
\includegraphics[width=8.5cm]{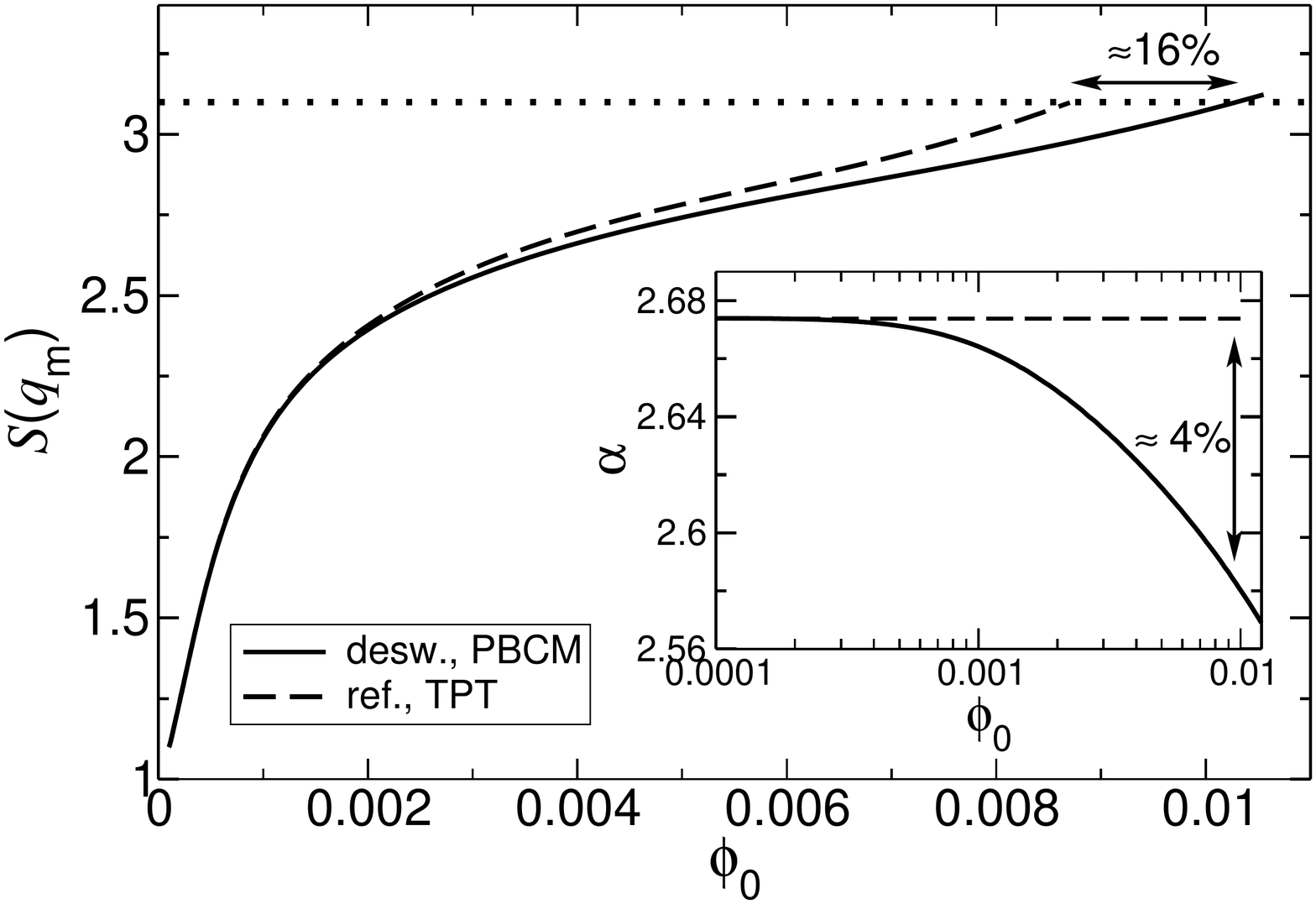}
\caption{RY peak height, $S(q_\text{m})$, for strongly repelling microgels with $Z=500$ and $c_\text{res}=50\;\mu$M. Results are presented for deswelling (solid curve) and reference system constant-size microgels (dashed curve), 
based on the 
charge-renormalized PBCM. The dotted horizontal line marks 
the freezing criterion $S(q_\text{m})=3.1$. 
Inset: Swelling ratio $\alpha =a/a_0$ in charge-renormalized PBCM.}
\label{figure9}
\end{figure}

Rogers-Young results for the concentration dependence of $S(q_\text{m})$ of strongly charged microgels are displayed in Fig.~\ref{figure9}. The effective pair potential parameters are determined here using the PBCM charge-renormalization method described above. The inset shows the size ratio $a(\phi_0)/a_\text{ref}$ with $a_\text{ref}=a(\phi_0=0.005)$, as predicted by the nonlinear PBCM method. 
The RY values for $g(2a)$ are practically zero (i.e. $g(2a)<0.001$) for all considered $\phi_0$,
so that $S(q_\text{m})= 3.1$ qualifies as a freezing indicator. 
Figure~\ref{figure9} illustrates that, for strongly charged microgels, deswelling significantly increases the freezing transition concentration by about $16\;\%$, corresponding to a 
$4\;\%$ decrease in the swelling ratio $\alpha$ (see inset). 

For deswelling microgels, the freezing transition concentration determined by $S(q_\text{m})=3.1$ is $\phi_0 \approx 0.01$. The RY rdf peak height is here  $g(r_\text{m})=2.7$, which is close to the freezing transition value $2.6$ holding for suspensions of colloids interacting by a repulsive hard-core-Yukawa pair potential, whose state points in the phase diagram are located on the fluid-bcc part of the freezing transition line, characterized by $\kappa_\text{coll} n^{-1/3} \lesssim 7$ \cite{Gapinski_JCP_2014}. If we identify $\kappa_\text{coll}$ by $\kappa^\text{ren}$, where  $\left(\kappa^\text{ren}\right)^2 = 4\pi \lambda_\text{B} n Z_\text{app}^\text{ren}$ for the present counterion-dominated microgel system, we obtain $\kappa_\text{coll} n^{-1/3} \approx 6.3$, consistent with a fluid-bcc freezing transition quite close to the fluid-bcc-fcc triple point \cite{Gapinski_JCP_2014}.

%	\textcolor{blue}{After analyzing the influence of the deswelling effect on $\phi$, $Z_\text{net}$ and $\kappa$, we concluded that the deswelling does not have effect on $\kappa$ and any influence of deswelling on quantities, which only depend on $u_\text{eff}(r)$, is through the influence on $\phi$ and $Z_\text{net}$. Closer analysis of $S(0)$ vs. $\phi_0$ indicates that the influence of the deswelling on $\phi$ dominates over the influence on $Z_\text{net}$. Now I wonder, If I plot $S(0)$ vs. $\phi$, for deswelling and fixed-size case, is the possible difference between the curves only due to the deswelling influence on $Z_\text{net}$. In Fig.~\ref{figure7b}, I plot the amplitude of $u_\text{Y}(r)$ versus both dry volume fraction and swollen volume fraction.}

\subsection{Diffusion and Rheological Properties}

We explore next dynamic properties of ionic microgel suspensions, using the one-component model of pseudo-microgels interacting via $u_\text{eff}(r;n)$. The deswelling ratio $\alpha(\phi_0)$, net valence $Z_\text{net}(\phi_0)$, and Debye screening constant $\kappa(\phi_0)$ in this model are determined using the TPT and PBCM methods. As explained in Subsec.~\ref{Subsec:Methods}, the employed methods for calculating dynamic properties depend on $u_\text{eff}(r;n)$ only implicitly via the radial distribution function $g(r)$ and static structure factor $S(q)$. On taking into account that solvent permeability effects are very small for non-overlapping ionic microgels \cite{Holmqvist_PRL_2012,Riest_SoftMatter_2015}, we identify the hydrodynamic microgel radius $a_\text{H}$ with the equilibrium radius $a(\phi_0)$.

\begin{figure}[h!]
\includegraphics[width=8.2cm]{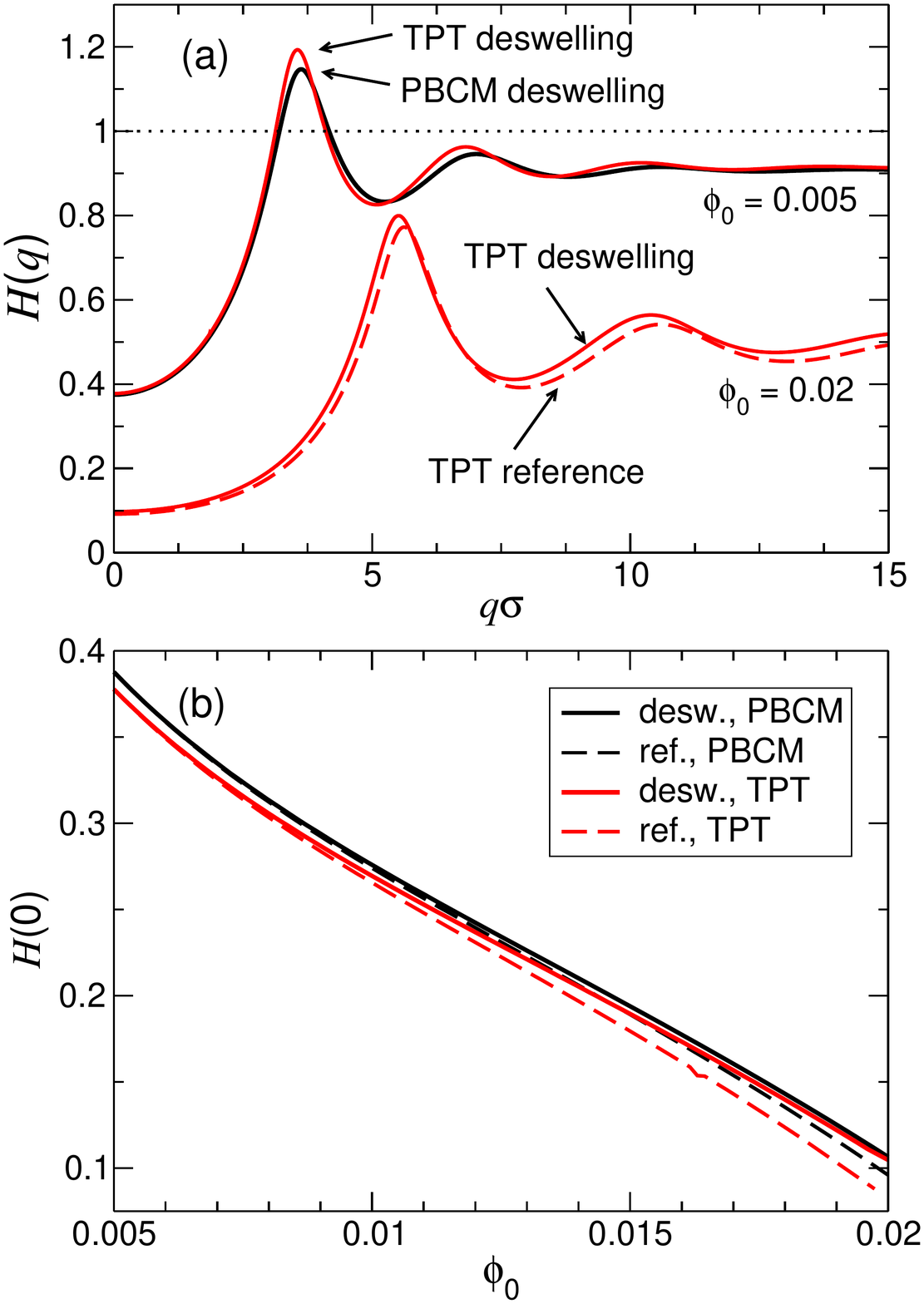}
\caption{(a) BM-PA results for the hydrodynamic function $H(q)$ as function of reduced wavenumber, $q\sigma$, for two concentrations $\phi_0$ as indicated. The lower one, $\phi_0\approx 0.005$, is the concentration where the collective diffusion coefficient attains its maximum. At $\phi_0=0.02$, the $H(q)$ of deswelling particles with TPT calculated radius is compared with that of the reference system.  
(b) Concentration dependence of sedimentation coefficient $K=H(q\to 0)$.  
Solid curves: deswelling particles in the TPT (red) and PBCM (black). 
System parameters: $Z=200$ and $c_\text{res}=100\;\mu$M.}
\label{extra1}
\end{figure}

Figure~\ref{extra1}(a) displays our results for the positive definite hydrodynamic function $H(q)$ at concentration $\phi_0=0.005$, calculated using the BM-PA hybrid scheme described in Subsec.~\ref{Subsec:Methods}, which requires $S(q)$ as the only input. This input is calculated using the RY scheme, which gives somewhat different results 
in the TPT and PBCM, respectively, owing to their different predictions for $a$ and $Z_\text{net}$. For example, at $\phi_0=0.005$, we find $\phi=0.067$ in the PBCM and $\phi=0.073$ in the TPT. The differences in $S(q)$ cause less pronounced differences in $H(q)$, since the latter depends on $S(q)$ only in a global (functional) way \cite{Heinen_Rheo_JCP_2011,Riest_SoftMatter_2015}. The differences in $H(q)$ are greatest at the peak, which is located at practically the same wavenumber $q_\text{m}$ as the principal peak of $S(q)$. There are pronounced undulations in $H(q)$ due to strong HIs between the microgels. In the absence of HIs, $H(q)=1$ independent of $q$. The peak height $H(q_\text{m})$ exceeds unity for $\phi_0=0.005$, a feature characteristic also of charge-stabilized suspensions at low salinity and low volume fractions $\phi$, where the hard core of the colloidal particles is masked by the strong and long-range electrostatic repulsion \cite{Westermeier_JCP_2012,Banchio_JCP_2018}.
\begin{figure}[h!]
\includegraphics[width=8.5cm]{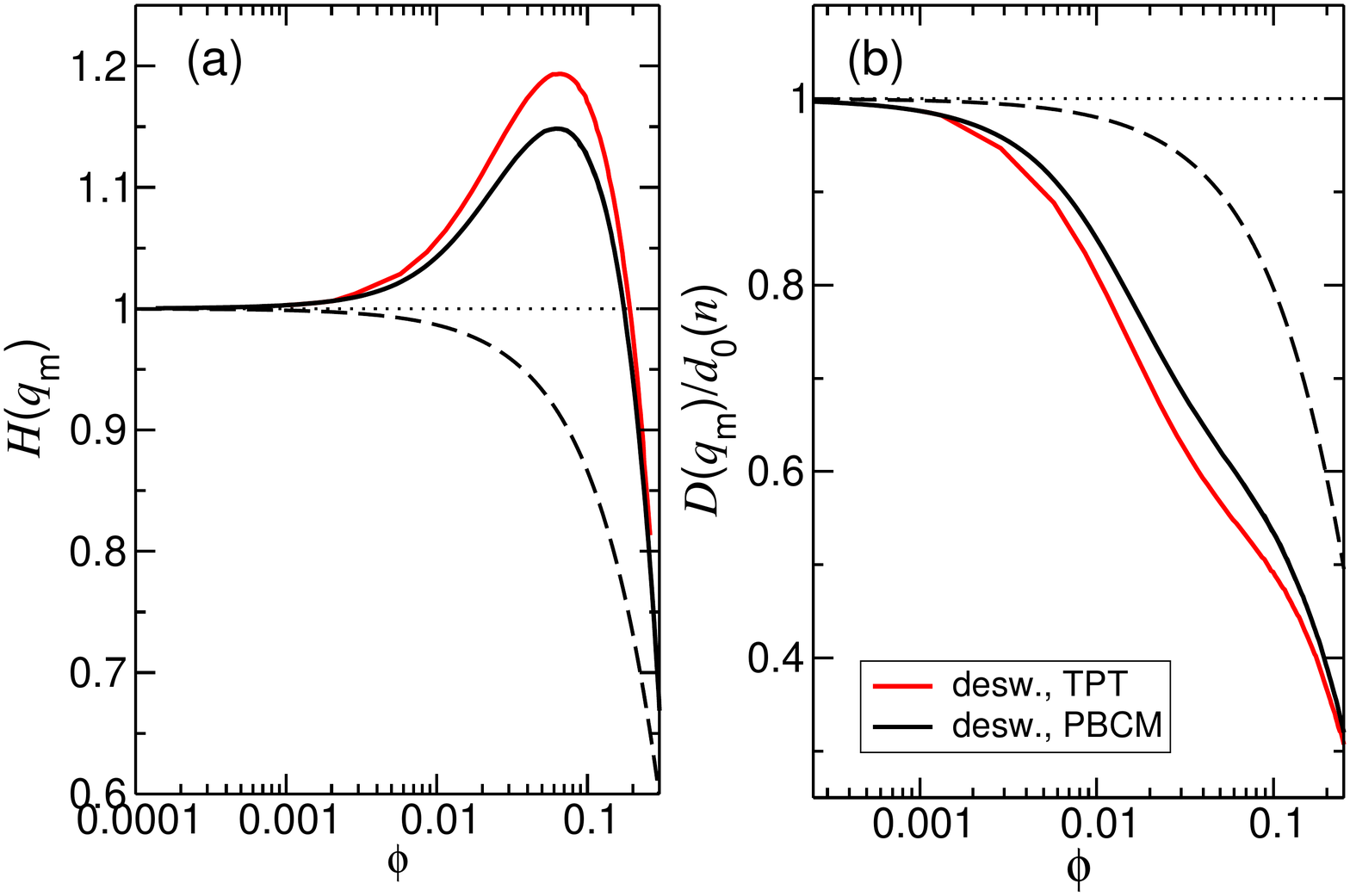}
\caption{BM-PA results for (a) hydrodynamic function peak height, $H(q_\text{m})$,
and (b) reduced cage diffusion coefficient, $D(q_\text{m})/d_0(n)$,  
as functions of volume fraction $\phi=\phi_0\;\!\alpha(n)^3$ using TPT (solid red curve) and PBCM (sold black curve) for $\alpha(n)$ and compared with 
hard-sphere results (dashed curves). 
System parameters: $Z=200$ and $c_\text{res}=100\;\mu$M.}
\label{extra1b}
\end{figure}

Also displayed in Fig.~\ref{extra1}(a) is the hydrodynamic function (with TPT input for $a$) of a more concentrated suspension at $\phi_0=0.02$, which corresponds in the TPT to the volume fraction $\phi=0.26$. The principal peak of $H(q)$ is here significantly below one. The reduced short-time self-diffusion coefficient, $d_\text{s}/d_0(a) = H(q\sigma\gg 1)$, is accordingly significantly lower than its value for $\phi_0=0.005$, which can be attributed to the enhanced hydrodynamic hindrance of self-diffusion for higher concentrations (cf. Eq.~(\ref{eq:HDist})). 

The differences in the $H(q)$'s of deswelling and reference microgels (solid and dashed curves, respectively, in Figs.~\ref{extra1} and \ref{extra1b}) are small, and basically due to the higher volume fraction of the reference system. This is also the reason for the slight downshift of the reference-system $H(q)$ relative to the one of the deswelling system. The microgel $H(q)$ bears a qualitative similarity to the one of colloidal hard spheres (hs) at the same volume fraction $\phi=0.26$, in particular regarding its peak value and location. The hydrodynamic function of hard spheres, $H^\text{hs}(q)$, is likewise characterized by a peak height below one, and the peak is located at $q_\text{m}\sigma \approx 2\pi$. 
TPT based explicit values are $H(q_\text{m})=0.81$ ($0.65$) for the peak height, $d_\text{s}/d_0=0.52$ ($0.51$) for the short-time self-diffusion coefficient, and $K=H(q\to 0)=0.11$ ($0.18$) for the sedimentation coefficient, where given in brackets are the respective values for colloidal hard spheres, 
obtained using the analytic expressions \cite{Riest_SoftMatter_2015,Pamvouxoglou_JCP_2019}
%
%\begin{eqnarray}\label{eq:hs}
%H^\text{hs}(q_\text{m})&=&1-\phi/\phi_\text{cp}=1- 1.35\;\!\phi\nonumber \\
%d^\text{hs}_\text{s}/d_0&=&1 %-1.8315\;\!\phi\left(1+0.12\;\!\phi-0.70\;\!\phi^2\right)\nonumber\\
%K^\text{hs}&=& 1 - 6.5464\;\!\phi\times\nonumber\\
%&& \!\!\!\!\!\!\!\!\!\left(1-3.348\phi +7.426\phi^2 -10.034\phi^3 %+5.882\phi^4 \right)\nonumber\\
%g^\text{hs}(r_\text{m}=2a^+)&=&\frac{1-0.5\;\!\phi}{\left(1-\phi\right)^3}\,,
%\end{eqnarray}
\begin{equation}
H^\text{hs}(q_\text{m})=1-\phi/\phi_\text{cp}=1-1.35\;\!\phi\nonumber
\end{equation}
\begin{equation}
d^\text{hs}_\text{s}/d_0=1-1.8315\;\!\phi\left(1+0.12\;\!\phi-0.70\;\!\phi^2\right)\nonumber
\end{equation}
\begin{eqnarray}
&&K^\text{hs}= 1 - 6.5464\;\!\phi\times\nonumber\\[1ex]
&&\left(1-3.348\phi +7.426\phi^2 -10.034\phi^3 +5.882\phi^4 \right)\nonumber
\end{eqnarray}
\begin{equation}\label{eq:hs}
g^\text{hs}(r_\text{m}=2a^+)=\frac{1-0.5\;\!\phi}{\left(1-\phi\right)^3}\,,
\end{equation}
which are accurate for volume fractions up to the hard-sphere freezing transition value $\phi=0.494$. Notice the strictly linear decline of $H^\text{hs}(q_\text{m})$ with increasing volume fraction, which holds to high  
accuracy for the complete liquid-phase concentration range. In the above expression for $H^\text{hs}(q_\text{m})$, $\phi_\text{cp}=\pi/\left(3\sqrt{2}\right)\approx 0.74$ is the highest possible volume fraction, attained for monodisperse hard spheres in close-packed fcc and hcp crystalline structures.

Eq.~(\ref{eq:hs}) quotes also the accurate Carnahan-Starling expression for the height, $g^\text{hs}(2a^+)$, of the principal peak of the hard-sphere rdf, located at the contact distance $r_\text{m}=2a^+$. The BM-PA values of $H(q)$ at $q=q_\text{m}$ and in the $q\to\infty$ limit are somewhat higher than the corresponding hard-sphere values. There are also differences between the microgel $g(r)$ and the hard-sphere $g^\text{hs}(r)$ (not shown here). The microgel rdf for $\phi_0=0.02$ has the peak height $g(r_\text{m})=2.50$ at pair distance $r_\text{m} = 1.32\sigma$, whereas $g^\text{hs}(\sigma^+)=2.15$. The differences from the hard-sphere values are due to the electrostatic repulsion between the microgels, which is here of shorter range $1/\kappa=0.4 a$. The pair potential contact value, $\beta u_\text{Y}(2a)\approx 18$, is still significantly higher, however, than the thermal energy $k_\text{B}T$ (see Fig.~\ref{figure7b}(b)), 
reflected in a nearly zero probability, $g(2a) < 10^{-3}$, of finding two microgels in contact.

Owing to HIs, two microparticles in contact sediment faster than at larger separations. This underlies  
the fact that the sedimentation coefficient $K$ for a homogeneous ionic microgel suspension is lower than the one 
for hard spheres at the same $\phi$. 
The monotonic decline of $K=V_\text{sed}/V_\text{sed}^0$ with increasing concentration is shown in Fig.~\ref{extra1}(b). Owing to stronger solvent backflow, the sedimentation velocity, $V_\text{sed}(\phi)$, is lower in a more concentrated suspension than in a less concentrated one. The maximal sedimentation velocity, $V_\text{sed}(\phi=0)=V_\text{sed}^0$, is thus attained at infinite dilution, where $K=1$. Since the major effect of deswelling is to lower $\phi$, $K$ is higher for deswelling microgels than for the constant-size reference particles, which explains the slightly higher values of $K$ in the PBCM, since $a_\text{PBCM}<a_\text{TPT}$.

As seen in Fig.~\ref{extra1b}(a), the $H(q_\text{m})$ of ionic microgels has a non-monotonic volume fraction dependence. Starting from a value of one at infinite dilution, with increasing $\phi$, $H(q_\text{m})$ increases towards its maximal value $\sim 1.2$ at $\phi\approx 0.07$ corresponding to $\phi_0\approx 0.005$, but thereafter declines monotonically, reaching values below one for $\phi \gtrsim 0.2$. This behavior should be contrasted with the strictly linear decrease of $H(q_\text{m})$ for hard spheres (curved, dashed line on the lin-log scale). In contrast to the non-monotonic $H(q_\text{m})$, both $K$ and $d_\text{s}/d_0$ (latter not shown here) decrease monotonically with increasing $\phi$. Furthermore, unlike the swollen radius $a$, which decreases with increasing $\phi_0$, 
the reduced Debye screening constant $\kappa a$ increases monotonically from $\kappa a \approx 1.24$ at $\phi=0.005$ to $\kappa a \approx 3.4$ at $\phi_0=0.05$.

For rigid charged particles interacting via a repulsive Yukawa-type potential, the order relations $H(q_\text{m};\phi)> H^\text{hs}(q_\text{m};\phi)$, $d_\text{s}(\phi)>d^\text{hs}_\text{s}(\phi)$, and $K(\phi)<K^\text{hs}(\phi)$ were previously demonstrated \cite{Heinen_JAC_2010,Gapinski_JCP2010}. These relations hold also for ionic microgels provided particle overlap is very unlikely, i.e., provided $g(2a) \approx 0$. 

For conditions not encountered in this paper, where overlap of microgels is likely and their softness matters, such as for low $Z$ or high salt content, the expected effect on $H(q)$ is a flattening of its oscillations at larger $q$, possibly to an extent that $H(q_\text{m}) \approx d_\text{s}/d_0$. Moreover, particle softness tends to enhance $K$, while $H(q_\text{m})$ is lowered. This behavior of $H(q)$ is observed indeed in a model system of particles interacting by the Hertz potential \cite{Riest_SoftMatter_2015}. Softness effects in non-ionic and weakly charged microgel systems will be the subject of a forthcoming study.

The short-time diffusion function, $D(q)$, measured in units of $d_0(n)$, is determined according to Eq.~(\ref{eq:dcShort}) by the ratio of the hydrodynamic factor $H(q)$ and $S(q)$, the latter being independent of HIs. 
The principal minimum of $D(q)$ is located, for repulsive interactions, at practically the same wavenumber $q_\text{m}$ at which $S(q)$ and $H(q)$ attain their respective maxima, with $S(q_\text{m})$ being in general distinctly higher than $H(q_\text{m})$. The so-called cage diffusion coefficient, $D(q_\text{m})$, quantifies the slow relaxation of concentration fluctuations of a wavelength $2\pi/q_\text{m}$ comparable with the diameter of the dynamic cage formed around each particle by its neighbors. For hard spheres, $D(q_\text{m})/d_0$ decreases monotonically with increasing $\phi$, which reflects a dynamical stiffening of the next-neighbor cage. The cage diffusion coefficient of hard spheres is quantitatively described, within $2\%$ of accuracy up to the freezing volume fraction, by the polynomial 
\begin{eqnarray}\label{eq:HSDqm}
\frac{D^\text{hs}(q_\text{m})}{d_0} = 1 - 2\;\!\phi -0.566\;\!\phi^2 + 2\;\!\phi^3\,,
\end{eqnarray}
according to which $D^\text{hs}(q_\text{m})$ follows closely a linear decline with slope $-2$ for volume fractions up to  $\phi\sim 0.3$. At freezing, where $H^\text{hs}(q_\text{m})\approx 0.33$ and $S^\text{hs}(q_\text{m})\approx 2.85$, $D^\text{hs}(q_\text{m})\approx 0.12\times d_0$.

The cage diffusion coefficient of ionic microgels is plotted in Fig.~\ref{extra1b}(b) as a function of $\phi$, where $D(q_\text{m})$ is normalized by the concentration-dependent single-microgel diffusion coefficient $d_0(n)$, allowing direct comparison with the reduced cage diffusion coefficient of hard spheres [Eq.~(\ref{eq:HSDqm})]. Unlike $H(q_\text{m})$, the reduced cage diffusion coefficient monotonically decreases with increasing $\phi$. The only remnant of the peak in $H(q_\text{m})$ is a shallow inflection point in $D(q_\text{m})/d_0(n)$ at $\phi \approx 0.07$. Owing to the electrostatic repulsion, the next-neighbor cage of microgels is more structured than that of hard spheres at the same $\phi$, reflected in an accordingly higher structure factor peak and lower cage diffusion coefficient. The distinctly higher values of $S(q_\text{m})$ in the TPT, in comparison to the PBCM, lead to lower values of $D(q_\text{m})/d_0(n)$ in the TPT, which explains the reverse order in the curves of $H(q_\text{m})$ and $D(q_\text{m})/d_0(n)$ in Figs.~\ref{extra1b}(a) and (b), respectively.  

\begin{figure}[h!]
\includegraphics[width=8.5cm]{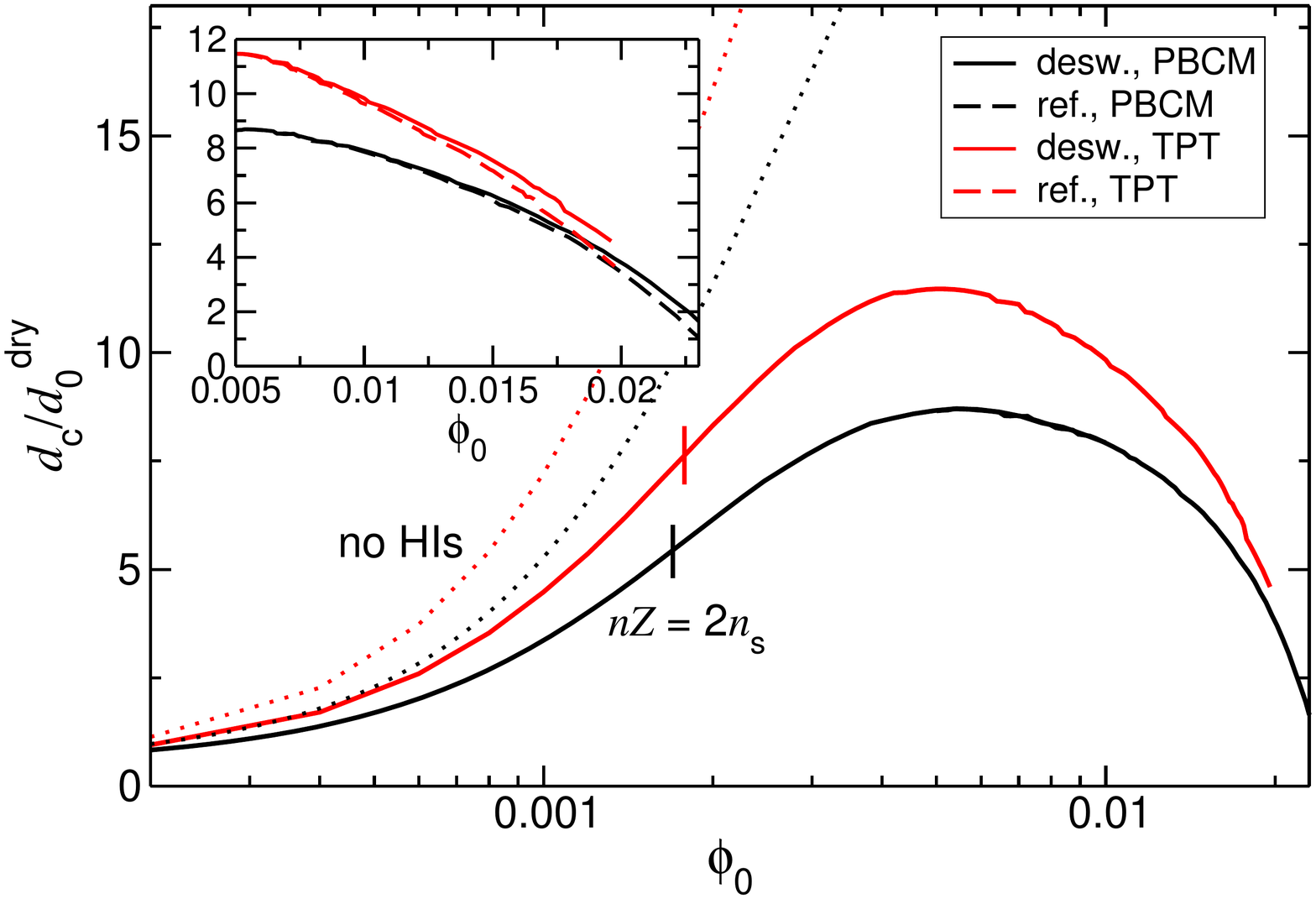}
\caption{BM-PA results for the reduced collective diffusion coefficient of deswelling microgels $d_\text{c}/d_0^\text{dry}$ versus $\phi_0$ (solid curves) for swollen radius $a$ calculated using TPT (red) and PBCM (black). Dotted curves are results without HIs where $H(0)=1$. Vertical line segments mark concentrations at which $\kappa_\text{c}=\kappa_\text{s}$ [cf. Eq.~(\ref{eq:KappaTwo})]. System parameters: 
$Z=200$ and $c_\text{res}=100\;\mu$M. Inset: Comparison with reference system results (dashed lines) for concentrations exceeding peak position value $\phi_0\approx 0.005$.}
\label{figure10}
\end{figure}

While $D(q)$ is minimal at $q_\text{m}$, it attains its maximum at $q=0$ where, according to Eq.~(\ref{eq:dcShort}), it has the physical meaning of a collective diffusion coefficient, denoted as $d_\text{c}=D(q\to 0)$. The maximum reflects the fast relaxation of long-wavelength concentration fluctuations by a collective diffusive motion of particles. In this context, recall that $H(q)$ and $S(q)$ are both minimal at $q=0$. At a given $\phi_0$, however, $S(0)$ appearing in the denominator of $d_\text{c}=d_0(n) K/S(0)$ is clearly below $K=H(0)$, as noticed from Figs.~\ref{figure8} and \ref{extra1}(b), with a consequentially high value of $d_\text{c}$. 
   
Figure~\ref{figure10} displays the $d_\text{c}$ for deswelling ionic microgels, obtained using the BM-PA method with respective TPT and PBCM input for $a$. To uncover its genuine concentration dependence, $d_\text{c}$ is divided, in lieu of $d_0(n)$, by the concentration independent single particle diffusion coefficient, $d_0^\text{dry}$, of collapsed microgels, implying that $d_\text{c}/d_0^\text{dry}\rightarrow a_0/a$ for $\phi_0 \to 0$. 

Akin to low-salinity suspensions of impermeable charge-stabilized particles, a non-monotonic concentration dependence of $d_\text{c}$ is observed, with a pronounced maximum of $d_\text{c}$ at $\phi_0 \approx 0.005$, i.e., at the same concentration where $H(q_\text{m})$ is greatest. The non-monotonic concentration dependence of $d_\text{c}$ is explained on noting first that $K$ and $S(0)$ are both monotonically decreasing with increasing $\phi_0$. At low $\phi_0$, the decrease of $S(0)$ with increasing $\phi_0$ is stronger than that of $K$, giving rise to a growing $d_\text{c}$. At higher concentrations, the slowing influence of HIs on $K$ is strong enough that the increase of $d_\text{c}$ is turned into a monotonic decline. 
To show explicitly that the maximum of $d_\text{c}$, and its decline at higher $\phi_0$, are due to HIs, results for $d_\text{c}$ without HIs are included in the figure for comparison. Without HIs, $K=1$ holds independent of concentration. The curves for $d_\text{c}(\phi_0)$ without HIs are monotonically increasing, and they converge to the ones with HIs at very low concentrations only. It is further noticed that the higher values of $d_\text{c}$ in the TPT are due to the lower osmotic compressibility values predicted by this method, and this even though $d_0(n)\propto1/a$ 
in the TPT is lower than in the PBCM. Quite interestingly, the concentration in Fig.~\ref{figure10} where the number of backbone-released counterions equals the number of salt counterions marks an inflection point, where the shape of the curve of $d_\text{c}(\phi_0)$ changes from convex to concave.         

The influence of deswelling on $d_\text{c}$ at higher $\phi_0$ is assessed in the inset of Fig.~\ref{figure10}, in comparison with the reference system predictions (dashed lines). Deswelling slightly enhances collective diffusion, as predicted by both the TPT and the PBCM. This enhancement can be attributed to weaker HIs between deswelling microgels, with a corresponding increase in $K$ overcompensating the increase in $S(0)$.   
	
A short discussion is in order regarding the BM-PA scheme results for $H(q)$ at wavenumbers $q \ll q_\text{m}$, where its accuracy is known to worsen with increasing concentration, up to a degree where non-physical negative values for $K$ are predicted \cite{Heinen_JAC_2010,Heinen_Rheo_JCP_2011}. This is mainly due to the self-diffusion contribution to $H(q)=H_\text{d}(q)+d_\text{s}/d_0$, which in the hybrid scheme is calculated using the pairwise additivity (PA) approximation. The PA method fully accounts for two-body HIs but neglects three-body and higher-order contributions. These complicated higher-order contributions account for the reduction in the strength of the HIs between two particles, due to a hydrodynamic shielding by intervening particles. The disregard of this hydrodynamic shielding effect by the PA scheme leads at higher concentrations to an underestimation of $d_\text{s}$. The latter 
contributes to $H(q)$ most significantly at $q=0$ where the distinct part, $H_\text{d}(0)$, is negative. 
For this reason, we show BM-PA results for $K=H(0)$ and $d_\text{c}\propto K$ for concentrations up to $\phi_0=0.02$ only where the small-$q$ BM-PA predictions are trustworthy.

%	because the diffusion of the system is enhanced due to the effectively repulsive interaction between the particles. It reaches a maximum, and later monotonically decreases. This maximum is associated to the transition between salt-dominated and counterion-dominated regimes. The corresponding transition concentrations have been highlighted in the figure for each method. We observe that at that concentration the curves change convexity. Even though the theoretical models present limitations for $\phi_0<0.005$, similarly to charge-stabilized systems, both methods depict the same maximum behavior. In the studied concentration range for ionic microgels, $d_c$ decreases with increasing concentration. TPT exhibit both larger equilibrium size and larger $Z_\text{net}$ than PBCM. Since TPT predicts mainly larger $d_c$ than PBCM, we see that the mutual effective repulsion interaction is the prevailing interaction, which dominates the diffusion behavior in this time scale. The influence of deswelling on $d_c$ is shown in the inset of Fig. \ref{figure10}. By the introduction of the reference system, we observe that deswelling enhances the diffusion and this effect is predicted by methods. $d_c$ is given by the ratio $K/\chi_\text{osm}$, which are two quantities modified in the same way by the deswelling. At given concentration, $\chi_\text{osm}$ and $K$ increase for deswelling microgels, but the impact of the HI dominates. At larger concentrations, the hydrodynamic interactions ...
%
%
\begin{figure}[h!]
\includegraphics[width=8.5cm]{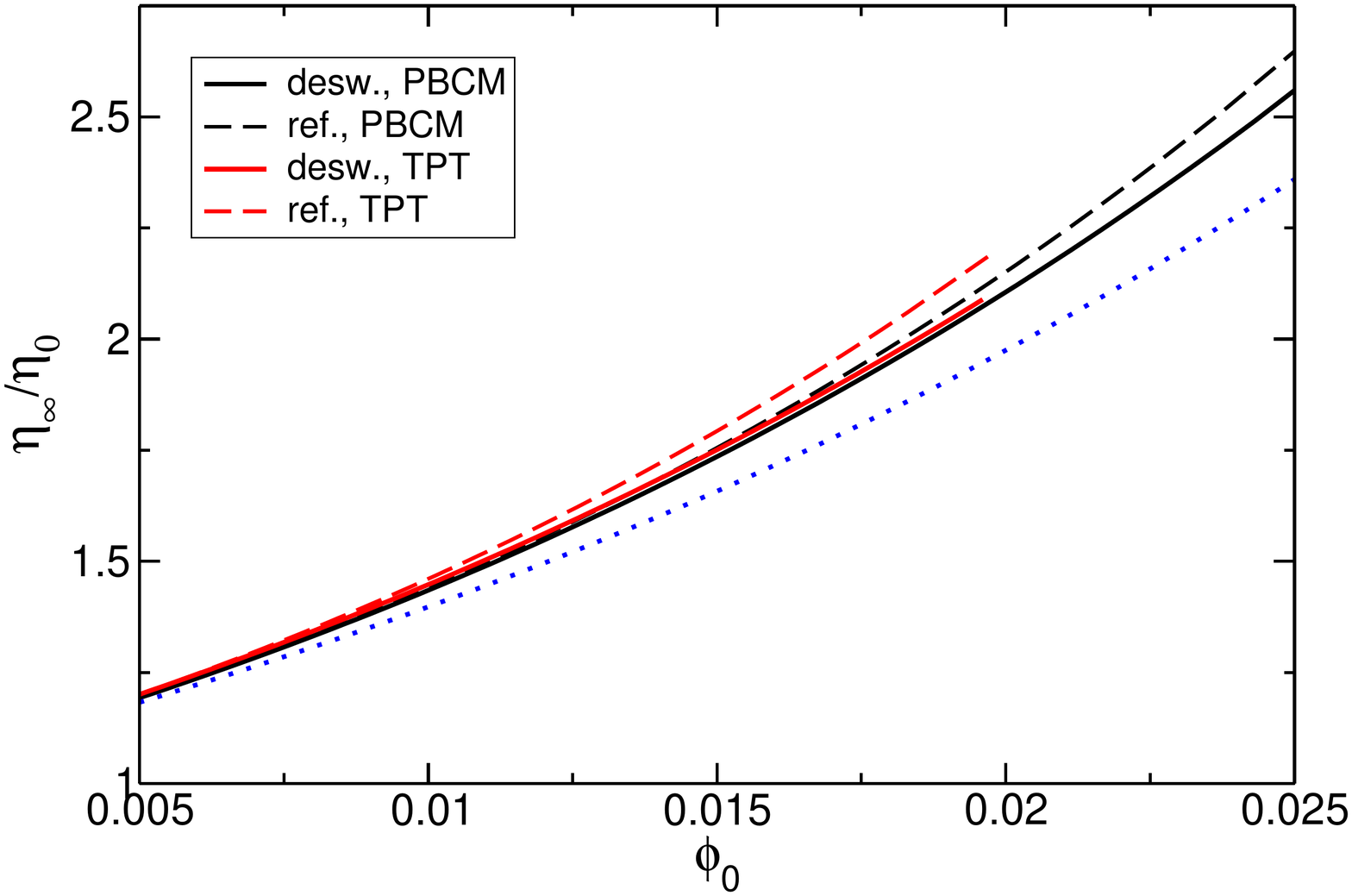}
\caption{Modified BM theory results for the reduced high-frequency viscosity, $\eta_\infty/\eta_0$, as function of $\phi_0$, for interaction parameters and $a$ calculated using TPT and PBCM. Solid curves are 
for deswelling microgels, while dashed curves are for the reference system. Dotted curve is the prediction by  Eq.~(\ref{visc_approx}) using $\phi =\phi_0\alpha^3(\phi_0)$, with $\alpha(\phi_0)$ calculated in the  
PBCM. System parameters: $Z=200$ and $c_\text{res}=100\;\mu$M.}
\label{figure11}
\end{figure}

Having discussed (short-time) diffusion properties of ionic microgel suspensions, we finally consider rheological properties, namely the high-frequency (short-time) viscosity $\eta_\infty$
and the zero-frequency viscosity $\eta$ introduced in Eqs.~(\ref{eq:EtaInfty}) and (\ref{eq:ViscoSum}), respectively. Our analysis is limited here to weakly sheared suspensions, where nonlinear phenomena such as shear thinning and the buildup of normal stress differences are negligible. Just as for the diffusion properties, we identify 
the hydrodynamic particle radius with $a$. As described in Subsec.~\ref{Subsec:Methods}, $\eta_\infty$ is calculated  using the modified Beenakker-Mazur (BM) expression in Eq.~(\ref{eq:BM-MOD}). The shear stress relaxation contribution  $\Delta\eta$ in $\eta=\eta_\infty + \Delta\eta$ is calculated using the simplified mode-coupling theory (MCT) expression in Eq.~(\ref{eq:MCT}). The only input to these methods is $S(q)$, which is calculated in RY approximation based on $u_\text{eff}(r;n)$, with $a$ obtained in the TPT and PBCM, respectively. 
HIs are incorporated into the simplified MCT expression via $H(q)$, determined using the BM-PA method. 

Figure~\ref{figure11} presents results for $\eta_\infty$ (in units of the solvent viscosity $\eta_0$) as a function of $\phi_0$. With increasing concentration, $\eta_\infty$ grows gradually, to a value at $\phi_0=0.03$ only three times higher than the solvent viscosity. Such a modest growth with increasing concentration is a characteristic feature of $\eta_\infty$. In addition, $\eta_\infty$ is known to be rather insensitive to the form of the pair potential \cite{Banchio_JCP_2008,Heinen_Rheo_JCP_2011} and hence to changes in the equilibrium radius $a$, as reflected in the nearly coincident curves for $\eta_\infty$ with $a$ obtained from the TPT and PBCM methods. At a given concentration, the reference microgel suspension has a higher volume fraction than the deswelling microgels system, which explains the mildly higher viscosity values. 

For comparison, we show the prediction for $\eta_\infty$ from the polynomial expression, 
\begin{equation}\label{visc_approx}
\frac{\eta_\infty}{\eta_0}\approx 1+\frac{5}{2}\phi(1+\phi)+7.9\phi^3\,,
\end{equation}
derived in \cite{Banchio_JCP_2008}. This expression is a good viscosity approximation for dilute suspensions of strongly repelling charge-stabilized spheres with prevailing two-body HIs and low values of $S(0)$. 
As shown in Fig.~\ref{figure11}, Eq.~(\ref{visc_approx}) is in qualitative accord with the modified BM results, but underestimates $\eta_\infty$ 
at higher concentrations. Note that Eq.~(\ref{visc_approx}), though not a virial expansion to third order in $\phi$, reduces to the linear Einstein viscosity formula for very low volume fractions where the particles are uncorrelated, and thus $\Delta\eta=0$. For the hypothetical case of vanishing HIs, the particles remain uncorrelated on short time scales for all fluid-phase volume fractions. In this case, $\eta_\infty/\eta_0 = 1 + [\eta]\;\!\phi$  holds for all $\phi$, with $[\eta]=5/2$ for no-slip spheres.
\begin{figure}[h!]
\includegraphics[width=8.5cm]{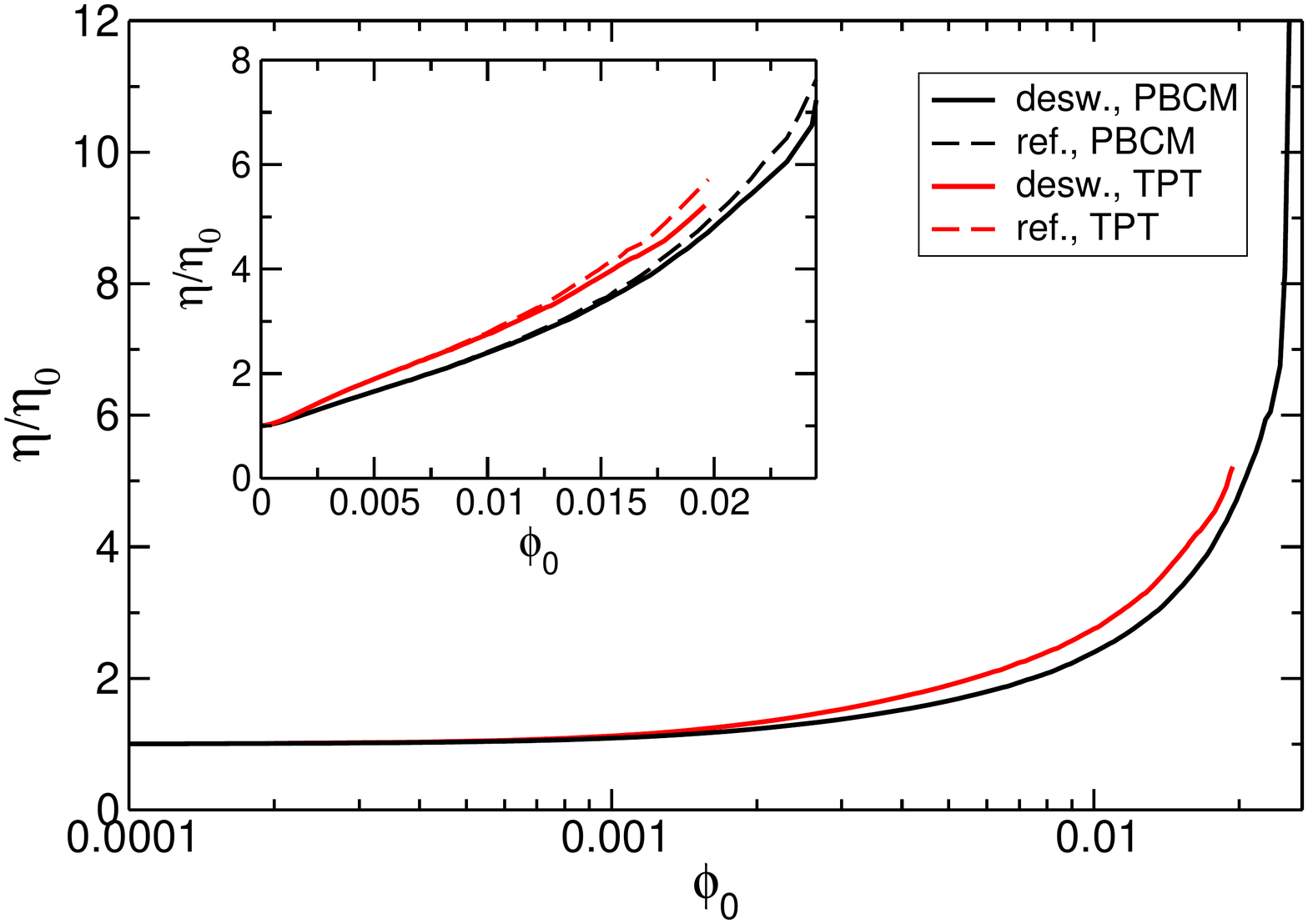}
\caption{Reduced zero-frequency viscosity, $\eta/\eta_0$, versus $\phi_0$ for system parameters 
$Z=200$ and $c_\text{res}=100\;\mu$M. The viscosity contribution $\eta_\infty$ is calculated using modified BM theory, and the shear stress relaxation contribution $\Delta\eta$ using simplified MCT. Inset: Comparison with reference system viscosity (dashed curves).}
\label{extra3}
\end{figure}

The reduced zero-frequency viscosity, $\eta/\eta_0$, of deswelling microgels is plotted in Fig.~\ref{extra3} (solid lines). The pronounced increase of $\eta$ at higher $\phi_0$ is mainly due to the shear stress relaxation part $\Delta\eta$. The latter is more sensitive to changes in the pair potential than $\eta_\infty$, as reflected in visibly higher values of $\eta$, for $\phi_0 \gtrsim 0.03$, when the TPT radius input is used. The higher volume fractions of the reference system in comparison to the system of deswelling microgels imply a lower zero-frequency viscosity for the deswelling particles, visible in the inset at higher concentrations.

%	In Fig. \ref{figure12}, we show the results for the zero-frequency viscosity versus concentration. 
%	
%	\begin{figure}[h!]
%		\includegraphics[width=8.5cm]{Fig11/Fig_extra3.pdf}
%		\caption{Reduced zero-frequency viscosity, $\eta/\eta_0$, versus dry volume fraction. Inset: Comparison against reference system. Dashed lines correspond to reference system for PBCM (black) and TPT (red). Other system parameters as Fig. \ref{figure4}.}
%		\label{figure12}
%	\end{figure}
%	
%	\subsection{Rescaling properties}
%	It is interesting to study the rescaling suggested by Prof. Magdaleno Medina-Noyola. He states that it is possible to map different soft -pair- potentials to an effective hard sphere potential. With that, it is possible to obtain master curves in terms of the new rescaled effective parameters for the long time transport properties such as low frequency viscosity $\eta/\eta_\infty=1+\Delta\eta_\text{MCT}/\eta_\infty$ and long-time collective diffusion coefficient. We can test whether this scaling also works for the case of Denton potential for microgels, since it works for Yukawa.
%	
%	This rescaling does not work for short-time transport properties, because the hydrodynamic interactions are relevant in this regime.

\section{Conclusions}
\label{Sec:Conclusions}

We have presented a comprehensive theoretical study of the influence of concentration on deswelling, thermodynamic, structural, and dynamic properties of suspensions of weakly cross-linked, ionic microgels dispersed in a good solvent and in osmotic equilibrium with an electrolyte reservoir.
To calculate microion density profiles, single-particle and bulk osmotic pressures, and state-dependent, equilibrium microgel swelling ratios, we implemented two mean-field methods and assessed their respective pros and cons. We consistently combined these methods -- a thermodynamic perturbation theory and a Poisson-Boltzmann spherical cell model -- with calculations of the net microion valence $Z_\text{net}$ and Debye screening constant $\kappa$, characterizing the electrostatic part of the effective one-component microgel pair potential derived from linear-response theory. On the basis of the effective one-component model of microion-dressed microgels, we determined static structural properties, including $S(q)$ and $g(r)$, by Monte-Carlo simulation and the self-consistent Rogers-Young integral-equation method, and used these properties as input to the calculation of dynamic suspension properties, with the salient hydrodynamic interactions included.

At salt concentrations high enough that salt ions contribute significantly to electrostatic screening, the microion distribution inside and outside the microgels is relatively uniform and counterion-induced deswelling is consequently weak. Therefore, our study focused on the counterion-dominated regime, with salt and microgel concentrations low enough, and microgel valences high enough, that deswelling is pronounced even without significant particle overlap.

The TPT method neglects nonlinear electrostatic effects, but accounts for inter-microgel correlations. In contrast, the PBCM method accounts for nonlinear screening by mobile microions, but neglects inter-microgel correlations, except for the remnant concentration dependence of the cell radius. Unlike impermeable surface-charged colloidal particles, ionic microgels are characterized by electrostatic interactions whose strength, as measured at mutual contact, increases with decreasing microgel concentration. This property restricts the applicability of the TPT method to non-vanishing microgel concentrations.

While both methods predict the same trends for the effective microgel pair potential, there are quantitative differences in the swelling ratio, net valence, and the potential value at contact, whose values are in general higher in the TPT than in the PBCM. In the counterion-dominated regime, the range $1/\kappa$ of the electrostatic repulsion is equal in both methods. 
The greatest differences in the pair potential parameters occur at very low concentrations and high backbone valences, which can be partially attributed to the linear-response approximation inherent in the TPT. The relative variation in the microgel radius with changing concentration is less pronounced in the PBCM, in which nonlinear response confines the counterions more strongly to the microgel interior. 

Differences in predictions of the TPT and PBCM methods are more pronounced for static (thermodynamic and structural) properties than for dynamic properties, which can be explained by the fact that dynamic properties depend only globally (i.e., functionally) on $S(q)$. The only exception is the collective diffusion coefficient $d_\text{c}$ which is directly proportional to the inverse of the static compressibility factor $1/S(0)$. 

Owing to the dominance of the electrostatic interactions in the considered microgel systems, their dynamic behavior resembles that of charged-stabilized suspensions of impermeable solid particles. In particular, the peak, $H(q_\text{m})$, of the hydrodynamic function has a non-monotonic concentration dependence, with a maximum higher than one at an intermediate concentration value, reflected in a concomitant inflection point of the cage diffusion coefficient. The collective diffusion coefficient, $d_\text{c}$, behaves likewise non-monotonically and has its maximum at the same concentration as $H(q_\text{m})$. This maximum was shown to arise from the slowing effect of HIs, which becomes more influential with increasing concentration. The electric repulsion between the microgels distinctly enhances the zero-frequency viscosity at higher concentrations, as compared to suspensions of uncharged particles. 

The comparison with corresponding results for the reference system of constant-sized microgels revealed that the major influence of deswelling on structural and dynamic properties is via the reduced volume fraction $\phi$, which grows only sublinearly with increasing concentration $n\propto \phi_0$. 
The effect of counterion-induced deswelling on structural and dynamic properties is overall quite weak, for valences where nonlinear electrostatic contributions to the microion distributions are negligible, 
and changes of $\alpha=a(\phi_0)/a_0$ with concentration are accordingly small.
 
At higher concentrations, deswelling slightly enhances $S(0)$ and the hydrodynamic function $H(q)$ for all wavenumbers $q$. Deswelling reduces the zero- and high-frequency viscosity and slightly enhances collective diffusion. From the behavior of $d_\text{c}=d_0 H(0)/S(0)$, one notices that the deswelling-induced enhancement of $d_0\propto 1/a$ is nearly counterbalanced by the accompanying de-enhancement of $H(0)/S(0)$. 

The most pronounced effect of deswelling is to shift the freezing (crystallization) transition to higher concentration values, as we have determined from an empirical freezing rule for the static structure factor peak height. This concentration shift is more pronounced for strongly charged microgels, in which case the nonlinear PBCM method can still be used to determine the swelling ratio $\alpha$. To determine the concentration shift, however, the PBCM must be combined with a charge renormalization procedure 
to determine renormalized values of the microgel net valence and screening constant from a linearized Poisson-Boltzmann equation in the cell model. The renormalized parameters are used in the linear-response pair potential [Eq.~(\ref{Yukawa_part})], where they summarily account for the enhanced accumulation of counterions inside and close to the spherical backbone region. We illustrated such a renormalization procedure using a linearization of the nonlinear PB equation around the potential at the cell boundary. While such a linearization is most commonly used in renormalization schemes applied to non-permeable and permeable colloidal particles, it is not the only choice. There are sound reasons to use instead a linearization around the mean (i.e., cell-volume-averaged) electrostatic potential value \cite{Trizac_Langmuir:2003,Deserno_Gruenberg:2002}. A proper assessment of the pros and cons of different charge renormalization schemes, formulated also in the framework of TPT, was not in the scope of the present work, but is the subject of a forthcoming paper \cite{Brito_tosubmit_2019}.      

Finally, in the presented generic study, we considered only uniformly cross-linked microgels, modeled by a uniformly distributed backbone charge. In future work, extensions to nonuniformly charged microgels can be explored where, e.g., the backbone charge is concentrated near the particle periphery. Such a nonuniform charge distribution is expected to significantly affect deswelling and the strength of the effective pair potential, and consequently also structural and dynamic properties.

\section*{Acknowledgments}%%%%%%%%%%%%%%%%%%%%%%%%%%%%%%%%%%%%%%%%%%%%%%%%%%%%%%%%%%%%%%%%%%%%%%%%%%%%%%%%%%%%%%%%%%%%%%%%%%%%%%%%%%%%%%%%%%%
M.B. and G.N. thank J. Riest (Viega Holding GmbH \& Co. KG, Attendorn, Germany) and G.-W. Park (FZ J{\"u}lich, Germany) for many helpful discussions. 
This work was under appropriation of funds from the Deutsche Forschungsgemeinschaft (SFB 985, project B6).

% Bibliography with titles:
%\bibliographystyle{ieeetr}
% Bibliography without titles:
\bibliographystyle{phcpc}
\bibliography{microgel_biblio}

\begin{thebibliography}{10}

\bibitem{Book_Fernandez_Nieves:2011}
Fern{\'{a}}ndez-Nieves, A., Wyss, H., Mattsson, J., and Weitz, D.~A., editors,
\newblock {\em Microgel Suspensions: Fundamentals and Applications}, Wiley-VCH
  Verlag GmbH \& Co. KGaA, Weinheim, 2011.

\bibitem{Lyon_Fernandez_Nieves:2012}
Lyon, L.~A. and Fern{\'{a}}ndez-Nieves, A.,
\newblock Annu. Rev. Phys. Chem. {\bf 63} (2012) 2.1.

\bibitem{Plamper_Richtering:2017}
Plamper, F.~A. and Richtering, W.,
\newblock Acc. Chem. Res. {\bf 50} (2017) 131.

\bibitem{Holmqvist_PRL_2012}
Holmqvist, P., Mohanty, P.~S., N{\"{a}}gele, G., Schurtenberger, P., and
  Heinen, M.,
\newblock Physical Review Letters {\bf 109} (2012) 1.

\bibitem{Nojd_SoftMatter_2018}
N{\"{o}}jd, S. et~al.,
\newblock Soft Matter {\bf 14} (2018) 4150.

\bibitem{Gasser_PRE_2019}
Gasser, U., Scotti, A., and Fernandez-Nieves, A.,
\newblock Phys. Rev. E {\bf 99} (2019) 042602.

\bibitem{weitz-jcp2012}
Romeo, G., Imperiali, L., Kim, J.-W., Fern{\'a}ndez-Nieves, A., and Weitz,
  D.~A.,
\newblock J. Chem. Phys. {\bf 136} (2012) 124905.

\bibitem{Colla_JCP_2014}
Colla, T., Likos, C.~N., and Levin, Y.,
\newblock J. Chem. Phys. {\bf 141} (2014) 234902.

\bibitem{DentonTang_JCP_2016}
Denton, A.~R. and Tang, Q.,
\newblock J. Chem. Phys. {\bf 145} (2016) 164901.

\bibitem{Weyer_SoftMatter_2018}
Weyer, T.~J. and Denton, A.~R.,
\newblock Soft Matter {\bf 14} (2018) 4530.

\bibitem{Hofzumhaus_SoftMatter_2018}
Hofzumahaus, C., Hebbeker, P., and Schneider, S.,
\newblock Soft Matter {\bf 14} (2018) 4087.

\bibitem{Urich_SoftMatter_2016}
Urich, M. and Denton, A.~R.,
\newblock Soft Matter {\bf 12} (2016) 9086.

\bibitem{Mohanty:2017}
Mohanty, P.~S. et~al.,
\newblock Scientific Reports {\bf 7} (2017) 1487.

\bibitem{Nir_SoftMatter_2016}
Nir, O., Trieu, T., Bannwarth, S., and Wessling, M.,
\newblock Soft Matter {\bf 12} (2016) 6512.

\bibitem{Roa_SoftMatter_2015}
Roa, R., Zholkovskiy, E.~K., and N{\"{a}}gele, G.,
\newblock Soft Matter {\bf 11} (2015) 4106.

\bibitem{Roa_SoftMatter_2016}
Roa, R. et~al.,
\newblock Soft Matter {\bf 12} (2016) 4638.

\bibitem{Park:2019}
Park, G.~W. and N{\"{a}}gele, G.,
\newblock to be submitted  (2019).

\bibitem{Denton_PRE_2003}
Denton, A.~R.,
\newblock Phys. Rev. E {\bf 67} (2003) 011804.

\bibitem{Hedrick_JCP_2015}
Hedrick, M.~M., Chung, J.~K., and Denton, A.~R.,
\newblock J. Chem. Phys. {\bf 142} (2015) 034904.

\bibitem{flory-rehner1943-I}
Flory, P.~J. and Rehner, J.,
\newblock J. Chem. Phys. {\bf 11} (1943) 512.

\bibitem{flory-rehner1943-II}
Flory, P.~J. and Rehner, J.,
\newblock J. Chem. Phys. {\bf 11} (1943) 521.

\bibitem{FloryBook}
Flory, P.~J.,
\newblock {\em Principles of Polymer Chemistry},
\newblock Cornell University Press, 1953.

\bibitem{Gottwald_JCP_2005}
Gottwald, D., Likos, C.~N., Kahl, G., and L{\"{o}}wen, H.,
\newblock J. Appl. Cryst. {\bf 122} (2005) 074903.

\bibitem{Riest_ZPhysChem_2012}
Riest, J., Mohanty, P.~S., Schurtenberger, P., and Likos, C.~N.,
\newblock Z. Phys. Chem. {\bf 226} (2012) 711.

\bibitem{Hansen-McDonald}
Hansen, J.-P. and McDonald, I.~R.,
\newblock {\em Theory of Simple Liquids},
\newblock Elsevier, 2013.

\bibitem{LandauLifschitzElasticity}
Landau, L.~D. and Lifschitz, E.~M.,
\newblock {\em Theory of Elasticity},
\newblock Elsevier, 1986.

\bibitem{Rovigatti_HertzModel2019}
Rovigatti, L., Gnan, N., Ninarello, A., and Zaccarelli, E.,
\newblock Macromolecules {\bf 52} (2019) 4895.

\bibitem{Louis2002}
Louis, A.~A.,
\newblock J. Phys.: Condens. Matter {\bf 14} (2002) 9187.

\bibitem{Hoffmann_JCP_2004}
Hoffmann, N., Likos, C.~N., and L{\"{o}}wen, H.,
\newblock J. Chem. Phys. {\bf 121} (2004) 7009.

\bibitem{Kirkwood_JCP_1951}
Kirkwood, J.~G. and Buff, F.~P.,
\newblock J. Chem. Phys. {\bf 19} (1951) 774.

\bibitem{Dobnikar2006}
Dobnikar, J., Castaneda-Priego, R., {Von Gr{\"{u}}nberg}, H.~H., and Trizac,
  E.,
\newblock New Journal of Physics {\bf 8} (2006).

\bibitem{Naegele_PhysRep_1996}
N{\"{a}}gele, G.,
\newblock Phys. Rep. {\bf 272} (1996) 215.

\bibitem{Henderson_PhysLettA_1974}
Henderson, R.~L.,
\newblock Phys. Lett. A {\bf 49} (1974) 197.

\bibitem{Henderson-PY:2009}
Henderson, D.,
\newblock Condensed Matter Physics {\bf 12} (2009) 127.

\bibitem{Denton_PRE_2006}
Denton, A.~R.,
\newblock Phys. Rev. E {\bf 73} (2006) 041407.

\bibitem{CollaLevinTrizacJCP2009}
Colla, T.~E., Levin, Y., and Trizac, E.,
\newblock J. Chem. Phys. {\bf 131} (2009) 074115.

\bibitem{Wennerstrom1982}
Wennerstrom, H., Jonsson, B., and Linse, P.,
\newblock J. Chem. Phys. {\bf 76} (1982) 4665.

\bibitem{Brito_tosubmit_2019}
Brito, M.~E., Riest, J., Denton, A.~R., and N{\"{a}}gele, G.,
\newblock Critical assessment of methods for calculating effective interactions
  and pressure in charge-stabilized dispersions,
\newblock to be submitted.

\bibitem{Banchio_JCP_2008}
Banchio, A.~J. and N{\"{a}}gele, G.,
\newblock J. Chem. Phys. {\bf 128} (2008) 104903.

\bibitem{Banchio_JCP_2018}
Banchio, A.~J., Heinen, M., Holmqvist, P., and N{\"{a}}gele, G.,
\newblock J. Chem. Phys. {\bf 148} (2018) 134902.

\bibitem{Naegele_Varenna2013}
N{\"{a}}gele, G.,
\newblock Colloidal hydrodynamics,
\newblock in {\em Physics of Complex Colloids}, edited by Bechinger, C.,
  Sciortino, F., and Ziherl, P., Proceedings of the International School of
  Physics "Enrico Fermi", 2013.

\bibitem{Riest_SoftMatter_2015}
Riest, J., Eckert, T., Richtering, W., and N{\"{a}}gele, G.,
\newblock Soft Matter {\bf 11} (2015) 2821.

\bibitem{Pamvouxoglou_JCP_2019}
Pamvouxoglou, A., Bogri, P., N{\"{a}}gele, G., Ohno, K., and Petekidis, G.,
\newblock J. Chem. Phys. {\bf 151} (2019) 024901.

\bibitem{Loewen_PRL_1993}
L{\"{o}}wen, H., Palberg, T., and Simon, R.,
\newblock Phys. Rev. Lett. {\bf 70} (1993) 1557.

\bibitem{Naegele_MolecPhys2002}
N{\"{a}}gele, G., Kollmann, M., Pesch{\'{e}}, R., and Banchio, A.~J.,
\newblock Molecular Physics {\bf 100} (2002) 2921.

\bibitem{AbadeVisc_JCP_2010}
Abade, G.~C., Cichocki, B., Ekiel-Jezewska, M.~L., N{\"{a}}gele, G., and
  Wajnryb, E.,
\newblock J. Chem. Phys. {\bf 133} (2010) 084906.

\bibitem{Szymczak_JStatMech_2008}
Szymczak, P. and Cichocki, B.,
\newblock J. Stat. Mech. {\bf 2008} (2008) P01025.

\bibitem{Naegele_Visco:1998}
N{\"{a}}gele, G. and Bergenholtz, J.,
\newblock J. Chem. Phys. {\bf 108} (1998) 9893.

\bibitem{Heinen_Rheo_JCP_2011}
Heinen, M., Banchio, A.~J., and N{\"{a}}gele, G.,
\newblock J. Chem. Phys. {\bf 135} (2011) 154504.

\bibitem{Das_SoftMatter_2018}
Das, S. et~al.,
\newblock Soft Matter {\bf 14} (2018) 92.

\bibitem{Braibanti_PRE_2016}
Braibanti, M., Haro-P{\'{e}}rez, C., Quesada-P{\'{e}}rez, M., Rojas-Ochoa,
  L.~F., and Trappe, V.,
\newblock Phys. Rev. E {\bf 94} (2016) 1.

\bibitem{Denton_JPCM_2008}
Denton, A.~R.,
\newblock J. Phys.: Condens. Matter {\bf 20} (2008) 494230.

\bibitem{Denton_JPCM_2010}
Denton, A.~R.,
\newblock J. Phys.: Condens. Matter {\bf 22} (2010) 364108.

\bibitem{Trizac2003}
Trizac, E., Bocquet, L., Aubouy, M., and {Von Gr{\"{u}}nberg}, H.~H.,
\newblock Langmuir {\bf 19} (2003) 4027.

\bibitem{Pianegonda2007}
Pianegonda, S., Trizac, E., and Levin, Y.,
\newblock J. Chem. Phys. {\bf 126} (2007) 014702.

\bibitem{Boon_PNAS_2015}
Boon, N., Guerrero-Garc{\'{i}}a, G.~I., van Roij, R., and {Olvera de la Cruz},
  M.,
\newblock Proceedings of the National Academy of Sciences {\bf 112} (2015)
  9242.

\bibitem{Baulin_SoftMatter_2012}
Baulin, V.~A. and Trizac, E.,
\newblock Soft Matter {\bf 8} (2012) 6755.

\bibitem{PhysRevE.100.032602}
Aguirre-Manzo, L.~A. et~al.,
\newblock Phys. Rev. E {\bf 100} (2019) 032602.

\bibitem{Gapinski_JCP2012}
Gapinski, J., N{\"{a}}gele, G., and Patkowski, A.,
\newblock J. Chem. Phys. {\bf 136} (2012) 024507.

\bibitem{Gapinski_JCP_2014}
Gapinski, J., N{\"{a}}gele, G., and Patkowski, A.,
\newblock J. Chem. Phys. {\bf 141} (2014) 124505.

\bibitem{Levin:2002}
Levin, Y., Diehl, A., Fern{\'{a}}ndez-Nieves, A., and Fern{\'{a}}ndez-Barbero,
  A.,
\newblock Phys. Rev. E {\bf 65} (2002) 036143.

\bibitem{Alexander1984}
Alexander, S. et~al.,
\newblock J. Chem. Phys. {\bf 80} (1984) 5776.

\bibitem{Trizac_Langmuir:2003}
Trizac, E., Bocquet, L., Aubouy, M., and {Von Gr{\"{u}}nberg}, H.~H.,
\newblock Langmuir {\bf 19} (2003) 4027.

\bibitem{Westermeier_JCP_2012}
Westermeier, F. et~al.,
\newblock J. Chem. Phys. {\bf 137} (2012) 114504.

\bibitem{Heinen_JAC_2010}
Heinen, M., Holmqvist, P., Banchio, A.~J., and N{\"{a}}gele, G.,
\newblock J. Appl. Cryst. {\bf 43} (2010) 970.

\bibitem{Gapinski_JCP2010}
Gapinski, J., Patkowski, A., and N{\"{a}}gele, G.,
\newblock J. Chem. Phys. {\bf 132} (2010) 054510.

\bibitem{Deserno_Gruenberg:2002}
Deserno, M. and von Gr\"unberg, H.-H.,
\newblock Phys. Rev. E {\bf 66} (2002) 011401.

\end{thebibliography}
\end{document}